\documentclass[12pt]{elsarticle}
\usepackage[usenames,dvips]{color}
\usepackage{enumitem}
\usepackage{amsmath,amsthm,latexsym,amssymb,ifthen,alltt,array}
\usepackage{moreverb}
\usepackage{epsfig}
\usepackage[all]{xy}
\usepackage{fancybox}
\usepackage{url}

\newtheorem*{example*}{Example}
\newtheorem{theorem}{Theorem}[section]

\newtheorem{proposition}[theorem]{Proposition}
\newtheorem{example}[theorem]{Example}
\newtheorem{corollary}[theorem]{Corollary}
\newtheorem{lemma}[theorem]{Lemma}
\newtheorem{fact}[theorem]{Fact}
\newtheorem{definition}[theorem]{Definition}
\newtheorem{remark}[theorem]{Remark}
\newenvironment{sketch}[0]{\noindent{\em Sketch of proof}.}{\hfill $\Box$ \medskip \noindent}

\newenvironment{itemize0}{
\begin{itemize}
\vspace{-0.75em}
\setlength{\itemsep}{0cm}
}
{\end{itemize}}

\newenvironment{enumerate0}{
\begin{enumerate}
\vspace{-0.75em}
\setlength{\itemsep}{0cm}
}
{\end{enumerate}}

\newcounter{teller}

\def\N{\mathbb{N}}

\newcommand{\myand}{\;\&\;}

\newcommand{\Card}{\msf{Card}}
\newcommand{\msf}[1]{\mathsf{#1}}
\newcommand{\mca}[1]{\mathcal{#1}}
\renewcommand{\AA}{\mca{A}}
\newcommand{\BB}{\mca{B}}
\newcommand{\CC}{\mca{C}}
\newcommand{\DD}{\mca{D}}

\newcommand{\FF}{\mca{F}}
\newcommand{\FUN}{\mca{F}}
\newcommand{\GG}{\mca{G}}

\newcommand{\PL}{\mca{PL}}
\newcommand{\PP}{\mca{P}}
\newcommand{\PR}{\mathsf{PR}}
\newcommand{\RR}{\mca{R}}

\renewcommand{\SS}{\mca{S}}
\newcommand{\TT}{\mca{T}}

\newcommand{\VV}{\mca{V}}

\newcommand{\TERM}{\TT(\FF,\VV)}

\newcommand{\GTERM}{\TT(\FF)}
\newcommand{\QTERM}{\TT(\FF \cup Q)}

\newcommand{\Language}{\msf{L}}
\newcommand{\last}{\msf{last}}

\newcommand{\Var}{\VV\msf{ar}}
\newcommand{\fr}{\msf{fr}}

\newcommand{\Tr}{\msf{Tr}}

\newcommand{\arit}{\msf{ar}}

\newcommand{\LFPO}{\msf{LFPO}}

\newcommand{\bu}{\msf{bu}}
\newcommand{\BU}{\msf{BU}}
\newcommand{\ID}{\msf{ID}}
\newcommand{\Path}{\msf{PATH}}
\newcommand{\sbu}{\msf{sbu}}
\newcommand{\SBU}{\msf{SBU}}
\newcommand{\MUP}{\msf{MUP}}

\newcommand{\wbu}{\msf{wbu}}

\newcommand{\barre}[1]{\overline{{#1}}}
\newcommand{\dbarre}[1]{\barre{\barre{{#1}}}}
\newcommand{\m}{\msf{m}}
\newcommand{\mar}[2]{{#1}^{#2}}
\newcommand{\ma}[1]{\msf{m}(#1)}
\newcommand{\mamin}[1]{\msf{mmin}(#1)}
\newcommand{\mamax}[1]{\mmax(#1)}

\newcommand{\mamaxu}[2]{\msf{mmax}^{\prec #1}(#2)}

\newcommand{\tm}{\msf{tm}}

\newcommand{\A}{\msf{A}}
\newcommand{\M}{\msf{M}}
\newcommand{\SG}{\msf{SG}}

\newcommand{\FFm}{\mar{\FF}{\N}}
\newcommand{\TTm}{\TT(\FFm)}
\newcommand{\TTqm}{\TT((\FF \cup Q))^\N}
\newcommand{\TTkm}{\TT(\FF^{\leq k})}
\newcommand{\TTqkm}{\TT((\FF \cup Q)^{\leq k})}
\newcommand{\TTqkmb}{\TT((\FF \cup Q)^{\leq k},\{\Box\})}
\newcommand{\TTvm}{\TT(\FFm,\VV)}

\newcommand{\torl}{~_{\msf rl}\!\!\to}
\newcommand{\tok}{~_k\!\!\to}

\newcommand{\toro}{~\circ\!\!\!\to}
\newcommand{\toroak}{~_{k+1}\!\!\toro}
\newcommand{\torok}{~_k\!\!\toro}

\newcommand{\Red}{\msf{Red}}
\newcommand{\Top}{\msf{Top}}
\newcommand{\Topd}{\msf{Topd}}

\newcommand{\seq}[2][n]{{#2_1},\dots,{#2_{#1}}}

\newcommand{\pos}{\msf{pos}}
\newcommand{\Pos}{\PP\msf{os}}
\newcommand{\Internal}{{\cal I}\msf{n}}
\newcommand{\Leaves}{{\cal L}\msf{v}}
\newcommand{\Posp}{\Pos^+}
\newcommand{\ov}{\overline{\VV}}
\newcommand{\Posv}{\Pos_{\VV}}

\newcommand{\Posnv}{\Pos_{\ov}}

\def\sys{\Gamma}

\def\nopos{\varepsilon}

\newcommand{\fta}{$f.t.a$ }
\newcommand{\nfta}{$f.t.a$ }
\newcommand{\nfa}{$n.f.a$ }

\newcommand{\CL}{\msf{CL}}

\newcommand{\dmax}{\msf{d}}

\newcommand{\SLHS}{\msf{SLHS}}

\newcommand{\Max}{\msf{Max}}
\newcommand{\Min}{\msf{Min}}
\newcommand{\mmax}{\msf{mmax}}

\def\sto#1#2{#1/#2}

\newcommand{\mA}{\msf{a}}

\newcommand{\mB}{\msf{b}}

\newcommand{\mF}{\msf{f}}
\newcommand{\mG}{\msf{g}}

\newcommand{\mH}{\msf{h}}
\newcommand{\mI}{\msf{i}}

\newcommand{\depth}{{\it dpt}}

\newcommand{\rrule}[2]{#1 \to #2}

\journal{Journal of Symbolic Computation}

\begin{document}

\begin{frontmatter}
\title{Bottom-up rewriting for words and terms}
\author{I. Durand and G. S{\'e}nizergues}
\ead{\{idurand,ges\}@labri.fr}
\address{LaBRI, Bordeaux 1 university, 351 cours de la lib\'eration,
33405, Talence, Cedex, France}
\begin{abstract}
For the whole class of linear term rewriting systems, we define \emph{bottom-up rewriting} which is a restriction of the usual notion of rewriting.  
We show that bottom-up rewriting  effectively inverse-preserves recognizability
and analyze the complexity of the underlying construction.\\
The \emph{Bottom-Up} class ($\BU$) is, by definition, the set of linear systems for 
which every derivation can be replaced by a bottom-up derivation.
Membership to $\BU$ turns out to be undecidable, we are thus lead to
define more restricted classes: the classes $\SBU(k), k \in \N$ of 
\emph{Strongly Bottom-Up$(k)$} systems for which we show that
membership is decidable. 
We define the class of \emph{Strongly Bottom-Up} systems by $\SBU = \bigcup_{k \in \N} \SBU(k)$.
We give a polynomial sufficient condition for a system to be in $\SBU$.
The class $\SBU$ contains (strictly) several classes of systems which were already known to inverse 
preserve recognizability:
the inverse left-basic semi-Thue systems (viewed as unary term rewriting systems),
the linear growing term rewriting systems, the inverse Linear-Finite-Path-Ordering systems.

\end{abstract}
\end{frontmatter}

\section{Introduction}
\paragraph{General framework}
An important concept in rewriting is the notion
of \emph{preservation of recognizability} through rewriting.
Each identification of a more general class of systems preserving
recognizability, yields almost directly a new decidable 
call-by-need~\cite{id-am:ic-2005}
class, decidability results for confluence, accessibility,
joinability.
Also, recently, this notion has been used to prove
termination of systems for which none of the already known termination
techniques work \cite{GHWZ05}.
Such a preservation property is also a tool for studying the recognizable/rational subsets
of various monoids which are defined by a presentation $\langle X,\RR\rangle$,
where $X$ is a finite alphabet and $\RR$ a Thue system 
(see for example \cite{LohSen05rat,Kam-Sil-Ste06}).
Consequently, the seek of new decidable classes of systems which preserve
(or inverse preserve) recognizability is worthwile.

Many such classes defined so far have been defined by imposing syntactical
restrictions on the rewrite rules. 
For instance, in \emph{growing} systems (\cite{J96,NT02})
variables at depth strictly greater than $1$ in the left-handside of a rule
cannot appear in the corresponding right-handside.
Finite-path Overlapping systems \cite{Tak-Kaj-Sek10} are also defined by syntactic restrictions
on the system.
The class of Finite-path Overlapping systems contains the class of growing systems 
\cite{NT02}.
Previous works on semi-Thue systems also prove recognizability preservation,
under syntactic restrictions: cancellation systems \cite{Ben-Sak86}, monadic systems \cite{BJW82}, basic systems \cite{Ben87}, and left-basic systems \cite{Sak79} (see \cite{Sen95} for a survey). 

Other works establish that some {\em strategies} i.e. restrictions 
on the derivations rather than on the rules, ensure preservation of recognizability. 
Various such strategies were studied in
\cite{Ful-Jur-Ste-Vag98}, \cite{Ret},\cite{Sey-Tis-Tom}.


We rather follow here this second approach: we define a new rewriting strategy which we call 
\emph{bottom-up rewriting} 
for linear term rewriting systems. The bottom-up derivations are,
intuitively, those derivations in which the rules are applied, roughly speaking, 
from the bottom of the term
towards the top (this set of derivations contains strictly the bottom-up derivations of \cite{Ret}
and the one-pass leaf-started derivations of \cite{Ful-Jur-Ste-Vag98}).
An important feature of this strategy, as opposed to the ones quoted above,  
is that it allows {\em overlaps} between successive applications of rules.
A class of systems is naturally associated with this strategy: it consists of the systems
$\RR$ for which the binary relation $\to^*_\RR$ coincides with its restriction
to the bottom-up strategy. We call ``bottom-up'' such systems and denote by $\BU$ the set of
all bottom-up systems.

\paragraph{Overview of the paper}
Most of the results proved in this paper were announced in \cite{Dur-Sen07}, which can thus be 
considered as a medium-scale overview of this paper. Let us give here a large-scale 
overview, section by section, of the contents of the paper.\\
In section \ref{preliminaries}, we have gathered all the necessary recalls and notation
about words, terms, rewriting and automata.\\
In Section~\ref{s-bottom-up-rewriting}, we define \emph{bottom-up rewriting} 
for linear term rewriting systems using marking techniques.
We first define \emph{bottom-up$(k)$} derivations for $k \in \N$ ($\bu(k)$ derivations for short)
and the classes Bottom-up$(k)$ ($\BU(k)$ for short)
of linear systems which consists of those systems
which admit $\bu(k)$ rewriting, i.e. such that every derivation between two terms can be replaced by
a $\bu(k)$ derivation, and the \emph{Bottom-up} class ($\BU$) of \emph{bottom-up} systems
which is the infinite union of the $\BU(k)$ (for $k$ varying in $\N$).

In Section~\ref{s-preservation}, we prove Theorem \ref{t-kbu-inverse-preserving} 
which is the main result of 
the paper: bottom-up rewriting inverse-preserves recognizability. 
Our proof consists of a reduction to the preservation of recognizability by
finite ground systems, shown in \cite{Bra69},\cite{Dau-Tis90}. 
The proof is constructive \emph{i.e} gives an algorithm for computing
an automaton recognizing the antecedents of a recognizable set of terms.
We estimate the complexity of the algorithm:
a separate tight upper-bound is given for $\BU^-(1)$ semi-Thue systems; another upper-bound is given
for $\BU^-(1)$ term rewriting systems; finally, a general upper-bound is given for $\BU^-(k)$ term rewriting systems. We then give a lower bound for $\BU^-(1)$ term rewriting systems showing that
some of our upper-bounds cannot be easily improved.\\ 
In Section~\ref{s-testing},
we show that $\BU$ contains all the classes of semi-Thue systems quoted above (once translated into
term rewriting systems in which all symbols have arity $0$ or $1$), and also the linear growing systems
of \cite{J96}. 
 We study the decidability of membership to
the $\BU(k)$ classes. We show that membership to $\BU(k)$ is undecidable for $k \geq 1$
even for semi-Thue systems.

In Section~\ref{s-strongly-bottom-up}, 
we define the restricted class of \emph{strongly bottom-up$(k)$} systems ($\SBU(k)$) for which we show
decidable membership. We define the class of \emph{strongly bottom-up} systems $\SBU = \bigcup_{k\in \N} \SBU(k)$.
Based on the results of \cite{Kna-Cal99}, it seems likely that 
the property [$\exists k \geq 0$ such that $\RR \in \SBU(k)$] (so membership to $\SBU$)
is undecidable.
We give a polynomial sufficient condition for a system to be in $\SBU$. 
We finally show that $\LFPO^{-1} \subsetneq \SBU$.

\tableofcontents

\section{Preliminaries}
\label{preliminaries}
This section is mostly devoted to recalling some classical notions and making precise our notation.
The reader is referred to \cite{TATA} for more details on the subject of tree-automata and to 
\cite{K92} for term rewriting.
\subsection{Sets, binary relations}
\label{sets_binrel}
\paragraph{Abstract rewriting}
Given a set $E$, we denote by ${\cal P}(E)$ its powerset i.e. the set of all its subsets.
For every sets $E,F$ and every binary relation $\rightarrow \subseteq E \times F$, and every subsets $E'\subseteq E, F' \subseteq F$,
we denote by $E' \rightarrow F'$ the fact that $\exists e \in E', \exists f \in F', e \rightarrow f$.
We sometimes abusively note $e \rightarrow F$ for what should be written 
$\{e\}\rightarrow F$.\\
The inverse binary relation $\rightarrow^{-1}$ is defined by
$$\forall f\in F,\forall e \in E, f \rightarrow^{-1} e 
\Leftrightarrow e \rightarrow f.$$
We note $\rightarrow^{0}= {\rm Id}_E, \rightarrow^{1}= \rightarrow$, for every $n \geq 1$
$\rightarrow^{n+1}=\rightarrow \circ \rightarrow^{n}$ and finally:
$$\rightarrow^{*}:= \bigcup_{n=0}^\infty \rightarrow^{n}.$$ 
The relation $\rightarrow^{*}$ is the reflexive and transitive closure of
the binary relation $\rightarrow$. 
A  finite {\em derivation} w.r.t. the relation $\to$, is a sequence
\begin{equation} 
D=(t_0 ,t_1,\cdots t_i, t_{i+1}, \cdots t_n)
\label{def_derivation}
\end{equation}
such that, for every $i \in [0,n-1]$, $t_i \to t_{i+1}$.\\

Given a subset $T \subseteq E$,
we define
\begin{equation}
(\to^*)[T] = \{ s \in E\mid s \to^* t \text{ for some~} t \in T \}
\label{e-ancestors-of-T}
\end{equation}
and
\begin{equation}
[T](\to^*) = \{ s \in E \mid t \to^* s \text{ for some~} t \in T \}
\label{e-descendants-of-T}
\end{equation}
\paragraph{Simulation}
Let $E,F$ be two sets endowed with binary relations
$\rightarrow_E \subseteq E \times E,\rightarrow_F \subseteq F \times F$. 
\begin{definition}
A binary relation
$R \subseteq E \times F$ is called a \emph{simulation} of the structure $(E,\rightarrow_E)$ by the structure
$(F,\rightarrow_F)$ iff \\
$
\forall e_1,e_2 \in E,\forall f_1 \in F, 
[((e_1 \rightarrow_E e_2) \wedge (e_1 R f_1)) \Rightarrow 
(\exists f_2 \in F,(f_1 \rightarrow_F f_2) \wedge (e_2 R f_2))]
$
\label{d-simulation}
\end{definition}
\vspace{-1em}
This is essentially the classical notion of simulation defined in \cite{Par81},
excepted that we do not impose on $R$ to be everywhere defined.

\subsection{Words and Terms}
A finite {\em word} over an alphabet $A$ is a map $u:[0,\ell-1] \rightarrow A$ for some 
$\ell \in \N$. The integer $\ell$ is the {\em length} of the word $u$ and is denoted by $|u|$.
The set of words over $A$ is  denoted by $A^*$ and endowed with the usual {\em concatenation}
operation $u,v \in A^* \mapsto u \cdot v \in A^*$. The {\em empty} word is denoted by $\varepsilon$.
A word $u$ is a prefix of a word $v$ iff there exists some $w \in A^*$ such that $v = uw$.
We denote by $u \preceq v$ the fact that $u$ is a prefix of $v$ and by $u \perp v$
the fact that $u,v$ are incomparable for the ordering $\preceq$ i.e. 
$$u \perp v \;\Leftrightarrow\;\neg (u \preceq v) \myand \neg (v \preceq u).$$
The incomparability relation is extended to sets of words by:
for every $P,Q \subseteq A^*$,
$$P \perp Q \;\Leftrightarrow\; [ \forall u \in P,\forall v \in Q, u \perp v] .$$
Given a total order on $A$,
we denote  by $u \leq_{lex} v$ the fact
that $u$ is lexicographically smaller than ( or equal to ) $v$.

Given $w \in A^* \setminus \{\varepsilon\}$,
we denote by $\last(w)$ the last (i.e. rightmost) letter of $w$.

We call \emph{signature} a set of symbols $\FF$ with fixed arity $\arit: \FF \rightarrow \N$.
The subset of symbols of arity $m$ is denoted by $\FF_m$.

As usual, a set $P \subseteq \N^*$ is called a {\em tree-domain} (or, domain, for short)
iff, for every $u \in \N^*, i \in \N$
$$(u \cdot i \in P \Rightarrow u \in P) \;\& \; (u \cdot (i+1) \in P \Rightarrow u \cdot i \in P).$$
We call $P' \subseteq P$ a {\em subdomain} of $P$ iff, $P'$ is a domain and,
for every $u \in P, i \in \N$
$$(u \cdot i \in P'\;\& \; u \cdot (i+1) \in P) \Rightarrow u \cdot (i+1) \in P'.$$
Given $Q \subseteq P$, the closure of $Q$ in the tree-domain $P$, denoted $\CL(Q,P)$, is the
smallest superset of $Q$ which is a subdomain of $P$.
A {\em chain} of a tree-domain $P$ is a  subset $C \subseteq P$ which is linearly ordered by $\preceq$.
A subset $P' \subseteq P$ is called a {\em path} of $P$ iff
it is a chain, which is an interval i.e.: $\forall x,z \in P',\forall y \in P,
x \leq y \leq z \Rightarrow y \in P'$.
A subset $B \subseteq P$ is called a {\em branch} of $P$ iff
it is a chain, which is maximal for inclusion (note that every branch is also a path).
An {\em antichain} of $P$ is a  subset $X \subseteq P$ such that, for every $u,u' \in X$,
$ u \preceq u' \Rightarrow u=u'$. 
We often denote a finite antichain by the sequence of its elements in increasing lexicographic
order. We sometimes do not distinguish between the antichain and this sequence. 
A subset $T \subseteq P$ is called a {\em transversal} of $P$ iff
it is an antichain, which is maximal for inclusion.
The ordering $\preceq$ is extended to transversals in the following way:
$T \preceq T'$ iff, $\forall u \in T, \exists u' \in T', u \preceq u'$.

\begin{lemma}
Let $P \subseteq \N^*$ be a tree domain. Let $Y \subseteq P$ be an antichain.
There exists a transversal $Z$ of $P$ such that\\
1- $Y \subseteq Z$\\
2- for every transversal $T$ of $P$, $Y \subseteq T \Rightarrow Z \preceq T$\\
3- $\forall z \in Z, \forall v \in P,\exists y \in Y, (v \prec z \Rightarrow v \prec y)$.
\label{l-smallest-transversal} 
\end{lemma}
\begin{sketch}
Let
$$Z := \{ z \in P\mid \exists y \in Y, \exists u \in P, \exists \alpha \in \N, u \prec y \myand z= u\cdot \alpha \myand (\forall y' \in Y, z \not\prec y')\}.$$
This set $Z$ fulfills points {\it (1)(2)(3)}.
\end{sketch}$\;\;$\\
After Lemma \ref{l-smallest-transversal}, we denote by $\Tr(Y,P)$ the transversal $Z$ 
determined by $Y$ and $P$ and we call it the {\em smallest transversal} containing the antichain $Y$ of the tree-domain $P$.\\
Given $Q \subseteq P \subseteq \N^*$ we call {\em frontier } of $Q$ in $P$, the set
$$\fr(Q,P):= \{ u\cdot i \mid u \in Q, i \in \N, u \cdot i \in P\}.$$
A (first-order) {\em term} on a signature $\FF$ is a partial map $t: \N^* \rightarrow \FF$
whose domain is a tree-domain and which respects the arities.
We denote by $\TERM$ the set of first-order terms built upon
the signature $\FF \cup \VV$, where $\FF$ is a denumerable signature and $\VV$ is a  denumerable 
set of variables of arity $0$. 

The domain of $t$ is also called its set of {\em positions} and denoted by $\Pos(t)$.
The set of variable positions (resp. non variable positions) of a term $t$ is denoted by $\Posv(t)$ 
(resp. $\Posnv(t)$). The set of {\em leaves} of $t$ is the set of positions $u \in \Pos(t)$
such that $u \cdot \N \cap \Pos(t)= \emptyset$. It is denoted by $\Leaves(t)$.
The set of {\em internal nodes} of $t$ is the set of positions $u \in \Pos(t)$
such that $u\cdot \N \cap \Pos(t)\neq \emptyset$. It is denoted by $\Internal(t)$.
We write $\Posp(t)$ for $\Pos(t) \setminus \{ \nopos \}$.
If $u, v \in \Pos(t)$ and $u \preceq v$, we say that $u$ is an \emph{ancestor} of $v$
in $t$.
Given $v \in \Posp(t)$, its \emph{father} is the position $u$ such that $v=uw$ and $|w|=1$.
The \emph{depth} of a term $t$ is defined by: 
$$\depth(t):= \sup\{ |u| \mid u \in \Posnv(t) \}+1.$$
Given a term $t$ and $u \in \Pos(t)$
the \emph{subterm of $t$ at $u$} is denoted by $t/u$ and defined by
$\Pos(t/u) = \{ w \mid uw \in \Pos(t) \}$ and $\forall w \in \Pos(t/u)$, $t/u(w) = t(uw)$.
A term $s$ is a {\em prefix} of the term $t$ iff there exists a substitution $\sigma$ such that
$s \sigma= t$.
A term containing no variable is called \emph{ground}.
The set of ground terms is abbreviated to $\TT(\FF)$ or $\TT$ whenever
$\FF$ is understood.
A term which does not contain twice the same variable is called \emph{linear}.
Given a linear term $t \in \TERM$, 
$x \in \Var(t)$, we shall denote by $\pos(t,x)$ the position of $x$ in $t$.\\
Among all the variables, there is a special one designated by $\Box$.
A term containing exactly one occurrence of $\Box$ is called a {\em context}.
We denote by ${\cal C}_1(\FF)$ the set of all contexts over $\FF$.\\
A context is usually denoted as $C[]$. If $u$ is the position of $\Box$ in
$C[]$,  $C[t]$ denotes
the term $C[]$ where $t$ has been substituted at position $u$.
We also denote by $C[]_u$ such a context and by $C[t]_u$ the result of the substitution.
Intuitively, the symbol $\Box$ denotes a ``hole'' in $C$, while $C[t]$ denotes  what is obtained
by plugging the term $t$ in the hole of $C[]$.
A term $s$ is a {\em factor} of the term $t$ iff there exists a context $C[]_u$ and a substitution
$\sigma$ such that $t = C[s \sigma]_u$.
In this case, we call {\em occurrence} of $s$ in $t$ the subset of $\Pos(t)$ which corresponds to the non-variable positions of $s$ i.e. $u\Posnv(s)$. Note that the only occurrence of a variable
$v \in \VV$ in a term $t$, is $\emptyset$ (by the above definition).
Note also that the frontier of the occurrence of $s$ in $t$ is, by definition, 
$\fr(u\Posnv(s),\Pos(t))$: it is equal to $u\Posv(s)$ i.e. to the ``positions 
of the variable of $s$ inside t'' (but these variables need not label these positions 
in the term $t$).
We denote by $|t|:=\Card(\Pos(t))$ the {\em size} of a term $t$.\\
Two terms $t,t' \in \TERM$ are called $\alpha$-equivalent iff, there exists a substitution
$\sigma: \VV \rightarrow \VV$ which is a permutation of the set $\VV$, and such that
$t\sigma=t'$. In this case we note $t \equiv_\alpha t'$.
\subsection{Semi-Thue systems}
Let $A$ be a set that we take as alphabet. A \emph{rewrite rule} over the alphabet $A$ is a pair $\rrule{u}{v}$ of words in $A^*$. We call $u$ (resp. $v$) the \emph{left-handside} (resp. \emph{right-handside}) of the rule (\emph{lhs} and \emph{rhs} for short).
A \emph{semi-Thue system} is a pair $(S,A)$
where $A$ is an alphabet and $S$ a set of rewrite rules built upon
the alphabet $A$. 
When $A$ is clear from the context or contains exactly the symbols occurring in $S$, 
we may omit $A$ and write simply $S$.
We call {\em size} of the set of rules $S$ the number
$\|S\|:= \sum_{\rrule{u}{v} \in S} |u| + |v|$.
The one-step derivation generated by $S$
(which is denoted by $\to_S$) is  defined by: 
for every  $f, g \in A^*, f \to_S  g$ iff there  exists $\rrule{u}{v} \in S$ and
$\alpha,\beta \in A^*$ such that $f = \alpha u \beta$ and $g = \alpha v \beta.$
The relation $\to_S^*$  (defined in section \ref{sets_binrel}) is also called the 
{\em derivation generated by S}.

The semi-Thue system $(S,A)$ is called {\em length-increasing} (resp. strict) iff, for every $\rrule{u}{v} \in S, |u| \leq |v|$ (resp. $|u| < |v|$).
\subsection{Term rewriting systems}
\label{ss-trs}
A \emph{rewrite rule} built over the signature $\FF$ is a pair $\rrule{l}{r}$ of terms in $\TERM$
which satisfy $\Var(r) \subseteq \Var(l)$.
We call $l$ (resp. $r$) the \emph{left-handside} (resp. \emph{right-handside}) 
of the rule (\emph{lhs} and \emph{rhs} for short).
A rule is \emph{ground} if both its left and right-handsides are ground.
A rule is \emph{linear} if both its left and right-handsides are linear.
A rule is \emph{left-linear} if its left-handside is linear.

A \emph{term rewriting system} (\emph{system} for short) 
is a pair 
$(\RR,\FF)$
where $\FF$ is a signature and $\RR$ a set of rewrite rules built upon
the signature $\FF$. 
When $\FF$ is clear from the context or contains exactly the symbols of $\RR$, 
we may omit $\FF$ and write simply $\RR$.
We call {\em size} of the set of rules $\RR$ the number
$\|\RR\|:= \sum_{\rrule{l}{r} \in \RR} |l| + |r|$.
We define the maximum arity of $\RR$ as the number
$$\A(\RR):= \max\{\Card(\Pos_\VV(l)) \mid \rrule{l}{r} \in \RR\}.$$
A system is \emph{ground} (resp. \emph{linear}, \emph{left-linear}) if each of its
rules is ground (resp. linear, left-linear). A system $\RR$
is \emph{shallow}~\cite{God-Tiw-Ver03} if, in  every side of rule, variables can occur only at 
depth $0$ or $1$. 
A system $\RR$ is \emph{growing}~\cite{J96} if every variable of a right-handside is
at depth at most $1$ in the corresponding left-handside.
Rewriting is defined as usual: for every $t,t' \in \TERM$, $t \to_\RR t'$ means that
there exists $C \in {\cal C}_1(\FF \cup \VV), l\to r \in \RR, \sigma: \VV \rightarrow \TERM$
such that
\begin{equation}
t = C[l\sigma]_u,\;\;t' = C[r\sigma]_u.
\label{e-one-step-RR}
\end{equation}
For this step: $\rrule{l}{r}$ is the rule used, $l\sigma$ is the {\em redex} and $r\sigma$ is the
{\em contractum}.
Let us fix some one-step derivation (\ref{e-one-step-RR}) and denote by $u$ the position of the
hole in $C$. Let $v \in \Pos(t), v' \in \Pos(t')$.
We call $v'$ a {\em residue} of $v$, w.r.t. the one-step derivation (\ref{e-one-step-RR}), iff\\
there exists $x \in \Var(l) \cap \Var(r),w \in \Pos(x \sigma), v_1 \in \Pos(l),
v'_1 \in \Pos(r)$ such that
\begin{equation}
l(v_1)=x,r(v'_1)=x, v= uv_1w, v' = uv'_1w,
\label{moved_by_substitution}
\end{equation}
or 
\begin{equation}
v=v', v \in \Pos(C) \mbox{ and } v \perp  u.
\label{independant_of_substitution}
\end{equation}
This notion extends to sets of positions in the following way:
a subset $P' \subseteq \Pos(t')$ is a {\em residue} of a subset $P \subseteq \Pos(t)$, 
w.r.t. the one-step derivation (\ref{e-one-step-RR}), iff\\
there exists $x \in \Var(l) \cap \Var(r), Q \subseteq \Pos(x \sigma), v_1 \in \Pos(l),
v'_1 \in \Pos(r)$ such that
\begin{equation}
l(v_1)=x,r(v'_1)=x, P= uv_1Q, P'= uv'_1Q.
\label{set_moved_by_substitution}
\end{equation}
or 
\begin{equation}
P=P' \subseteq \Pos(C)\mbox{ and } \{u\} \perp P .
\label{set_independant_of_substitution}
\end{equation}
Given a derivation
$$D: t_0 \to_\RR t_1 \to_\RR \cdots t_i \to_\RR t_{i+1} \to_\RR \cdots
t_n$$ and $v \in \Pos(t_0), v' \in \Pos(t_n)$, we call $v'$ a {\em residue
of $v$, w.r.t. derivation $D$}, iff, there exists positions
$v_i \in \Pos(t_i)$ for $0 \leq i \leq n$ such that, $v=v_0$, for
every $i \in [0,n-1]$, $v_{i+1}$ is a residue of $v_i$ w.r.t. the
$i$-th step of derivation $D$ and $v_n=v'$.

\begin{remark}
\label{rem-on-derivations}
The notion of derivation used here consists, in fact, not merely in a sequence of terms 
(as is defined in \S \ref{def_derivation}) but in a sequence of rewriting steps, each of them
being defined by a rule and a position.
\end{remark}  

Similarly, if $P \subseteq \Pos(t_0)$ 
and $P' \subseteq \Pos(t_n)$, we call $P'$ a residue of $P$, w.r.t. derivation
$D$, iff, there exist subsets $P_i \subseteq \Pos(t_i)$ for $0 \leq i \leq n$ such that, 
$P=P_0$, for every $i \in
[0,n-1]$, $P_{i+1}$ is a residue of $P_i$ w.r.t. the $i$-th step of
derivation $D$ and $P_n=P'$.
When $P \subseteq \Pos(t_0)$ is an occurrence of a term $s$
we say that the subterm $t_n/v'$ is a residue of the subterm $t_0/v$ w.r.t. $D$.
A notion of {\em descendant of $v$, w.r.t. derivation $D$} is obtained by removing the 
incomparability restriction in condition (\ref{independant_of_substitution});
and similarly , a notion of {\em descendant of $P$, w.r.t. derivation $D$} by removing the 
incomparability restriction in condition (\ref{set_independant_of_substitution}).
\subsection{Words viewed as Terms}
\label{words_as_terms}
In order to transfer every definition (or statement) about Term Rewriting Systems into a similar
one about Semi-Thue systems, we define here precisely an embedding of the set of words (resp. semi-Thue systems) over an alphabet $A$ into the set of terms (resp. Term Rewriting Systems) over some signature $\FF$.

Let $A$ be some alphabet. We define the signature $\FUN(A)$ by
$$\FUN(A) := A \cup \{\#_0\},\;\;\forall a \in A, \arit(a)=1 \mbox{ and } \arit(\#_0)=0.$$
We define two mappings $\FUN_i: A^* \rightarrow \TT(\FUN(A),\{\Box\})$ ($i \in \{0,1\}$) by
setting:
$$\FUN_1(\varepsilon)=\Box,\;\;\FUN_1(a_1 a_2 \cdots a_n) = a_1(a_2( \ldots(a_n(\Box))\ldots)),$$
$$\FUN_0(\varepsilon)=\#_0,\;\;\FUN_0(a_1 a_2 \cdots a_n) = a_1(a_2( \ldots(a_n(\#_0))\ldots)),$$
Note that, for every word $w$,  $\FUN_1(w)$ is a context while $\FUN_0(w)$ is a ground term.
We associate with every rewriting rule $\rrule{u}{v}$ , the (term) rewriting rule
$$\FUN(\rrule{u}{v}) := \rrule{\FUN_1(u)}{\FUN_1(v)},$$
and with every semi-Thue system $(S,A)$ the term-rewriting system
$$(\FUN(S),\FUN(A)) \mbox{ where } \FUN(S):= \{ \FUN(\rrule{u}{v})\mid \rrule{u}{v}\in S\}.$$
The following lemma is straightforward
\begin{lemma}
Let $(S,A)$ be a semi-Thue system and $w,w' \in A^*$. Then\\
$w \to_S w' \Leftrightarrow \FUN_0(w) \to_{\FUN(S)} \FUN_0(w') \Leftrightarrow \FUN_1(w) \to_{\FUN(S)} \FUN_1(w').$
\label{l-words_as_terms}
\end{lemma}
In the sequel, the explicit application of $\FUN_1$ will be sometimes omitted:
if $w \in A^*$ and $t \in \TT(\FUN(A),\{\Box\})$, the expression $w(t)$
will denote the unary term $\FUN_1(w)[t]$.
\subsection{Automata}
We shall consider bottom-up finite term (tree) automata only \cite{TATA} 
(which we abbreviate to \fta).
A \fta is a 4-tuple $\AA:= (\FF, Q, Q_f, \sys)$
where $\FF$ is the signature, $Q$ is a finite set of symbols of arity $0$, called the set of 
{\em states}, 
$Q_f$ is the set of {\em final} states, $\sys$ is the set of {\em transitions}.
Every element of $\sys$ has the form
\begin{equation}
q \to q'
\label{e-spontaneous-rule-of-A}
\end{equation}
for some $q,q' \in Q$, or
\begin{equation}
f(q_1,\ldots,q_m) \to q
\label{e-rule-of-A}
\end{equation}
for some $m \geq 0, f \in \FF_m, q_1,\ldots, q_m \in Q$.
The {\em size} of $\AA$ is defined by:
$\|\AA\| := \Card(\sys) + \Card(Q)$.
The set of rules $\Gamma$ can be viewed as a rewriting system over the signature 
$\FF \cup Q$. We then denote by $\to_\Gamma$ or by $\to_\AA$ 
(resp. by $\to^*_\Gamma$ or by $\to^*_\AA$) the one-step rewriting relation
(resp. the rewriting relation) generated by $\Gamma$.

Given an automaton $\AA$, the set of terms accepted by
$\AA$ is defined by:
$$L(\AA):= \{ t \in \TT(\FF) \mid \exists q \in Q_f, t \to_\AA^* q \}.$$
A set of terms $T$ is \emph{recognizable} if there exists a finite term automaton $\AA$
such that $T = L(\AA)$.

The automaton $\AA$ is called {\em deterministic} iff\\
D1- $\sys$ posesses no rule of the form (\ref{e-spontaneous-rule-of-A})\\
D2- for every $t,u,u' \in \TT(\FF \cup Q)$,
$$ (t \to u \in \sys \myand t \to u' \in \sys) \Rightarrow (u=u').$$
The automaton $\AA$ is called {\em complete} iff
for every $m \geq 0$, $f \in \FF_m$ and $m$-tuple of states
$(q_1,\ldots,q_m) \in Q^m$, either ($m=0$ and $f \in Q$) or, there exists $q \in Q$ such that
$$ f(q_1,\ldots,q_m) \rightarrow q \in \sys.$$

Beside the above usual properties we introduce here the notion of 
standard automaton as follows:
\begin{definition}
A \fta $\AA= (\FF, Q, Q_f, \sys)$ is called {\em standard} iff it fulfills the four conditions:\\
1- $\sys$ posesses no rule of the form {\rm (\ref{e-spontaneous-rule-of-A})}\\
2- $\FF_0 \subseteq Q$\\
3- every rule {\rm (\ref{e-rule-of-A})} of $\AA$ is such that $m \geq 1$\\
4- for every $m \geq 1, f \in \FF_m, q_1,\ldots,q_m \in Q$ there exists a unique $q \in Q$
such that $f(q_1,\ldots,q_m) \to_\AA q$.
\label{d-standard_fta}
\end{definition}
Note that our definition of the notion of \fta corresponds to the notion of generalized 
finite term automaton of \cite{TATA} (it is slightly more general than the usual one)
while the above notion of standard \fta is more restricted than the usual notion of 
deterministic and complete \fta (we have, in some sense, removed the ``initial'' rules 
of the form $f \to_\AA q$ for symbols $f \in \FF_0$ and included the alphabet $\FF_0$ in $Q$).
Note that, for a standard \fta $\AA$, the relation $\to_\AA$ 
strictly reduces the size of terms.
We give later on, in \S \ref{p-Automaton-hatA}, a precise procedure transforming any \fta 
$\AA$ into a standard \fta $\hat{A}$ with ``similar'' rewriting relation, 
hence recognizing the same language. Therefore most theorems will assert properties for 
\fta's while most proofs will only manipulate standard \fta's.

\subsection{Automata and rewriting}
A system $\RR$ is \emph{recognizability preserving} if 
$[T](\to_\RR^*)$ is recognizable for every recognizable $T$.

A system $\RR$ is \emph{inverse recognizability preserving} if $(\to_\RR^*)[T]$
is recognizable for every recognizable $T$ or equivalently if $\RR^{-1}$ is
recognizability preserving.
\paragraph{Some technical notions}
The following lemma extends the property of determinism to tree-domains larger than just a single point. 
\begin{lemma}
\label{l-generalized-determinism}
Let $\AA$ be some standard \fta over the signature $\FF$.
Let $t,t_1,t_2 \in \TT(\FF \cup Q)$. If $t \to^*_\AA t_1,t \to^*_\AA t_2$ 
and $\Pos(t_1) = \Pos(t_2)$, then $t_1=t_2$.
\end{lemma}
We extend to subdomains the usual notion of state reached by some deterministic complete
\fta from a given term $t$: we call it the {\em reduct} of $t$ over the subdomain $P$.
\begin{definition}[$\AA$-reduct]
\label{d-Red}
Let $\AA$ be some standard \fta over the signature $\FF$.
Let $t \in \QTERM$ and let $P$ be some subdomain of $\Pos(t)$. We define $\Red(t,P) = t'$ as 
the unique element of $\QTERM$ such that\\
1- $\Pos(t')=P$\\
2- $t \to^*_\AA t'$\\
\end{definition}
The existence and unicity of such a term $\Red(t,P)$ follows from the technical conditions 
imposed by Definition \ref{d-standard_fta}.

\begin{lemma}
\label{l-confluence-commondomain}
Let $\AA$ be some standard \fta over the signature $\FF$.
Let $t,t_1,t_2 \in \QTERM$. If $t \to^*_\AA t_1,t \to^*_\AA t_2$ and $\Pos(t_1) \subseteq \Pos(t_2)$,
then $t_2 \to^*_\AA t_1$.
\end{lemma}
\begin{proof}
Since $t \to^*_\AA t_1$,
$t \to^*_\AA t_2 \to^*_\AA \Red(t_2, \Pos(t_1))$ 
and
$\Pos(t_1) = \Pos(\Red(t_2, \Pos(t_1))$,  
by Lemma~\ref{l-generalized-determinism}, 
$t_1= \Red(t_2, \Pos(t_1))$ which implies that $t_2 \to^*_\AA t_1$.
\end{proof}

\section{Bottom-up rewriting}
\label{s-bottom-up-rewriting}
In order to define \emph{bottom-up rewriting}, we need some marking tools.
In the following we assume that $\FF$ is a signature.
We shall illustrate many of our definitions with the following system $(\RR_1,\FF)$
\begin{example}
\label{ex-1}

$\RR_1 = \{ \mF(x) \to \mG(x), \mG(\mH(x)) \to \mI(x), \mI(x) \to \mA \}$,
$\FF= \{\mA,\mF,\mG,\mH,\mI\}$ with $\arit(\mA)= 0,\arit(\mF)= 1,\arit(\mG)= 1,\arit(\mH)= 1,\arit(\mI)=1$.
\end{example}

\subsection{Marking}
As in \cite{GHW04}, we may mark the symbols of a term using natural integers.

\subsubsection*{Marked symbols}
\begin{definition}
We define the (infinite) \emph{signature of marked symbols}:

$\mar{\FF}{\N} = \{ \mar{f}{i} \mid f \in \FF, i \in \N\}$.\\
For every integer $k \geq 0$ we note:
$\mar{\FF}{\leq k} = \{ \mar{f}{i} \mid f \in \FF, 0 \leq i \leq k\}$.
The mapping $\m:\mar{\FF}{\N} \rightarrow \N$ maps every marked symbol into its mark: 
$\ma{\mar{f}{i}} = i$.
\end{definition}

\subsubsection*{Marked terms}
\begin{definition}
The terms in $\TT(\mar{\FF}{\N},\VV)$ are called \emph{marked} terms.
\end{definition}
\noindent
The mapping $\m$ is extended to marked terms by:\\ 
if $t \in \VV,\ma{t} = 0$, otherwise, $\ma{t} = \ma{t(\varepsilon)}$.\\
For every $f \in \FF$, we identify $\mar{f}{0}$ and $f$;
it follows that $\FF \subset \FF^{\N}$,
$\TT(\FF) \subset \TT(\mar{\FF}{\N})$ and $\TT(\FF,\VV) \subset \TT(\mar{\FF}{\N},\VV)$.

\begin{example*}
$\ma{\mar{\mA}{2}} = 2, \ma{\mI(\mar{\mA}{2})} = 0, \ma{\mar{\mH}{1}(\mA)} = 1,
\ma{\mar{\mH}{1}(x)} = 1, \ma{x} = 0$.
\end{example*}

\begin{definition}
Given $t \in \TT(\FF^\N,\VV )$ and $i \in \N$, we define the marked term $\mar{t}{i}$ whose
marks are all equal to $i$:
\[
\begin{array}{ll}
\text{if~} t  \text{~is ~a ~variable~} x & \mar{t}{i} = x\\
\text{if~} t  \text{~is ~a ~constant~} c & \mar{t}{i} = \mar{c}{i}\\
\text{otherwise~} (t = f(\seq{t})) \text{where~} n \geq 1 & \mar{t}{i} = \mar{f}{i}(\mar{t_1}{i}, \ldots, \mar{t_n}{i})
\end{array}
\]
\end{definition}
This marking extends to sets of terms $S$ ($\mar{S}{i} = \{ \mar{t}{i} \mid t \in S \}$) 
and substitutions $\sigma$ ($\mar{\sigma}{i}: x \mapsto \mar{(x\sigma)}{i}$).

We use $\mamax{t}$ (resp. $\mamin{t}$) to denote the maximal (resp. minimal) mark of a marked term $t$.
\[
\begin{array}{c}
\mamax{t} := \text{max}\{ \ma{\sto{t}{u}} \mid u \in \Pos(t) \}\\
\mamin{t} := \text{min}\{ \ma{\sto{t}{u}} \mid u \in \Pos(t) \}\\
\end{array}
\]
For $u \in \Posp(t)$, $\mamaxu{u}{t} := \max \{ \ma{\sto{t}{v}} \mid v \prec u \}$. 

\begin{example*}
$\mamax{\mI(\mar{\mA}{2})} = 2, \mamin{\mI(\mar{\mA}{2})} = 0, \mamaxu{1.1}{\mG(\mar{\mH}{1}(\mar{\mA}{2}))} = 1.$
\end{example*}

\noindent
Notation: in the sequel, given a term $t \in \TERM$, $\barre{t}$ will always refer to
a term of $\TTvm$ such that $\mar{\barre{t}}{0} = t$. The same rule will apply to
substitutions and contexts.

\paragraph{Finite automata and marked terms.}
Given a \fta $\AA=(\FF, Q, Q_f, \sys)$ we extend it over  
the signature $\FF^{\leq k}$, by setting 
$$\sys^{\leq k}:=\{ (f^j(q_1^{j_1},\ldots,q_n^{j_n}) \rightarrow q^j) \mid
(f(q_1,\ldots,q_n \rightarrow q) \in \Gamma, j,j_1,\ldots,j_n \in [0,k] \},$$ and
$$\AA^{\leq k}:=(\FF^{\leq k}, Q^{\leq k}, Q_f^{\leq k}, \sys^{\leq k}).$$
Since, for every integers $k,k'$, $\AA^{\leq k}$ and $\AA^{\leq k'}$ have the same action on terms
 with marks not greater than $\min(k,k')$,
we often denote by $\AA$ any extension $\AA^{\leq k}$ with a sufficiently large $k$ w.r.t. the terms
under consideration.
\paragraph{$\N$ acts on marked terms.} 
We define a right-action $\odot$ of the monoid $(\N,\max,0)$ over the set $\FFm$ which just consists
in applying the operation $\max$ on every mark: for every $\bar{t} \in \FFm, n \in \N$,
$$\Pos(\bar{t}\odot n) := \Pos(\bar{t}),\;\;\forall u \in \Pos(\bar{t}),
\ma{(\bar{t}\odot n)/u}:= \max(\ma{\bar{t}/u},n),\;\;
(\bar{t}\odot n)^0= \bar{t}^0$$
Since a marked term can be viewed as a map from its domain to the direct product $\FF \times \N$,
and since the operation $\odot$ acts on the second component only while every \fta acts 
on the first component only, the following statement is straightforward.
\begin{lemma}
\label{l-odot-is_Acompatible}
Let $\AA$ be some finite term automaton over $\FF$, $\bar{s},\bar{t} \in \TTm$ and $n \in \N$.
If $\bar{s} \to^*_\AA \bar{t}$ then $(\bar{s}\odot n) \to^*_\AA (\bar{t}\odot n)$.
\end{lemma}
\subsubsection*{Marked rewriting}
\begin{figure}[htb]
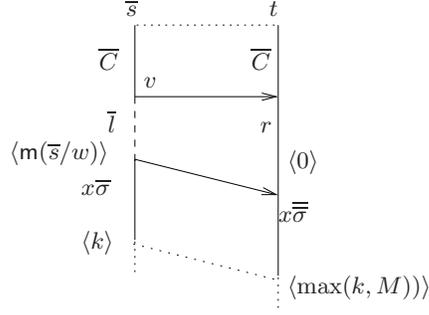

\centering
\input marked-step.pstex_t
\caption{A marked rewriting step}
\label{marked-step}
\end{figure}

We define here the rewrite relation $\toro$ between marked terms.
For every linear marked term $\bar{t} \in \TTvm$
and variable $x\in \Var(\bar{t})$, 
we define:
\begin{equation}
\label{d-M}
\M(\bar{t},x) = \sup\{ \m(\barre{t}/w) \mid w \prec \pos(\barre{t},x)\} +1.
\end{equation}
Let $\RR$ be a left-linear system, $\barre{s} \in \TTm$ and $t \in \TT$.
Let us suppose that $\barre{s} \in \TTm$ decomposes as
\begin{equation}
\barre{s}=\barre{C}[\barre{l}\barre{\sigma}]_v,\;\;\mbox{ with } \;\;(l,r) \in \RR,
\label{start_marqued_onestep}
\end{equation}
for some marked context $\barre{C}[]_v$ and substitution 
$\barre{\sigma}$.
We define a new marked substitution $\dbarre{\sigma}$ (such that 
$\dbarre{\sigma}^0=\barre{\sigma}^0$) by: for every $x \in \Var(r)$, 
\begin{equation}
x\dbarre{\sigma}:= (x \barre{\sigma}) \odot M(\barre{C}[\barre{l}],x).
\label{barbarsigma}
\end{equation}
We then write $\barre{s} \toro \barre{t}$ where
\begin{equation}
\barre{s}=\barre{C}[\barre{l}\barre{\sigma}],\;\; \barre{t}=\barre{C}[r \dbarre{\sigma}].
\label{marqued_onestep}
\end{equation}
(This is illustrated by Figure~\ref{marked-step}, where $M$ denotes $M(\barre{C}[\barre{l}],x)$
and the marks are written between brackets $\langle \ldots \rangle$).
More precisely, an ordered  pair of marked terms $(\barre{s},\barre{t})$ is linked by the relation $\toro$
iff, there exists $\barre{C}[]_v,(l,r),\barre{l},\barre{\sigma}$ and $\dbarre{\sigma}$ fulfilling equations
(\ref{start_marqued_onestep}-\ref{marqued_onestep}).
The intuitive idea behind the above definition is that the marks are storing
the relevant information concerning the {\em ordering} of successive positions of redexes during 
the derivation. A mark $k$ will roughly mean that there were $k$ successive applications of rules, 
each one with a leaf of the left-handside at a position strictly greater than a leaf of the 
previous right-handside.

The map $\barre{s} \mapsto \barre{s}^0$ (from marked terms to unmarked terms) extends into a map from marked derivations to unmarked derivations: 
every 
\begin{equation}
\barre{s}_0 = \barre{C}_0[\barre{l}_0\barre{\sigma}_0]_{v_0} \toro 
\barre{C}_0[r_0\dbarre{\sigma}_0]_{v_0} = \barre{s}_1 \toro \ldots  
\barre{C}_{n-1}[r_{n-1}\dbarre{\sigma}_{n-1}]_{v_{n-1}} = \barre{s}_n
\label{marked_derivation}
\end{equation}
is mapped to the derivation
\begin{equation}
s_0 = C_0[l_0\sigma_0]_{v_0} \to C_0[r_0\sigma_0]_{v_0} = s_1 \to \ldots C_{n-1}[r_{n-1}\sigma_{n-1}]_{v_{n-1}} = s_n.
\label{usual_derivation}
\end{equation}
The context $\barre{C}_i[]_{v_i}$, the rule $(l_i,r_i)$, the marked version $\bar{l_i}$ of $l_i$ 
and the substitution $\barre{\sigma}_i$ completely determine $\barre{s}_{i+1}$. Thus, for every fixed
pair $(\barre{s_0},s_0)$, this map is
a bijection from the set of derivations (\ref{marked_derivation}) starting from $\barre{s_0}$, 
to the set of derivations (\ref{usual_derivation}) starting from $s_0$.
\begin{example}
\label{ex-toro}
With the system $\RR_1$ of Example~\ref{ex-1} we get the following marked derivation:
\[
\begin{array}{l}
\mF(\mH(\mF(\mH(\mA)))) \toro \mF(\mH(\mG(\mar{\mH}{1}(\mar{\mA}{1})))) \toro 
\mF(\mH(\mI(\mar{\mA}{2}))) \toro \mF(\mH(\mA)) \toro \\ 
\mG(\mar{\mH}{1}(\mar{\mA}{1})) \toro \mI(\mar{\mA}{2}) \toro \mA
\end{array}
\]
\end{example}

From now on, each time we deal with a derivation $s \to^* t$ 
between two terms $s,t \in \TT(\FF,\VV)$, we may implicitly decompose it 
as (\ref{usual_derivation}) where $n$ is the length of the derivation, $s = s_0$ and $t = s_n$.

\subsection{Bottom-up derivations}
\begin{definition}
\label{d-marked-wbu_derivation}
The marked derivation (\ref{marked_derivation}) is \emph{weakly bottom-up} if,
for every $0 \leq i < n$,
\begin{eqnarray}
l_i \notin \VV & \Rightarrow &\m(\barre{l_i})=0,\label{wbu-nonemptycase}\\
l_i \in \VV & \Rightarrow &\sup\{\m(\barre{s_i}/u) \mid u \prec v_i \}=0.\label{wbu-emptycase}
\end{eqnarray}
\end{definition}
(Handling the case where some lhs are just variables is worthwhile: for example the systems of
\cite{Ben-Sak86}, when viewed as term rewriting systems, have all their lhs in $\VV$).
\begin{definition}
The derivation~(\ref{usual_derivation}) is \emph{weakly bottom-up}
if the corresponding marked derivation (\ref{marked_derivation}) starting on the same
term $\barre{s}=s$ is \emph{weakly bottom-up} (following the above definition).
\label{d-wbu_derivation}
\end{definition}

\begin{figure}[htb]
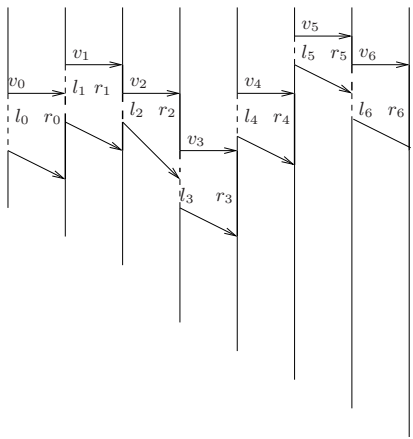

\centering
\scalebox{0.80}{
\input wbu-derivation.pstex_t
}
\caption{A wbu derivation}
\label{wbu-derivation}
\end{figure}

\begin{figure}[htb]
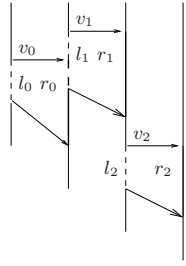

\centering
\scalebox{0.80}{
\input nonwbu-derivation.pstex_t
}
\caption{A non wbu derivation}
\label{nonwbu-derivation}
\end{figure}



\begin{remark}
An alternative formulation for defining a weakly bottom-up derivation is to say that
no redex $l_j\sigma_{j}$ is contracted
at a position $v_{j}$ strictly greater 
than a variable of a previous $r_i$. This means, in some sense, that the 
reductions are made in a ``bottom-up'' fashion, see Figure~\ref{wbu-derivation}
and Figure~\ref{nonwbu-derivation}.
\end{remark}

We shall abbreviate ``weakly bottom-up'' to $\wbu$.
Note that the notion of $\wbu$ marked derivation is defined step by step.
It is thus clear that the composition of two $\wbu$ marked derivations is $\wbu$ too.
This might be false for $\wbu$ unmarked derivations.
In the following we thus mainly handle \emph{marked} $\wbu$ derivations.

The next lemma shows that in the case of a linear system,
a derivation can always be replaced by a $\wbu$-derivation.
\begin{lemma}
Let $\RR$ be a linear system.
If $s \to_\RR^* t$ then there exists a $\wbu$-derivation between $s$ and $t$.
\label{l-wbualwaysexist}
\end{lemma}
\begin{sketch}
We prove by induction on the integer $n$, that,
for every derivation $s \to^n t$, there exists a $\wbu$-derivation from $s$ to $t$,
with the same length $n$ and reducing the same redexes of $s$\\
{\bf Basis}: $n = 0$\\
 then $s = t$; the empty derivation is $\wbu$.\\
{\bf Induction step}: $n >0$\\
As $\RR$ is linear, every redex
may have at most one descendant in each term of the derivation.
We choose a redex $l\sigma$ ($\rrule{l}{r} \in \RR)$ of $s$, whose position is maximal (w.r.t. $\preceq$) among the set of positions of redexes 
contracted somewhere 
in the derivation $s \to^n t$; let $u$ be the position of this maximal  redex in $s$.
A new derivation can be obtained by transferring the contraction of $l \sigma$ at the beginning
of the derivation: we obtain a derivation of equal length
\begin{equation}
s = C[l\sigma]_u \to C[r\sigma]_u \to^{n-1} t.
\label{e-complete_derivation}
\end{equation}
By induction hypothesis, the derivation $C[r\sigma]_u \to^{n-1} t$ can be made $\wbu$ while
preserving its length $n-1$ and the set of redexes of $C[r\sigma]_u$ that are contracted.
Let us  consider the unique marked derivation associated to (\ref{e-complete_derivation}):
\begin{equation}
s = C[l\sigma]_u \toro C[r\dbarre{\sigma}]_u \toro^{n-1} \dbarre{t}.
\label{e-complete_marked_derivation}
\end{equation}
and the unique marked derivation associated to the $\wbu$-derivation $C[r\sigma]_u \to^{n-1} t$:
\begin{equation}
C[r\sigma]_u \toro^{n-1} \barre{t}.
\label{e-shorter_marked_derivation}
\end{equation}
By assumption and preservation of the redexes, $\sigma$ does not contain any redex which is 
contracted inside the derivation $C[r\sigma] \to^{n-1} t$. 
Hence, the $(j+1)$th step of derivation (\ref{e-complete_marked_derivation}) uses a lhs with
a root that possesses the same mark as the root of the lhs of the $j$th step of derivation (\ref{e-shorter_marked_derivation}). Since this mark is always null in (\ref{e-shorter_marked_derivation}), it is also null in (\ref{e-complete_marked_derivation}).
This shows that (\ref{e-complete_derivation}) is $\wbu$. 
\end{sketch}\\
Note that, what the above lemma shows, is that the condition $\wbu$ is not a restriction on
the associated ``derivation-graph'' (this notion is defined in \cite{Bos-Dau-War88} in the
case of context-sensitive word grammars but could be extended to arbitrary linear 
term rewriting systems) but, rather on the {\em traversal} of this graph corresponding to the 
particular order in which the reductions are performed. 
\begin{definition}
A marked term $\barre{s}$  is said $\m$-{\em increasing} iff, 
for every $u,v \in \Pos(\barre{s}),
u \preceq v \Rightarrow \m(\barre{s}/u) \leq \m(\barre{s}/v)$. 
\label{d-mincreasing}
\end{definition}

\begin{lemma}
Suppose that $\barre{s}$ is a $\m$-{\em increasing} marked term, 
$\barre{t} \in \TT(\FF^\N,\VV)\setminus \VV$,
$\ma{\barre{t}}=0$, $\barre{C}[]_v$ is a marked context,  
$\barre{\sigma}$ is a marked substitution
and $\barre{s}= \barre{C}[\barre{t} \barre{\sigma}]_v$.\\
Then,  $\barre{C}[]_v$ has no mark above the position $v$.
\label{f-context-isunmarked}
\end{lemma}
\begin{proof}
Let $u \in \Pos(\barre{C})$ such that $u \prec v$ and $\barre{C}(v)=\Box$.\\
Since $\barre{C}[\barre{t} \barre{\sigma}]$ is $\m$-increasing, 
$$\m(\barre{C}[\barre{t} \barre{\sigma}]/u) \leq \m(\barre{C}[\barre{t} \barre{\sigma}]/v)$$
But $\m(\barre{C}[\barre{t} \barre{\sigma}]/v)= \m(\barre{t})=0$.
\end{proof}
\begin{lemma}
\label{l-onestep-marks_increase}
Let $\barre{s} \toro \barre{t}$ be a $\wbu$ marked derivation-step between $\barre{s}, \barre{t} \in \TT(\FF^\N)$. 
If $\barre{s}$ is $\m$-increasing, then $\barre{t}$ is $\m$-increasing too.
\end{lemma}
\begin{proof}
Suppose that $\barre{s}$ is $\m$-increasing and that $\barre{s},\barre{t}$ fulfill (\ref{start_marqued_onestep}-\ref{marqued_onestep}).
Let us consider $v_1,v_2 \in \Pos(\barre{t})$ such that $v_1 \preceq v_2$.
Let us show that 
\begin{equation}
\m(\barre{t}/v_1) \leq \m(\barre{t}/v_2).
\label{e-tbarre_is_increasing}
\end{equation}



We distinguish 3 cases depending on the relative positions of $v, v_1,v_2$ .\\
{\bf Case 1}: $v_1 \prec v$.\\
Since the derivation step is $\wbu$, we have $\m(\barre{s} \slash v)=0$. According to the definition of a marked derivation-step, we have $\m(\barre{s} \slash v_1)=\m(\barre{t} \slash v_1)$. Moreover, $\barre{s}$ is $\m$-increasing. Hence, $\m(\barre{s} \slash v)=\m(\barre{s} \slash v_1)=0 \leq \m(\barre{t}/v_2)$. \\
{\bf Case 2}: $v \preceq v_1$.\\ If $v_1 \in v \cdot \Pos(r)$ then we have $\m(\barre{t}/v_1)=0 \leq \m(\barre{t}/v_2)$. Otherwise, $v_1=v\cdot w \cdot w_1,v_2=v\cdot w \cdot w_2$, 
where $x$ is the label of $w$ in $r$ and $w_1,w_2 \in \Pos(x \sigma)$.
Let
$$v'_1 := v\cdot w' \cdot w_1,\;\;v'_2:=v\cdot w' \cdot w_2$$
where $w'= \pos(\ell,x)$.\\
Since $\barre{s}$ is $\m$-increasing, $\m(\barre{s}/v'_1)\leq \m(\barre{s}/v'_2)$,
hence
$$\max(\m(\barre{s}/v'_1),\M(C[\barre{l}],x))\leq \max(\m(\barre{s}/v'_2),\M(C[\barre{l}],x))$$
i.e. $\m(\barre{t}/v_1)\leq \m(\barre{t}/v_2)$.\\
{\bf Case 3}: $v_1 \perp v$.\\
In this case we also have $v_2 \perp v$.
It follows that for every $i \in \{1,2\}$,  $\m(\barre{s}/v_i)= \m(\barre{t}/v_i)$,
and we can conclude as in case 1.\\
In all cases we have established that (\ref{e-tbarre_is_increasing}) holds.
\end{proof}
The previous Lemma generalizes to a sequence.
\begin{lemma}
\label{l-marks_increase}
Let $\barre{s} \toro^* \barre{t}$ be a $\wbu$ marked derivation between $\barre{s}, \barre{t} \in \TT(\FF^\N)$. 
If $\barre{s}$ is $\m$-increasing, then $\barre{t}$ is $\m$-increasing too.
\end{lemma}
\begin{proof}
Straightforward induction on the length $n$ of the derivation based on Lemma \ref{l-onestep-marks_increase}.
\end{proof}
\begin{remark}
Let us examine the value of $\M(\barre{C}[\barre{l}],x)$ when 
$s' \toro^* \barre{s} = \barre{C}[\barre{l}\barre{\sigma}] \toro \barre{C}[r\barre{\sigma}] = \barre{t}$,
and $s' \to^* s$ is $\wbu$:\\
- if $C$ is the empty context and $l=x$ then $\M(\barre{C}[\barre{l}],x)=1$\\
- otherwise, by Lemma~\ref{l-marks_increase}, $\M(\barre{C}[\barre{l}],x)=\ma{\barre{C}[\barre{l}]/f_x}+1$,
where $f_x$ is the father of $\pos(\barre{C}[\barre{l}],x)$.
\label{max_marks_above2}
\end{remark}

We classify the derivations according to the maximal value of the marks.
We abbreviate ``bottom-up'' to $\bu$.

\begin{definition}
\label{d-bottom-up-deriv}
A derivation is \emph{$\bu(k)$} (resp. \emph{$\bu^-(k)$})
if it is $\wbu$ and, in the corresponding
marked derivation  $\forall i \in [0,n], \mamax{\barre{s_i}} \leq k$
(resp. $\forall i \in [0,n-1], \mamax{\barre{l_i}} < k$).
\end{definition}
Let us introduce a convenient notation.
\begin{definition}Let $k \geq 1$. The binary relation $\torok_\RR^*$ over $\TT(\FF^\N)$ is defined by:\\
$\barre{s} \torok_\RR^* \barre{t}$ if and only if  there exists a
$\wbu$ marked derivation from $\barre{s}$ to $\barre{t}$ where all the marks belong to $[0,k]$.\\
The binary relation $\tok_\RR^*$ over $\TT(\FF)$ is defined by:\\
$s \tok_\RR^* t$ if and only if there exists a
$\bu(k)$-derivation from $s$ to $t$.
\end{definition}
\begin{example}
\label{ex-not-sbu}
For the system $\RR_0 = \{ \mF(\mF(x)) \to \mF(x) \}$ with the signature 
$\FF = \{ \mA^{(0)}, \mF^{(1)} \}$, 
although for every $k$ we may get a $\bu(k)$-derivation for a term of the form $\mF( \ldots \mF(\mA) \ldots)$ with
$k+1$ $\mF$ symbols:
\[
\mF(\mF(\mF(\mF(\mA)))) \toro \mF(\mar{\mF}{1}(\mar{\mF}{1}(\mar{\mA}{1}))) \toro \mF(\mar{\mF}2{}(\mar{\mA}{2})) \toro \mF(\mar{\mA}{3})
\]
we can always achieve a $\bu(1)$-derivation:
\[
\mF(\mF(\mF(\mF(\mA)))) \toro \mF(\mF(\mF(\mar{\mA}{1}))) \toro \mF(\mF(\mar{\mA}{1})) \toro \mF(\mar{\mA}{1})
\]
\end{example}
\begin{example*}[comparison with innermost derivation]
\label{ex-diff_from_innermost}
Let us consider the signature $\FF := \{ \mA^{(0)}, \mF^{(1)},\mG^{(1)} \}$ and the 
rewriting system $\RR := \{ \mF\mG(x) \to \mG\mF(x),  \mG\mF(x) \to \mH(x)\}.$ 
The derivation $$ \mF \mG \mG \mG (\mA) \to \mG \mF \mG \mG (\mA) \to \mG \mG \mF \mG (\mA)
\to \mG \mG \mG \mF (\mA)$$
corresponds to the marked derivation
$$ \mF \mG \mG \mG (\mA) \to \mG \mF \mG^1 \mG^1 (\mA^1) \to \mG \mG \mF \mG^2 (\mA^2)
\to \mG \mG \mG \mF (\mA^3)$$
which is not $\BU(2)$; note however, that this derivation is {\em innermost} i.e.
each derivation step rewrites the innermost redex of the given term.\\
 The derivation $$ \mF \mG \mF \mG \mF \mG (\mA) \to \mF \mG \mF \mH \mG (\mA)
\to \mF \mH \mH \mG (\mA)$$
corresponds to the marked derivation
$$ \mF \mG \mF \mG \mF \mG (\mA) \toro \mF \mG \mF \mH \mG^1 (\mA^1)
\toro \mF \mH \mH^1 \mG^1 (\mA^1)$$
which is $\BU(1)$; note however, that this derivation is {\em not innermost} 
since the innermost redex of the first term is  $\mF \mG (\mA)$,  which is not rewritten
in this derivation.\\
\end{example*}

\subsection{Bottom-up systems}
\label{sub-bottom-up-systems}
We introduce here a hierarchy of classes of rewriting systems \footnote{
a {\em class} of TRS is a subset of the set of all TRS (over a fixed denumerable ranked alphabet) which is closed under alphabet isomorphism (i.e. renaming the symbols).}
and show that several
well-known classes of rewriting systems are included in the low levels of this hierarchy:
namely the right-ground systems, the left-basic semi-Thue systems, the linear shallow systems and
the linear growing systems.
\begin{definition}
Let $P$ be some property of derivations w.r.t. Term Rewriting Systems.\\ 
1- A Term Rewriting System $(\RR,\FF)$ is called $P$ if for every $s, t \in \TT(\FF)$ such that $s \to_\RR^* t$ there exists a $P$-derivation from $s$ to $t$.\\
2- A semi-Thue system $(S,A)$ is called $P$ if the Term Rewriting System $(\FUN(S),\FUN(A))$
is called $P$.
\label{d-property-P}
\end{definition}
We shall use the convention that, for a property $P$ denoted by a lower-case acronym for derivations, we use the same acronym, but in upper-case, to denote the property $P$ extended to systems
by Definition \ref{d-property-P}.
For example, a Term Rewriting System $(\RR,\FF)$ is called $\BU(k)$ if for every $s, t \in \TT(\FF)$ such that $s \to_\RR^* t$ there exists a $\bu(k)$-derivation from $s$ to $t$.\\
We denote by $\BU(k)$ the class of $\BU(k)$ systems, by $\BU^-(k)$ the class
of $\BU^-(k)$ systems. We define the class of \emph{bottom-up systems}, denoted  $\BU$, by: 
\[
\BU := \bigcup_{k \in \N} \BU(k)
\]
\begin{lemma}
For every  $k > 0$, $\BU(k-1) \subsetneq \BU^-(k) \subsetneq \BU(k)$.
\end{lemma}

\begin{lemma}
\label{l-ground_are_bu0}
Every right-ground system is $\BU(0)$.
\end{lemma}
\begin{proof}
The right-handsides being ground no mark ($> 0$) is ever introduced by $\toro$.
\end{proof}
\begin{lemma}
\label{l-basic_are_bu-1}
Every inverse of a left-basic semi-Thue system is $\BU^-(1)$.
\end{lemma}
\begin{proof}
Let $(S,A)$ be a semi-Thue system such that $S^{-1}$ is left-basic. 
The combinatorial restrictions  defining the property ``left-basic''
(see, for example, conditions C1,C2 of section 2.5 in \cite{Sen95}) imply that, in {\em every} marked $\wbu$-derivation
$$w\# \toro^*_{\FUN(S)} \barre{\alpha}\barre{u}\barre{\beta}\# \toro_{\FUN(S)} \barre{\alpha}\barre{v}\barre{\beta}\#,$$ 
with $w \in A^*, \rrule{u}{v} \in S, \barre{\alpha},\barre{\beta} \in (A^\N)^*$, we must have
$$\mamax{\barre{u}}=0.$$
Hence $\FUN(S) \in \BU^-(1)$, so that, by point 2 of Definition \ref{d-property-P}, $S \in \BU^-(1)$.
\end{proof}

\begin{lemma}
\label{l-shallow_are_bu1}
Every shallow system is $\BU^-(1)$.
\end{lemma}
\begin{proof}
Let $\RR$ be a shallow set of rules. This implies that every rule $\rrule{l}{r} \in \RR$ is such 
that $l$ is a variable or all the occurrences of variables in $l$ have depth $1$.
Let us consider a $\wbu$-derivation of the form (\ref{marked_derivation}) starting on some unmarked term $s$.
Every marked term $\bar{l_i}$ either has no mark (because it has depth $0$)
or has only one mark above each occurrence of variable: the mark of the root of $\bar{l_i}$. 
In this case $\m(\bar{l_i})=0$,  
by Definition~\ref{d-marked-wbu_derivation} and  because,
by Lemma~\ref{l-marks_increase}, $\barre{C_i}[\barre{l_i}\barre{\sigma_{i}}]$ is $\m$-increasing.
Hence Definition~\ref{d-bottom-up-deriv} is fulfilled by the given derivation.
\end{proof}
\begin{lemma}
\label{l-growing_are_bu1}
Every growing linear system is $\BU(1)$.
\end{lemma}
\begin{proof}
Let $\RR$ be a linear growing system over a signature $\FF$ and $s,t \in \TT(\FF)$.
We prove by induction on the integer $n$ that:
if $\barre{s_0} \toro \ldots \toro \barre{s_n}$ is a $\wbu$ marked derivation starting on an
unmarked term $\barre{s_0} \in \TT(\FF)$, 
$$\forall i \in [0,n],\;\;\mamax{\barre{{s_i}}} \leq 1.$$
{\bf Basis}: $n = 0$. Then $\mamax{\barre{s_0}} = 0$ (by hypothesis).\\
{\bf Induction step}: Suppose that
$$\barre{s_0} \toro \ldots \toro \barre{s_{n+1}}
$$
with $\barre{s_0} \in \TT(\FF)$.
By induction hypothesis, $\forall i \in [0,n], \mamax{\barre{s_i}} \leq 1$.
$$
\barre{s_n} = \barre{C_n}[\barre{l_n}\barre{\sigma_n}]_{v_n} \toro \barre{C_n}[r_n\dbarre{\sigma_n}]_{v_n} = \barre{s_{n+1}}
$$
From $\mamax{\barre{s_n}} \leq 1$, we get $\mamax{\barre{C_n}} \leq 1$ and $\forall x \in \Var(l_n), \mamax{x\barre{\sigma_n}} \leq 1$.\\
We also have $\mamax{r_n} = 0$.\\
If $\Var(r_n) = \emptyset$ 
then $\barre{s_{n+1}} = \barre{C_n}[r_n]$ and $\mamax{s_{n+1}} \leq 1$.\\
Let us assume now that $\Var(r_n) \neq \emptyset$.\\ 
Since $\RR$ is growing, $x$ is at depth $0$ or $1$ in $l_n$.
Since the derivation is $\wbu$, $\ma{\bar{l}_n}=0$.
By Lemma \ref{l-marks_increase}, for every position $v$ of $\barre{s_n}$,
$$v \preceq v_n \Rightarrow \ma{v}=0$$
It follows that
$$\M(\bar{C}_n[\bar{l}_n],x) \leq 1$$
By definition of relation $\toro$
$$x\dbarre{\sigma_n}:= (x \barre{\sigma_n}) \odot M(\barre{C_n}[\barre{l_n}],x)$$
where both $\mamax{x \barre{\sigma_n}}\leq 1$ and $M(\barre{C_n}[\barre{l_n}],x)\leq 1$.
Hence 
$$\mamax{x \dbarre{\sigma_n}}\leq 1.$$
Finally, all the marks whether in $\barre{C_n}$ or in $r_n$ or in $x \dbarre{\sigma_n}$
(for $x \in \Var(r_n)$) are bounded by $1$, hence
$\mamax{\barre{s_{n+1}}}  \leq 1$.
\end{proof}

\begin{example}
\hspace*{1em}{~}\\
The system $\RR_0 = \{ \mF(\mF(x)) \to \mF(x) \} \in \BU^-(1)$ and $\RR_0$ is not growing.\\
The system $\RR_1$ of Example~\ref{ex-1} belongs to $\BU^-(2)$ and $\RR_1$ is not growing.\\
The system $\RR_2 = \{ \mF(x) \to \mG(x), \mH(\mG(\mA)) \to \mA \}$ is growing and belongs to $\BU^-(1)$.\\
The system $\RR_3 = \{ \mF(x) \to \mG(x), \mG(\mH(x)) \to \mA \}$ is growing and belongs to $\BU(1)$.
\end{example}

\begin{corollary}
${\msf Linear Growing} \subsetneq \BU(1)$.
\end{corollary}

\section{Inverse-preservation  of recognizability}
\label{s-preservation}
Let us recall the following classical result about ground rewriting systems
\begin{theorem}[\cite{Bra69}]
\label{t-ground-inverse-preserving}
Every ground system is inverse-recognizability preserving.
\end{theorem}
This theorem was further refined and extended in \cite{Der-Gil89,Dau-Tis90,Dau-HLT90}, see \cite{TATA} for an exposition.
The main theorem of this section (and of the paper) is the following extension of
Theorem~\ref{t-ground-inverse-preserving} to $\bu(k)$ derivations of linear rewriting systems

\begin{theorem}
\label{t-kbu-inverse-preserving}
Let ${\cal R}$ be some linear rewriting system over the signature $\FF$ , 
let $T$ be some recognizable  
subset of $\TT(\FF)$ and let $k \geq 0$. Then, the set $(\tok_\RR^*)[T]$ is recognizable too.
\end{theorem}

\subsection{Basic construction}
\label{s-basic-construction}
In order to prove Theorem~\ref{t-kbu-inverse-preserving} we have to introduce some technical definitions, and to prove some technical lemmas.
Let us fix, from now on and until the end of the subsection, a linear system $(\RR,\FF)$, 
a language $T\subseteq \TT(\FF)$ recognized by a finite automaton over the extended signature 
$\FF \cup \{\Box\}$, $\AA = (\FF\cup\{\Box\},Q,Q_f,\sys)$ and an integer $k \geq 0$.
In order to make the proofs easier, we assume in this subsection that:\\
\begin{equation}
\label{no-empty-lhs}
\forall \rrule{l}{r} \in \RR, l \notin \VV,
\end{equation}
\begin{equation}
\label{A-deterministic}
\AA \mbox{ is standard}.
\end{equation}
We postpone to \S \ref{s-general-construction} the proof that these restrictions are not a loss of generality.
Let us define the integer
\begin{equation}
\label{def_dmax}
\dmax := \max 
\{\depth(l) \mid \rrule{l}{r} \in \RR\}.
\end{equation}
\begin{example}For the system $\RR_1$ of example \ref{ex-1}, $\dmax = 2$.
\end{example}
We introduce now a notion of {\em top } part of a term $\barre{t}$, which is, intuitively, the only part of 
$\barre{t}$ which can be used in a $\bu(k)$-derivation starting on $\barre{t}$. 
Everything below this
part is merely included in the substitutions used by the derivation-steps, and thus copied
(eq.(\ref{def_dmax}) and Def.\ref{d-Topd} are tuned for this property). 
Such copied parts of $\barre{t}$ can be handled just by a state of the f.t.a. $\AA$.
The replacement of terms by their top-part will be used subsequently to show that
the full derivations w.r.t  $\tok_\RR^*$ can be simulated by derivations w.r.t 
some ``approximating'' ground rewriting system ( introduced by Def. \ref{d-ssystem}): 
the top-part of a real derivation is a derivation for the ground rewriting system (Lemma \ref{l-RprojectstoSn}) and, conversely, every derivation for the ground rewriting system is the top-part of some real derivation (Lemma \ref{l-SliftedtoR}).

We define, at first, the {\em top domain } of a term and, later on, the top of a term.
\begin{definition}[Top domain of a term]
\label{d-Topd}
Let $\barre{t} \in \TTqkmb$. We define the {\em top domain} of $\barre{t}$, denoted by $\Topd(\barre{t})$
as: $u \in \Topd(\barre{t})$ iff\\
1- $u \in \Pos(\barre{t})$\\
2- $\forall u_1,u_2 \in \N^*$ such that $u = u_1 \cdot u_2$, either $\m(\barre{t}/u_1)=0$ or
$|u_2| \leq (k+1-\m(\barre{t}/u_1))\dmax$.
\end{definition}
\begin{lemma}
For every $\barre{t} \in \TTqkmb$, $\Topd(\barre{t})$ is a subdomain of $\Pos(\barre{t})$.
\end{lemma}
\begin{proof}
1- Let $u  \in \Topd(\barre{t})$ and let $v \preceq u$. Let $w \in \N^*$ such that 
$v \cdot w = u$. Suppose that $v = v_1 \cdot v_2$
and that $\m(\barre{t}/v_1) \neq 0$.\\
Since $u = v_1 \cdot v_2 \cdot w$ and $u$ belongs to $\Topd(\barre{t})$, the inequality
$|v_2w| \leq (k+1-\m(\barre{t}/u_1))\dmax$ holds. But $|v_2|\leq |v_2w|$, hence
$$|v_2| \leq (k+1-\m(\barre{t}/u_1))\dmax.$$
2- Let $u \in\N^*$ and $i,j \in \N$ such that $u \cdot i \in \Topd(\barre{t})$ and
$u \cdot j \in \Pos(\barre{t})$.
Suppose $u_1,u_2 \in \N^*$ such that $u \cdot j = u_1 \cdot u_2$:\\ 
- If $u_2 \neq \varepsilon$, since $u \cdot i = u_1 \cdot u'_2 $,
where $u'_2:= u_2 (j)^{-1} i$, we know that:\\
$$\m(\barre{t}/u_1)=0\mbox{ or }|u'_2| \leq (k+1-\m(\barre{t}/u_1))\dmax.$$
which implies, since $|u_2|= |u'_2|$ that:
$$\m(\barre{t}/u_1)=0\mbox{ or }|u_2| \leq (k+1-\m(\barre{t}/u_1))\dmax.$$
- If $u_2= \varepsilon$ the required inequality for $|u_2|$ is obvious.
\end{proof}

\begin{definition}[Top of a term]
\label{d-Top}
For every $\barre{t} \in \TTqkmb$, $\Top(\barre{t}) = \Red(\barre{t},\Topd(\barre{t}))$.
\end{definition}
Note that, since $\Topd(\barre{t})$ is a subdomain of $\Pos(\barre{t})$ and is written over
the alphabet of the standard automaton $\AA$, $\Top(\barre{t})$ is well-defined.\\
This definition extends naturally, in a pointwise manner, to substitutions.
\begin{lemma}[$\Top$ is morphic]
\label{l-Top-is-morphic}
Let $\barre{C}[]$ be a context with no mark above the symbol $\Box$ and let $\barre{t}$ be any
marked term in $\TTqkm$. Then 
$\Top(\barre{C}[\barre{t}])=\Top(\barre{C})[\Top(\barre{t})]$.
\end{lemma}
\noindent The proof is easy and therefore omitted.
\begin{lemma}[$\Top$ preserves unmarked terms]
\label{l-Top-preserves-unmarked}
If $t \in \TT(\FF \cup Q,\VV)$ and $\barre{\sigma}: \VV \rightarrow \TTqm$ then
$\Top(t\barre{\sigma})=t\Top(\barre{\sigma})$.
\end{lemma}
\begin{proof}
 The proof is easy and therefore omitted.
\end{proof}
\begin{lemma}[$\Top$ is decreasing]
\label{l-Top-is-decreasing}
Let $\barre{s},\barre{t} \in \TT((\FF \cup Q)^\N)$ be such that 
$\Pos(\barre{s}) = \Pos(\barre{t})$ and, such that, for every $u \in \Pos(\barre{s}),
\;\;\m(\barre{s}/u) \leq \m(\barre{t}/u)$.\\
 Then $\Pos(\Top(\barre{s})) \supseteq \Pos(\Top(\barre{t}))$.
\end{lemma}
\begin{definition}
\label{d-ssystem}
We consider the following ground rewriting system $\SS$ 
over $\TT((\FF \cup Q)^{\leq k})$ consisting of all the rules of the form:\\
\begin{equation}
\label{e-rule-of-SS}
\barre{l}\barre{\tau} \rightarrow r \dbarre{\tau}
\end{equation}
where $\rrule{l}{r}$ is a rule of $\RR$ 
\begin{equation}
\label{e-wbu-rule}
\ma{\barre{l}}=0
\end{equation}
and $\barre{\tau}: \VV \rightarrow \TTqkm$ is a marked substitution such that, 
$\forall x \in \Var(l)$
\begin{equation}
\label{e-short-substitution}
x\dbarre{\tau}= x\barre{\tau}\odot M(\barre{l},x),\;\;
\depth(x\bar{\tau}) \leq k \cdot \dmax.
\end{equation}
\end{definition}
(recall that the number $\dmax$ was defined by (\ref{def_dmax})).
\begin{lemma}[lifting $\SS \cup \AA$ to ${\cal R}$]$\;\;$\\
\label{l-SliftedtoR}
Let $\barre{s},\barre{s'}, \barre{t}\in \TTqkm$ such that $\barre{s'}$ is $\m$-increasing.
If $\barre{s}' \to^*_\AA \barre{s}$ and $\barre{s} \to^*_{\SS \cup \AA}\barre{t}$ then, there exists a term $\bar{t'} \in \TTqkm$ 
such that\\
$\barre{s}' \torok^*_\RR \barre{t'}$ and $ \barre{t'}\to^*_\AA \bar{t}$.
\end{lemma}
\begin{figure}[htb]
\centering
\input lift_and_project.pstex_t
\caption{Lemma~\ref{l-SliftedtoR} and \ref{l-RprojectstoSn}}
\label{lift_and_project}
\end{figure}
\begin{proof}
1- Let us prove that the lemma holds for $\barre{s} \to_{\SS \cup \AA}\barre{t}$.
Let us suppose that $\barre{s}' \to^*_{\AA} \barre{s} \to_{\AA}\barre{t}$. 
Let us then choose $\barre{t}':= \barre{s}'$.
It satisfies:
$\barre{s}' \torok^0_\RR \barre{t'}$ and $ \barre{t'}= \barre{s}'\to^*_\AA \barre{s} \to_{\AA} \bar{t}$. Hence the conclusion of the lemma holds.\\
Suppose now that $\barre{s} \to_{\SS}\barre{t}$. This means that
$$\barre{s} = \barre{C}[\barre{l}\barre{\tau}],\;\;\barre{t} = \barre{C}[r\dbarre{\tau}]$$
for some rule $\rrule{l}{r} \in \RR$, marked context $\barre{C}$, and marked substitution $\barre{\tau}$,
satisfying (\ref{e-wbu-rule}-\ref{e-short-substitution}).\\
Since $\barre{s}' \to^*_\AA \barre{s}$ it must have the form
$$\barre{s}' = \barre{C}[\barre{l}\barre{\tau}']$$ 
where, for every $x \in \Var(l)$, $x \barre{\tau}'\to^*_\AA x \barre{\tau}$.
Let us set
$$ x \dbarre{\tau}':= x \barre{\tau}'\odot M(\barre{l},x),\;\;
\barre{t}' := \barre{C}[r\dbarre{\tau}'].$$
Since $\barre{s'}$ is $\m$-increasing, $M(\barre{l},x)= M(C[\barre{l}],x)$.
Hence, by definition of $\toro$, $\barre{s}' \toro_\RR \barre{t}'$
 and by condition (\ref{e-wbu-rule}) this step is $\wbu$, i.e.
$$\barre{s}' \torok_\RR \barre{t}'.$$
By Lemma~\ref{l-odot-is_Acompatible}, for every $x$,
$$x \dbarre{\tau}' =x \barre{\tau}' \odot M(\barre{l},x)  \to^*_\AA 
x \barre{\tau} \odot M(\barre{l},x)= x \dbarre{\tau}.$$
Hence $\barre{t}' = \barre{C}[r\dbarre{\tau}'] \to^*_\AA \barre{C}[r\dbarre{\tau}]= \barre{t}$.\\
2- 
Let us prove, by induction over the integer $n \geq 0$,  the statement 
$$\forall n \in \N, \forall \barre{s},\barre{s'}, \barre{t}\in \TTqkm$$
\begin{equation}
\label{e-implication_for-n}
(\barre{s'}\; \m\mbox{-increasing} \myand \barre{s}' \to^*_\AA \barre{s} \myand \barre{s} \to^n_{\SS \cup \AA}\barre{t})
\Rightarrow  \exists \bar{t'}, 
(\barre{s}' \torok^*_\RR \barre{t'} \myand  \barre{t'}\to^*_\AA \barre{t}).
\end{equation}
(here $\barre{t'}$ is implicitly quantified over $\TTqkm$).\\
{\bf Basis}: $n=0$.\\
In this case $\barre{s}' =\barre{s}$.
Choosing $\barre{t'} := \barre{t}$, the conclusion of implication (\ref{e-implication_for-n}) holds.
{\bf Induction step}: $n \geq 1$.\\
Let us suppose that the hypothesis of implication (\ref{e-implication_for-n}) holds.
There exists a term $\hat{t} \in \TTqkm$ such that
$$\barre{s} \to^{n-1}_{\SS \cup \AA}\hat{t} \to^1_{\SS \cup \AA}\barre{t}.$$
By induction hypothesis, there exists some $\hat{t'} \in \TTqkm$ such that
\begin{equation}
\label{e-inductionstep_partn-1}
\barre{s}' \torok^*_\RR \hat{t'} \myand  \hat{t'}\to^*_\AA \hat{t}.
\end{equation}
By Lemma \ref{l-marks_increase} $\hat{t'}$ is $\m$-increasing
and by point 1 of this proof, there exists some $\barre{t'} \in \TTqkm$ such that
\begin{equation}
\label{e-inductionstep_lastpart}
\hat{t'} \torok^*_\RR \barre{t'} \myand  \barre{t'}\to^*_\AA \barre{t}.
\end{equation}
Putting together statements (\ref{e-inductionstep_partn-1}) and (\ref{e-inductionstep_lastpart}), 
we obtain the conclusion of implication (\ref{e-implication_for-n}).
\end{proof}
\begin{remark}
The assumption that $\AA$ is standard (\ref{A-deterministic}) is not used in the above proof. 
Hence Lemma \ref{l-SliftedtoR} also holds without this restriction.
\label{r-Anonstandard-allowed-forlifting}
\end{remark}
\begin{lemma}[projecting one step of $\RR$ on $\SS \cup {\cal A}$]$\;\;$\\
\label{l-RprojectstoS1}
Let $\barre{s},\barre{t} \in \TTqkm$ such that:\\
1- $\barre{s} \toro_\RR \barre{t}$,\\
2- The marked rule $(\barre{l},r)$ used in the above rewriting-step is such that $\m(\barre{l})=0$.\\
3- $\barre{s}$ is $\m$-increasing.\\
Then, 
$\Top(\barre{s}) \to^*_\AA \to_\SS \Top(\barre{t})$.
\end{lemma}
\begin{proof}
Let us assume hypotheses (1,2,3) of Lemma~\ref{l-RprojectstoS1}. In particular:
$$\barre{s}=\barre{C}[\barre{l}\barre{\sigma}],\;\;\barre{t}=\barre{C}[r\dbarre{\sigma}] $$
for some $\barre{C},\barre{\sigma},\barre{l},r,\dbarre{\sigma}$ fulfilling (\ref{start_marqued_onestep}-\ref{marqued_onestep}) and $\m(\barre{l})=0$.
Let us then define a context $\barre{D}$ and marked substitutions $\barre{\tau},\dbarre{\tau}$ by: 
\begin{equation}
\label{d-barD}
\barre{D}[]=\Top(\barre{C}[]).
\end{equation}
\begin{equation}
\label{d-bartau}
\forall x \in \VV,\;x\dbarre{\tau}=\Top(x\dbarre{\sigma}),\;\;
x\barre{\tau}=\Red(x\barre{\sigma},\Pos(x\dbarre{\tau})).
\end{equation}
We claim that
\begin{equation}
\Top(\barre{s}) \to^*_\AA \barre{D}[\barre{l}\barre{\tau}] \to_\SS 
\barre{D}[r\dbarre{\tau}] = \Top(\barre{t}).
\label{THE-claim}
\end{equation}
We cut into four facts the detailed verification of this claim.
\begin{fact}
\label{f-domain-inclusion}
$\Pos(\barre{l}\Top(\dbarre{\sigma})) \subseteq \Pos(\Top(\barre{l}\dbarre{\sigma}))$.
\end{fact}
\begin{figure}[htb]
\centering
\input domain_inclusion.pstex_t
\caption{Fact~\ref{f-domain-inclusion}}
\label{schema_domain_inclusion}
\end{figure}
\noindent Let $u \in \Pos(\barre{l}\Top(\dbarre{\sigma}))$.\\
{\bf Case 1}: $u \in \Pos_{\barre{\VV}}(\barre{l})$.\\
In this case $|u| \leq \dmax$. Hence, for every factorization $u = u_1 \cdot u_2$,
since $\m(\barre{t}/u_1) \leq k$, 
$$ |u_2| \leq |u| \leq \dmax \leq (k+1 - \m(\barre{t}/u_1))\dmax.$$
{\bf Case 2}:
$$u = v \cdot w $$
for some $x \in \Var(l), v=\pos(\barre{l},x),w \in \Topd(x\dbarre{\sigma})$.
Let us consider any decomposition $u= u_1 \cdot u_2$ and show it fulfils 
condition (2) of Definition~\ref{d-Topd}.\\
We use the notation
$$m_1=\ma{\barre{l}\dbarre{\sigma}/u_1},\;\;m=\ma{\barre{l}\dbarre{\sigma}/f}$$
where $f$ is the father of $v$.
If $m_1=0$ this condition (2) is clearly true. Let us assume that $m_1 \geq 1$.\\
{\bf Case 2.1}: $u_1 \preceq v$.\\
In this case there exists $u'_1$ such that
$$v=u_1u'_1,\;\;u_2=u'_1w,\;\;|u'_1| \geq 0.$$
As $w \in \Topd(x\dbarre{\sigma})$,
\begin{equation}
|w| \leq (k+1-\ma{x\dbarre{\sigma}})\dmax 
\label{w_in_topd}
\end{equation}
but $\ma{x\dbarre{\sigma}} \geq M(\barre{l},x)=m+1$, hence
\begin{equation}
|w| \leq (k+1-m-1)\dmax.
\label{w_is_small}
\end{equation}
Using the fact that $|u'_1| \leq \depth(\barre{l})\leq \dmax$ we obtain that
\begin{equation}
|u'_1w| \leq (k+1-m-1)\dmax+\dmax=(k+1-m)\dmax
\label{u2_is_small11}
\end{equation} 
and, since the marks increase from top to leaves, $m \geq m_1$, so that
\begin{equation}
|u'_1w| \leq (k+1-m_1)\dmax
\label{u2_is_small12}
\end{equation}
which can be reformulated as
\begin{equation}
|u_2| \leq (k+1-\ma{\barre{l}\dbarre{\sigma}/u_1})\dmax.
\label{u2_is_small1}
\end{equation}
{\bf Case 2.2}: $v \prec u_1$.\\
In this case there exists $u'_1$ such that
$$u_1=vu'_1,\;\;u'_1u_2=w,\;\;|u'_1| \geq 1.$$
As $w \in \Topd(x \dbarre{\sigma})$
\begin{equation}
|u_2| \leq (k+1-\ma{x\dbarre{\sigma}/u'_1})\dmax
\label{u2_is_small21}
\end{equation}
which can be rewritten
\begin{equation}
|u_2| \leq (k+1-\ma{\barre{l}\dbarre{\sigma}/u_1})\dmax.
\label{u2_is_small22}
\end{equation}
Since in all cases condition (2) of Definition~\ref{d-Topd} is fulfilled, 
Fact~\ref{f-domain-inclusion} is established.
\begin{fact}
\label{f-sigma-Aderivation}
$\Top(\barre{l}\barre{\sigma}) \to^*_\AA \barre{l}\barre{\tau}$.
\end{fact}
\noindent We know that 
\begin{equation}
\barre{l}\barre{\sigma} \to^*_\AA \Top(\barre{l}\barre{\sigma})
\label{e-smallreduction}
\end{equation} 
(by definition of $\Top$) and that 
\begin{equation}
\barre{l}\barre{\sigma} \to^*_\AA \barre{l}\barre{\tau}
\label{e-bigreduction}
\end{equation}
because, by (\ref{d-bartau}), every $x \barre{\tau}$ is a reduct of the corresponding 
$x \barre{\sigma}$.
Moreover, by Fact~\ref{f-domain-inclusion}, 
$$\Pos(\barre{l}\barre{\tau})= \Pos(\barre{l}\Top(\dbarre{\sigma})) \subseteq \Pos(\Top(\barre{l}\dbarre{\sigma})),$$
and by Lemma \ref{l-Top-is-decreasing}  $ \Pos(\Top(\barre{l}\dbarre{\sigma})) \subseteq 
\Pos(\Top(\barre{l}\barre{\sigma})),$
so that
\begin{equation}
\Pos(\barre{l}\barre{\tau}) \subseteq \Pos(\Top(\barre{l}\barre{\sigma})). 
\label{e-domaininclusion}
\end{equation}
Lemma~\ref{l-confluence-commondomain} applied to (\ref{e-smallreduction}-\ref{e-domaininclusion})
 shows that
$\Top(\barre{l}\barre{\sigma}) \to^*_\AA \barre{l}\barre{\tau}$.
\begin{fact}
\label{f-tau-Sderivation}
$\barre{D}[\barre{l}\barre{\tau}] \to_\SS \barre{D}[r\dbarre{\tau}]$.
\end{fact}
\noindent By hypothesis ($2$) of the lemma, $\ma{\barre{l}}=0$.\\
By the general assumption (\ref{no-empty-lhs}) and hypothesis ($3$) of the lemma,
$$\forall x \in \Var(l), \;\M(\barre{l},x)= \M(\barre{C}[\barre{l}],x),$$
hence
$$x\dbarre{\tau}:= x \barre{\tau} \odot \M(\barre{l},x).$$
Moreover, $\depth(x\barre{\tau})\leq (k+1-\M(\barre{l},x)) \cdot \dmax \leq k \cdot \dmax$, 
since $\M(\barre{l},x)\geq 1$.
Hence $\rrule{\barre{l}\barre{\tau}}{r\dbarre{\tau}}$ is a rule of $\SS$.
\begin{fact}
\label{f-context-and-tau}
$\barre{D}[r \dbarre{\tau}] =\Top(\barre{t})$.
\end{fact}
\noindent This fact follows from Lemma~\ref{l-Top-is-morphic} and Lemma~\ref{l-Top-preserves-unmarked}.\\
Using these facts we obtain that
\begin{eqnarray*}
\Top(\barre{s}) & = \barre{D}[\Top(\barre{l}\barre{\sigma})]
& \mbox{( by Lemma~\ref{f-context-isunmarked} and Lemma~\ref{l-Top-is-morphic})}\\
\barre{D}[\Top(\barre{l}\barre{\sigma})]& \to^*_{\AA} \barre{D}[\barre{l}\barre{\tau}]
& \mbox{( by Fact~\ref{f-sigma-Aderivation})}\\
\barre{D}[\barre{l}\barre{\tau}] &\to_{\SS} \barre{D}[r \dbarre{\tau}] 
& \mbox{( by Fact~\ref{f-tau-Sderivation})}\\
\barre{D}[r \dbarre{\tau}] & = \Top(\barre{t}) 
& \mbox{( by Fact~\ref{f-context-and-tau})}.
\end{eqnarray*}
Thus claim (\ref{THE-claim}) is verified, which proves the lemma.
\end{proof}
\begin{lemma}[projecting $\RR$ on $\SS \cup {\cal A}$]$\;\;$\\
\label{l-RprojectstoSn}
Let $\barre{s}, \barre{t} \in \TT(\FF^{\leq k})$ and assume that $\barre{s}$ is $\m$-increasing.
If $\barre{s} \torok^*_\RR \barre{t}$ then, there exist terms $\barre{s}', \barre{t}' \in \TT((\FF \cup Q)^{\leq k})$ such that
$$\barre{s}\to^*_{\AA}\barre{s}' \to^*_{\SS \cup \AA} \barre{t'}\;\; \mbox{ and }\;\;
\barre{t}\to^*_{\AA}\barre{t}'.$$
\end{lemma}
\begin{proof}
The marked derivation $\barre{s} \toro^*_\RR \barre{t}$ is $\wbu$, hence it can be decomposed
into $n$ successive steps where the hypothesis 2 of Lemma~\ref{l-RprojectstoS1} is valid.
Hypothesis 3 of Lemma~\ref{l-RprojectstoS1} will also hold, owing to our assumption and to Lemma~\ref{l-marks_increase}.
We can thus deduce, inductively, from the conclusion of Lemma~\ref{l-RprojectstoS1}, that
$\Top(\barre{s})\to^*_{\SS \cup \AA} \Top(\barre{t} )$. The choice $\barre{s}':=\Top(\barre{s}),
\barre{t}':=\Top(\barre{t})$ fulfills the conclusion of the lemma.
\end{proof}
\begin{lemma}
\label{l-RviaSetA}
Let $s \in \TT(\FF)$. Then  
$s \tok^*_\RR T$ iff  $s \to^*_{\SS \cup \AA}Q_f^{\leq k}$. 
\end{lemma}
\begin{proof}$\;\;$\\
({\bf $\Rightarrow$}): Suppose $s \tok^*_\RR t$ and $t \in T$.
Let us consider the corresponding marked derivation
\begin{equation}
\label{e-initial-markedD}
\barre{s} \torok^*_\RR \barre{t}
\end{equation}
where $\barre{s} :=s$. Derivation (\ref{e-initial-markedD}) is $\wbu$ 
and lies in $\TTkm$. Let us consider the terms $\barre{s}',\barre{t}'$ given by Lemma~\ref{l-RprojectstoSn}:
\begin{equation}
\label{e-projectedD}
\barre{s} \to^*_\AA \barre{s}' \to^*_{\SS \cup \AA} \barre{t}'
\end{equation}
and $\barre{t} \to^*_\AA \barre{t}'$.
Since $\barre{t} \to^*_\AA Q_f^{\leq k}$, by Lemma~\ref{l-confluence-commondomain},
\begin{equation}
\label{e-projectedC}
\barre{t}' \to^*_\AA Q_f^{\leq k}.
\end{equation}
Combining (\ref{e-projectedD}) and (\ref{e-projectedC}) we obtain
$$s \to^*_{\SS \cup \AA} Q_f^{\leq k}.$$
({\bf $\Leftarrow$}): Suppose $s \to^*_{\SS \cup \AA} q^j \in Q_f^{\leq k}$.\\
The hypotheses of Lemma~\ref{l-SliftedtoR} are met by
$\barre{s}:=s,\barre{s}':=s$ and $\barre{t}:=q^j$. By Lemma~\ref{l-SliftedtoR} there exists some 
$\barre{t}'\in \TTqkm$ such that
$$\barre{s} \torok^*_\RR \barre{t'}\to^*_\AA q^j \in Q_f^{\leq k}.$$
These derivations are mapped (by removal of the marks) into:
$$ s \tok^*_\RR t'\to^*_\AA q\in Q_f,$$
which shows that $t' \in T$ hence that $s \tok^*_\RR T$.
\end{proof}
\noindent We can now prove {\bf Theorem~\ref{t-kbu-inverse-preserving}}.
\begin{proof}
By Lemma~\ref{l-RviaSetA},
$(\tok^*_\RR)[T] = (\to^*_{\SS \cup \AA})[Q_f^{\leq k}] \cap \TT(\FF)$. The rewriting systems $\SS$ and $\AA$ being ground are inverse-recognizability preserving (Theorem~\ref{t-ground-inverse-preserving}).
So $(\to^*_{\SS \cup \AA})[Q_f^{\leq k}]$ is recognizable and thus $(\tok^*_\RR)[T]$ is recognizable. \end{proof}
\begin{corollary}
\label{c-BU-inverse-preserving}
Every linear rewriting system of the class $\BU$ is inverse-recognizability preserving.
\end{corollary}
\begin{proof}
If $\RR$ belongs to $\BU(k)$, then $(\to^*_\RR)[T] = (\tok^*_\RR)[T]$. 
\end{proof}
\begin{remark}
In the above proof of corollary \ref{c-BU-inverse-preserving} we could use the ground
rewriting system ${\SS^0} \cup \AA$ over the signature $\FF$ (recall that ${\SS}^0$ is obtained
from $\SS$ by forgetting the marks):
when $\RR$ belongs to $\BU(k)$,
$$(\to^*_\RR)[T]= (\to^*_{{\SS^0} \cup \AA})[Q_f] \cap \TT(\FF). $$
This also gives an effective way for computing a $\fta$ recognizing $(\to^*_\RR)[T]$.
\end{remark}
\begin{example*}
With $\RR_1$ of example~\ref{ex-1} and $\AA = (\FF, \{ q_{\mA}\},\{ q_{\mA}\}, \{ \mA \to q_{\mA}\})$
recognizing $T = \{ \mA \}$, we obtain
\[
\SS^0 \supseteq \{ \mA \to q_{\mA}\} \cup 
 \{ \mF(q_{\mA}) \to g(q_{\mA}), g(h(q_{\mA})) \to \mI(q_{\mA}), \mI(q_{\mA}) \to \mA \}
\]
\end{example*}
\begin{example*}
The derivation $\mF(\mH(\mF(\mH(\mA)))) \toro^* \mA$ given in Example~\ref{ex-toro}
may be simulated by $\SS^0$:
$$
\begin{array}{l}
\mF(\mH(\mF(\mH(\mA)))) \to_{\SS^0} \mF(\mH(\mF(\mH(q_{\mA})))) \to_{\SS^0} \mF(\mH(\mG(\mH(q_{\mA}))))
\to_{\SS^0} \\
\mF(\mH(\mI(q_{\mA}))) \to_{\SS^0} \mF(\mH(q_{\mA}))\to_{\SS^0} \mG(\mH(q_{\mA})) 
\to_{\SS^0} \mI(q_{\mA}) \to \mA\\
\end{array}
$$
\end{example*}

\subsection{General construction}
\label{s-general-construction}
We show here that Theorem~\ref{t-kbu-inverse-preserving} still holds when the
restrictions (\ref{no-empty-lhs}-\ref{A-deterministic}) are removed.

\subsubsection{Allowing variable lhs}
\label{s-variable-lhs}
Let $\RR$ be some left-linear finite rewriting system over the signature $\FF$.
We show here how to reduce the properties of this TRS $\RR$ to properties of 
a TRS which has no variable left-handside nor any variable right-handside 
(this reduction is borrowed from \cite{Syl10}).\\
Let us introduce a new unary symbol $\#_1 \notin \FF$ and consider the signature
$\FF_1:=\FF \cup \{\#_1\}$. We consider the map $E_1:\TERM \rightarrow \TT(\FF_1,\VV)$ 
defined inductively by:
$$\forall v \in \VV, E_1(v) = v,\;\;
\forall a \in \FF_0, E_1(a) = \#_1(a),$$
$$\forall n \geq 1, \forall f \in \FF_n,\forall t_1,\ldots,t_n \in \TERM,
E_1(f(t_1,\ldots,t_n)) = \#_1(f(E_1(t_1),\ldots,E_1(t_n))).$$
It is clear that $E_1$ is an injective map and, since $E_1$ is a term-homorphism,
for every subset $T \subseteq \TERM$, $T$ is recognizable if and only if 
$E_1(T)$ is recognizable.
We define a new TRS
$$\RR_1:= \{\rrule{E_1(l)}{E_1(r)} \mid \rrule{l}{r} \in \RR\}.$$
The system $\RR_1$ is a left-linear finite rewriting system over the signature $\FF_1$
and every rule $(l_1,r_1) \in \RR_1$ is such that $l_1 \notin \VV, r_1 \notin \VV$.

\begin{lemma}[$\RR$ embeddable in $\RR_1$]
\label{l-RR-embedded-in-RR1}
For every $s,t \in \GTERM$ and integer $k \geq 0$,\\
1- $s \to^*_\RR t \Leftrightarrow  E_1(s) \to^*_{\RR_1} E_1(t)$\\
2- $s \tok^*_\RR t \Leftrightarrow  E_1(s) \tok^*_{\RR_1} E_1(t)$
\end{lemma}
\noindent In particular:
$s \tok^*_\RR T \Leftrightarrow  E_1(s) \tok^*_{\RR_1} E_1(T)$ and
$\RR$ is $\BU(k)$ iff $\RR_1$ is $\BU(k)$.\\
Hence Theorem~\ref{t-kbu-inverse-preserving} and Corollary~\ref{c-BU-inverse-preserving} 
still hold, without assuming (\ref{no-empty-lhs}).
\subsubsection{Allowing non-deterministic automata}
\label{s-non-deterministicA}
Let $\RR$ be some left-linear finite rewriting system over the signature $\FF$
fulfilling restriction (\ref{no-empty-lhs}) and let $\AA=(\FF, Q, Q_f, \sys)$ be some 
\fta recognizing a language $T$ (this \fta is not assumed standard, nor merely deterministic).
\paragraph{Automaton $\hat{\AA}$}
\label{p-Automaton-hatA}
Let us define $\hat{\AA}:=(\FF, \FF_0 \cup \hat{Q}, \FF_{0,f} \cup \hat{Q}_f, \hat{\sys})$ 
by:
\begin{eqnarray*}
\hat{Q} & := & {\cal P}(Q)\\
\FF_{0,f}& := &\{ a \in \FF_0 \mid a \in L(\AA) \}\\
\hat{Q}_f & := & \{ P \in \hat{Q} \mid P \cap Q_f \neq \emptyset \}\\
\hat{\Gamma}& := & \{ f(P_1,\ldots,P_m) \to P \mid m \geq 1,f \in \FF_m, P_1,\ldots,P_m  \in \FF_0 \cup \hat{Q},\\
 && P= Q \cap [f(P_1,\ldots,P_m)]({\to_\AA^*})\}.
\end{eqnarray*}
Some precisions about our notation:\\
- in the last definition the $P_i$ which are equal to an element $a_i \in \FF_0$ are identified with
the singleton $\{a_i\}$ in the notation  $[f(P_1,\ldots,P_m)]({\to_\AA^*})$.\\
- with this convention, $f(P_1,\ldots,P_m)$ denotes the set $\{ f(p_1,\ldots,p_m) \mid
p_1 \in P_1,\ldots, p_m \in P_m\}$.\\
- $[f(P_1,\ldots,P_m)]({\to_\AA^*})$ denotes the set of descendants of $f(P_1,\ldots,P_m)$,
as defined in equality (\ref{e-descendants-of-T}).

Note that this construction of $\hat{\AA}$ from $\AA$ is just a slight variant of the usual powerset-construction.
We still denote by $\SS$ the system deduced from $\RR$ and $\AA$ along Definition \ref{d-ssystem};
we denote by $\hat{\SS}$ the system deduced from $\RR$ and $\hat{\AA}$ along 
Definition \ref{d-ssystem}.
We shall show that the structures $(\TT(\FF\cup Q),\to_{\SS \cup \AA})$ and $(\TT(\FF\cup \hat{Q}),\to_{\hat{\SS} \cup \hat{\AA}})$ are very close to each other. A precise formulation will be given in terms 
of {\em simulation} (see Definition \ref{d-simulation}).
\begin{lemma}For every \fta $\AA$, the \fta $\hat{\AA}$ is standard.
\label{l-hat_is_standard}
\end{lemma}
\noindent This lemma follows immediately from the above definition.
\paragraph{Simulations for $\AA$ and $\hat{\AA}$}
We define a binary relation $\approx \subseteq (\FF_0 \cup Q \cup \VV) \times (\FF_0 \cup \hat{Q} \cup \VV)$ by:
\begin{eqnarray*}
\approx & := & \{ (a,a) \mid a \in \FF_0\} \cup\{ (v,v) \mid v \in \VV \}\cup
\{(p,P) \mid p \in P, P \in \hat{Q}\}\\
& \cup &
\{(q,a) \mid q \in Q, a \in \FF_0, a \to_\AA^*q\}.
\end{eqnarray*}
We extend $\approx$ into the binary relation $\sim \subseteq \TT((\FF \cup Q)^\N,\VV) \times \TT((\FF \cup \hat{Q})^\N,\VV)$ defined as follows
\begin{definition}For every $\bar{t},\hat{t} \in 
\TT((\FF \cup Q)^\N,\VV) \times \TT((\FF \cup \hat{Q})^\N,\VV)$,
$\bar{t} \sim \hat{t}$ if and only if\\
1- $\Pos(\bar{t}) = \Pos(\hat{t})$\\
2- $\forall u \in \Internal(\bar{t}), \bar{t}(u) = \hat{t}(u)$\\
3- $\forall u \in \Leaves(\bar{t}), \ma{\bar{t}/u}=\ma{\hat{t}/u} \myand \bar{t}^0(u) \approx 
\hat{t}^0(u).$
\label{d-sim}
\end{definition}
\begin{lemma}$\;\;$\\
\noindent{1}- $\sim$ is a simulation of $(\TT((\FF \cup Q)^\N,\VV), \to_\SS)$ by 
$((\TT(\FF \cup \hat{Q})^\N,\VV),\to_{\hat{\SS}})$.\\
2- $\sim^{-1}$ is a simulation of $(\TT((\FF \cup \hat{Q})^\N,\VV),\leftarrow_{\hat{\SS}})$ by 
$(\TT((\FF \cup Q)^\N,\VV), \leftarrow_\SS)$.\\
\label{l-Ssimulations}
\end{lemma}
\begin{figure}[htb]
\centering
\input simulation_S.pstex_t
\caption{Lemma \ref{l-Ssimulations}}
\label{diagram-simulation_S}
\end{figure}
\begin{proof}$\;\;$\\
{\bf Point 1} Let us suppose that $\bar{s},\bar{t} \in \TT((\FF \cup Q)^\N,\VV), \barre{s'} \in (\TT(\FF \cup \hat{Q})^\N,\VV)$ are such that $\bar{s} \sim \barre{s'}$ and $\barre{s} \to_{\SS} \barre{t}$.\\
Thus
\begin{equation*}
\barre{s}=\barre{C}[\barre{l}\barre{\tau}],\;\;\barre{t}=\barre{C}[r \dbarre{\tau}]
\end{equation*}
for some context $\barre{C}[]$, rule $\rrule{l}{r} \in \RR$ and substitutions 
$\barre{\tau},\dbarre{\tau}$ fulfilling (\ref{e-short-substitution}).
By Definition \ref{d-sim} the term $\barre{s'}$ has the form
\begin{equation*}
\barre{s'}=\barre{C'}[\barre{l'}\barre{\tau'}]
\end{equation*}
with
\begin{equation}
\bar{l} \sim \barre{l'},
\label{barl_sim_hatl}
\end{equation}
\begin{equation}
\bar{C} \sim \barre{C'},
\label{barC_sim_hatC}
\end{equation}
\begin{equation}
\forall x \in \Var(l), x\bar{\tau} \sim x\barre{\tau'}.
\label{bartau_sim_hattau}
\end{equation}
Every label of a leaf of $\barre{l}$ belongs to $\FF^\N_0 \cup \VV$. Relation 
(\ref{barl_sim_hatl})
thus implies that $\bar{l} = \barre{l'}$, hence that 
\begin{equation}
l' \to r \in \RR.
\label{hat_is_rule}
\end{equation}
Let us define
\begin{equation*}
\barre{t'}:=\barre{C'}[r \dbarre{\tau'}]
\end{equation*}
where the substitution $\dbarre{\tau'}$ is defined on every $x \in \Var(l)$ by
\begin{equation*}
x\dbarre{\tau'}:= x \barre{\tau'} \odot M(\barre{l'},x).
\end{equation*}
Relation (\ref{bartau_sim_hattau}) implies that, for every $x \in \Var(l)$,
$(x\barre{\tau})\odot M(\barre{l},x) \sim (x \barre{\tau'})\odot M(\barre{l'},x)$
i.e.
\begin{equation}
x\dbarre{\tau}\sim x \dbarre{\tau'}.
\label{dbtau_sim_hatdbtau}
\end{equation}
Statement (\ref{hat_is_rule}) shows that $\barre{s'} \to_{\hat{\SS}}\barre{t'}$ while
relations (\ref{barC_sim_hatC}) and (\ref{dbtau_sim_hatdbtau}) show that $\barre{t} \sim \barre{t'}$.
Point 1 of the lemma is thus proved .\\
{\bf Point 2}\\
Let us suppose that $\bar{t} \in \TT((\FF \cup Q)^\N,\VV), \barre{s'},\barre{t'} \in (\TT(\FF \cup \hat{Q})^\N,\VV)$ are such that $\bar{t} \sim \barre{t'}$ and $\barre{s'} \to_{\hat{\SS}} \barre{t'}$.\\
We know that
\begin{equation*}
\barre{s'}=\barre{C'}[\barre{l}\barre{\tau'}],\;\;\barre{t'}=\barre{C'}[r\dbarre{\tau'}]
\end{equation*}
for some context $\barre{C'}[]$, rule $\rrule{l}{r} \in \RR$ and substitutions 
$\barre{\tau'},\dbarre{\tau'}$ fulfilling (\ref{e-short-substitution}).
Since $\barre{t} \sim \barre{t'}$ we must have
\begin{equation*}
\barre{t}=\barre{C}[r\dbarre{\tau}]
\end{equation*}
for some $\barre{C}[]\sim \barre{C'}[]$ and some substitution $\dbarre{\tau}$ fulfilling
\begin{equation}
\forall x \in \Var(r), x \dbarre{\tau} \sim x {\dbarre{\tau'}}.
\label{dbarretau_sim_hatdbarretau}
\end{equation}
 Let us define
\begin{equation}
\barre{s} := \barre{C}[\barre{l} \barre{\sigma}]
\label{def_barre_t}
\end{equation}
where the substitution $\barre{\sigma}$ is built in the following way:
\begin{equation*}
\forall x \in \Var(l), \Pos(x \sigma) := \Pos(x \tau)
\end{equation*}
The labels of the unmarked underlying substitution are defined by
\begin{eqnarray*}
\forall x \in \Var(r), \forall u \in \Pos(x\barre{\tau}), 
x\sigma (u)& := & x \tau(u) \nonumber\\
\forall x \in \Var(l) \setminus \Var(r),\forall u \in \Internal(x\sigma), 
x\sigma(u) & := & x \tau'(u) \nonumber\\
\forall x \in \Var(l) \setminus \Var(r), \forall u \in \Leaves(x\sigma), 
x\sigma(u) & \approx & x \tau'(u)
\end{eqnarray*}
(this last choice can be made because, for every symbol $\alpha' \in \FF_0 \cup \hat{Q}$ there exists
some $\alpha \in \FF_0 \cup Q$, such that $\alpha \approx \alpha'$)\\
and the marks are defined by
\begin{equation}
\forall x \in \Var(l), \forall u \in \Pos(x \sigma),
\ma{x \barre{\sigma}/u} := \ma{x {\barre{\tau'}}/u}
\label{barretau_same marks_hatbarretau}
\end{equation}
From (\ref{def_barre_t}), the construction of $\barre{\sigma}$ and (\ref{dbarretau_sim_hatdbarretau})
follows the property that 
$$\barre{s} \sim \barre{s'}.$$
Since the marks of $\barre{\sigma}$ are taken from those of ${\barre{\tau'}}$ (see (\ref{barretau_same marks_hatbarretau})), using the hypothesis that $\barre{s'} \to_{\hat{\SS}} \barre{t'}$, we obtain that
$$\barre{s} \to_{\SS} \barre{t}.$$
\end{proof}
\begin{lemma}$\;\;$\\
\noindent{1}- $\sim$ is a simulation of $(\TT((\FF \cup Q)^\N,\VV), \to_\AA)$ by 
$(\TT((\FF \cup \hat{Q})^\N,\VV),\to^*_{\hat{\AA}})$.\\
2- $\sim^{-1}$ is a simulation of $(\TT((\FF \cup \hat{Q})^\N,\VV),\leftarrow_{\hat{\AA}})$ by 
$(\TT((\FF \cup Q)^\N,\VV), \leftarrow_\AA)$.
\label{l-Asimulations}
\end{lemma}
\begin{figure}[htb]
\centering
\input simulation_A.pstex_t
\caption{Lemma \ref{l-Asimulations}}
\label{diagram-simulation_A}
\end{figure}
\begin{proof}$\;\;$\\
{\bf Point 1}\\
Let us suppose that $\bar{s},\bar{t} \in \TT((\FF \cup Q)^\N,\VV), \barre{s'} \in (\TT(\FF \cup \hat{Q})^\N,\VV)$ are such that $\bar{s} \sim \barre{s'}$ and $\barre{s} \to_{\AA} \barre{t}$.\\
Thus
\begin{equation*}
\barre{s}=\barre{C}[\barre{l}],\;\;\barre{t}=\barre{C}[\barre{r}],
\;\;\barre{s'}=\barre{C'}[\barre{l'}]
\end{equation*}
for some contexts $\barre{C}[],\barre{C'}[]$ and rule $\rrule{l}{r} \in \Gamma$ fulfilling
$\barre{C}[] \sim \barre{C'}[], \barre{l} \sim \barre{l'}$.\\
{\bf case 1.1}: $l \to r= a \to q$ for some $a \in \FF_0, q \in Q$.\\
We define 
\begin{equation}
\barre{t'}:= \barre{s'}.
\end{equation}
Since $a \to_\AA q$ , the relation $q \approx a$ holds, hence $\barre{t} \sim \barre{t'}$.
It is clear that $\barre{s'} \to^*_{\hat{\AA}} \barre{t'}$.\\
{\bf case 1.2}: $l \to r= f(p_1,\ldots,p_m) \to p$ for some $m \geq 1, f \in \FF_m,p_1,\dots,p_m,p \in Q$.\\ 
We thus have
\begin{equation*}
\barre{l} = \barre{f}(\barre{p}_1,\ldots,\barre{p}_m),\;\;
\barre{r} = \barre{p},\;\
\barre{l'} = \barre{f}(\barre{P}_1,\ldots,\barre{P}_m)
\end{equation*}
for some $\barre{P}_i \in (\FF_0 \cup \hat{Q})^\N$ such that $\barre{p}_i \approx \barre{P}_i$.
Let us define $\barre{P'} \in \hat{Q}^\N$ by
\begin{equation}
P':= Q \cap [{f} (P_1,\ldots,P_m)]_{\to_\AA^*},\;\;
\ma{\barre{P'}} := \ma{\barre{f}}
\label{d-hatPprime}
\end{equation}
and finally
\begin{equation*}
\barre{t'} := \barre{C'}[\barre{P'}]. 
\end{equation*}
Since $p_i \subseteq P_i$ (if $P_i \in \hat{Q}$) or $P_i \to_\AA^* p_i$ (if $P_i \in \FF_0$),
$p \in P'$. It follows that $\barre{t} \sim \barre{t'}$.
The definition (\ref{d-hatPprime}) of $\barre{P'}$ also implies that
$\barre{f}(\barre{P}_1,\ldots,\barre{P}_m)\to_{\hat{A}} \barre{P'}$, hence that
$\barre{s'} \to_{\hat{A}} \barre{t'}$.\\
{\bf Point 2}\\
Let us suppose that $\bar{t} \in \TT((\FF \cup Q)^\N,\VV), \barre{s'},\barre{t'} \in (\TT(\FF \cup \hat{Q})^\N,\VV)$ are such that $\bar{t} \sim \barre{t'}$ and $\barre{s'} \to_{\hat{\AA}} \barre{t'}$.\\
We know that
\begin{equation*}
\barre{t}=\barre{C}[\barre{p}],\;\;\barre{s'}=\barre{C'}[\barre{l'}],
\;\;\barre{t'}=\barre{C'}[\barre{r'}]
\end{equation*}
for some contexts $\barre{C}[],\barre{C'}[]$, symbol $\barre{p} \in (\FF \cup Q)^\N\cup\VV$ and rule $\rrule{l'}{r'} \in \hat{\Gamma}$. Such a rule has the form
\begin{equation}
l'= f(P'_1,\ldots,P'_m) \to P'=r'
\label{rule_below}
\end{equation}
where $P'_i \in \FF_0 \cup \hat{Q},P' \in \hat{Q}$. Since $\bar{t} \sim \barre{t'}$, we must have
$\barre{C} \sim \barre{C'}$ and $p \in P'$. Since $p \in P'$ there exist
$p_1,\ldots,p_m \in \FF_0 \cup Q$ such that, $f(p_1,\ldots,p_m) \to_\AA^* p $ and,
for every $i \in [1,m]$, either $p_i=P'_i \in \FF_0$ or ($p_i \in Q , P'_i \in \hat{Q}, p_i \in P'_i$).
Let us define $\barre{s}$ by
\begin{equation*}
s := C[f(p_1,\ldots,p_m)],\;\;
\forall u \in \Pos(\barre{s}),\ma{\barre{s}/u} = \ma{\barre{s'}/u}.
\end{equation*}
Since $p_i \approx P'_i$ we get that $\barre{s} \sim \barre{s'}$ and since $f(p_1,\ldots,p_m) \to_\AA^* p $ we get that $\barre{s} \to_\AA^*  \barre{t}$.
\end{proof}
As a straightforward consequence of Lemma~\ref{l-Ssimulations} and 
Lemma~\ref{l-Asimulations} we get the following lemma.
\begin{lemma}$\;\;$\\
\noindent{1}- $\sim$ is a simulation of $(\TT((\FF \cup Q)^\N,\VV), \to_{\SS \cup \AA})$ by 
$(\TT(\FF \cup \hat{Q})^\N,\VV),\to^*_{\hat{\SS} \cup \hat{\AA}})$.\\
2- $\sim^{-1}$ is a simulation of $(\TT((\FF \cup \hat{Q})^\N,\VV),\leftarrow_{\hat{\SS}\cup \hat{\AA}})$ by 
$(\TT((\FF \cup Q)^\N,\VV), \leftarrow_{\SS \cup \AA})$.
\label{l-SSAsimulations}
\end{lemma}
Let us show that Theorem~\ref{t-kbu-inverse-preserving} still holds without assuming (\ref{no-empty-lhs}-\ref{A-deterministic}). 
By \S\ref{s-variable-lhs} we are reduced to treat the case of a system $\RR$ fulfilling (\ref{no-empty-lhs}).
Since $\hat{\AA}$ is standard, Lemma~\ref{l-RviaSetA} applies on $\RR$ and 
$\to^*_{\hat{\SS}\cup \hat{\AA}}$. Using then
Lemma~\ref{l-SSAsimulations} we get that
for every term $s \in \TT(\FF)$
$$s \tok^*_\RR T \Leftrightarrow s \to^*_{\SS \cup \AA}Q_f^{\leq k}$$
and we can conclude, as before, that $(\tok^*_\RR)[T]$ is recognizable.

\subsection{Complexity}
The proofs  that we gave for Theorem~\ref{t-kbu-inverse-preserving} are constructive i.e. give an 
{\em algorithm} for computing a non-deterministic \fta 
recognizing $(\tok^*_\RR)[T]$ from a non-deterministic \fta recognizing $T$ and a system
$\RR$ which belongs to the subclass $\BU(k)$.
We sketch here some estimation of the complexity of this algorithm: in \S \ref{s-complexity-bu1}
we treat in details the case of semi-Thue systems belonging to $\BU^-(1)$ and, later on, the
case of term rewriting systems in $\BU^-(1)$; in \S \ref{s-complexity-buk} we sketch  an analysis
of the more general case of systems in $\BU^-(k)$, for any natural integer $k$.
In \S \ref{s-lower-bound} we prove a NP-hardness lower-bound showing that
some of our upper-bounds 
cannot (presumably) be significantly improved.\\
\subsubsection{Upper-bounds for systems in $\BU^-(1)$}
\label{s-complexity-bu1}
Let us treat here the case where $\RR$ belongs to the subclass $\BU^-(1)$.
\paragraph{Semi-Thue systems}
\begin{theorem}
\label{t-complexityW}
Let $\FF$ be a signature with symbols of arity $\leq 1$,
let $\AA$ be some \nfta recognizing a language $T\subseteq \TT(\FF)$ and let $\RR$ be a finite rewriting system in $\BU^-(1)$.
One can compute a \nfta $\BB$ recognizing $(\to^*_\RR)[T]$ in time 
${\rm O}(|\FF|\cdot (\log(|\FF|))^3\cdot \|\AA\|^3\cdot \|\RR\|^3)$.
\end{theorem}
Our proof consists in reducing the above problem, via the computation of the ground system
$\SS$ of Section~\ref{s-basic-construction}, to the computation of a set of descendants 
modulo some
set of cancellation rules, which is achieved in cubic time in \cite{Ben-Sak86}.\\

Suppose $\FF$ is a fixed alphabet without arities and $\RR$ is a 
semi-Thue system over $\FF$.\\
We consider the {\em right-linear} one-step derivation relation generated by $\RR$:
for every $u,v \in \FF^*, u \torl_\RR v$ iff there exist $w \in \FF^*, \rrule{l}{r} \in \RR$ such that
$$u = w \cdot l,\;\;v= w \cdot r.$$
The binary relation $\torl^*_\RR$ is, as usual, the reflexive and transitive closure 
of $\torl_\RR$.
\begin{lemma}
\label{complexity_semithue_step1}
Let $T$ be a subset of $\FF^*$, recognized by a non-deterministic finite automaton $\AA$ and 
$\RR$ a semi-Thue system over $\FF$.
A \nfa recognizing $[T] (\torl^*_\RR)$ can be computed in time 
${\rm O}((\|\AA\| + \|\RR\|)^3)$.
\end{lemma}
\begin{proof}
Let us construct the symmetric alphabet associated with $X$ by adding a twin-letter
$x'$ for every letter $x \in X$: 
$$ X':= \{x' \mid x \in X\},\;\;\hat{X} := X \cup X'.$$
The map $x \mapsto x'$ is extended to $X^*$ by
$$ (x_1 \cdot x_2\cdots x_i \cdots x_n)' := x_n'\cdots x_i'\cdots x_2' \cdot x_1'.$$
We then define a rational set $\hat{\RR}$ and a semi-Thue system $\DD$ by:
$$\hat{\RR} := \{l'r \mid \rrule{l}{r} \in \RR \}^*,\;\; \DD:= \{ (xx',\varepsilon) \mid x \in X \}.$$
It is proved in \cite{Boa-Niv84} that: for every $u,v \in X^*$
\begin{eqnarray*}
u \torl^*_\RR v & \Leftrightarrow & u \hat{\RR}^* \to^*_\DD v\\
& \Leftrightarrow & v \in  [u \hat{\RR}^*] (\to^*_\DD).
\end{eqnarray*}
Hence
\begin{equation}
\label{e-RRvsDD}
[T] (\torl^*_\RR) = [T \hat{\RR}^*] (\to^*_\DD)
\end{equation} 
A \nfa recognizing $T \hat{\RR}^*$ 
can be computed in time ${\rm O}(\|\AA\| + \|\RR\|)$.\\
By the main result of \cite{Ben-Sak86}, for every recognizable set $R$, 
a \nfa recognizing 
$[R] (\to^*_\DD)$ can be computed in time 
${\rm O}(n^3)$ where $n$ is the size of a \nfa recognizing $R$.
Hence a \nfa recognizing 
$[T \hat{\RR}^*] (\to^*_\DD)$ can be computed in time 
${\rm O}((\|A\| + \|\RR\|)^3)$ and, by equality (\ref{e-RRvsDD}), the lemma is proved.
\end{proof}
\begin{lemma}
\label{complexity_semithue_step2}
Suppose that $\FF$ is an alphabet 
without arities (which is not fixed anymore), $T$ is a recognizable subset of $\FF^*$ and 
$\RR$ is a semi-Thue system over $\FF$.
Then, a (non-deterministic) finite automaton recognizing $[T](\torl^*_\RR)$ 
can be computed in time
${\rm O}(|\FF| \cdot (\log(|\FF|)(\|\AA\| + \|\RR\|))^3)$.
\end{lemma}
\begin{proof}
Suppose that $\FF :=\{x_1, \ldots, x_n \}$ and $n =2^p$.
Let $\varphi: \FF^* \rightarrow \{a,b\}$ be some suffix encoding. For example we can define
$\varphi(x_i)$ as the $i$-th word in $\{a,b\}^p$ for some total ordering over $\{a,b\}^p$.
One can check that
\begin{equation}
\label{e-RRvsvarphiRR}
[T](\torl^*_\RR)= \varphi^{-1}([\varphi(T)](\torl^*_{\varphi(\RR)})).
\end{equation}
A \nfa $\AA'$ recognizing $\varphi(T)$ can be computed from $\AA$ in time ${\rm O}(\|A\| \cdot p)$
i.e.\\ ${\rm O}(\|A\| \cdot \log(|\FF|))$.
Using the result of Lemma \ref{complexity_semithue_step1}, a \nfa  
$\AA''$  recognizing $[\varphi(T)](\torl^*_{\varphi(\RR)})$ can be computed in time \\ 
${\rm O}((\log(|\FF|)(\|\AA\| + \|\RR\|))^3)$. A \nfa $\BB$ can be obtained from $\AA''$ 
by the classical construction for the operation $\varphi^{-1}$. This gives a complexity:\\
${\rm O}(|\FF| \cdot (\log(|\FF|)(\|\AA\| + \|\RR\|))^3)$.
\end{proof}
\noindent{\bf Proof} of theorem \ref{t-complexityW}.\\
We suppose now that  $\FF$ is a signature  
with arities in $\{0,1\}$, $T$ is a subset of $\TT(\FF)$ recognized by a \fta $\AA$
and $(\RR,\FF)$ is a rewriting system in $\BU^-(1)$.\\
Let $\SS:= \{\rrule{l\tau}{r\tau} \mid \rrule{l}{r} \in \RR, \tau: \VV \rightarrow Q\}.$
Since $\RR$ is $\BU^-(1)$, by a small variation of Lemma~\ref{l-RviaSetA}, 
for every $s \in \TT(\FF)$,
$s \to^*_\RR T \Leftrightarrow s \to^*_{\SS \cup \AA} Q_f.$
Thus 
\begin{equation}
\label{e-RRvsSSAA-1wordcase}
(\to^*_{\RR})[T]= [Q_f](\to^*_{(\SS \cup \AA)^{-1}})\cap \TT(\FF).
\end{equation}
Let us denote by $\CC$ the ground rewriting system $(\SS \cup \AA)^{-1}$.
Let us notice that $\TT(\FF)$ is a subset of $\FF^*$. Moreover 
$\TT(\FF)$ is saturated by $\torl^*_\CC$ and 
the relation $\to^*_\CC$ restricted to  $\TT(\FF)$ coincides with
$\torl^*_\CC$ restricted to $\TT(\FF)$. 
We can thus apply the results of Lemma \ref{complexity_semithue_step2}:
as $\|\SS \cup \AA\| \leq (\|\RR\| \cdot \|\AA\|)$, 
a \fta recognizing $[Q_f](\to^*_{\CC}[T])$ can be computed in time\\
${\rm O}(|\FF| \cdot (\log(|\FF|)(\|\RR\| \cdot\|\AA\| + \|\RR\|))^3)$ hence in time
$${\rm O}(|\FF|\cdot (\log(|\FF|))^3\cdot \|\RR\|^3 \cdot\|\AA\|^3).$$
$\Box$\\
Let us recall that every left-basic semi-Thue system can be viewed as a $\BU^-(1)$
term rewriting system (Lemma~\ref{l-basic_are_bu-1}).
This Theorem~\ref{t-complexityW} thus extends \cite{Ben-Sak86}, where a cubic complexity
is proved for {\em cancellation systems} over a fixed alphabet, and improves \cite{Ben87},
where a degree 4 complexity is proved for {\em basic} semi-Thue systems.\\
\paragraph{Term rewriting systems}
Let us turn now to term rewriting systems over arbitrary signatures.
The following refinement of Theorem \ref{t-ground-inverse-preserving} has been proved in 
\cite{Der-Gil89}
\begin{theorem}
\label{t-grounddescendants_in_P}
Let $T$ be a finite set of terms and $\SS$ be a ground term rewriting system.
A \fta recognizing the set $[T]\to^*_{\SS}$ can be computed in time which is a {\em polynomial}
function of $\Card(T)+ \|\SS\|$.
\end{theorem}
Given a system $\RR$ we recall the maximum arity of $\RR$ was defined in \S \ref{ss-trs} by:
$$\A(\RR):= \max\{\Card(\Pos_\VV(l)) \mid \rrule{l}{r} \in \RR\}.$$
We extend the above complexity result into the following
\begin{theorem}
\label{t-complexityT}
Let $(\RR,\FF)$ be a finite rewriting system in $\BU^-(1)$ and let $\AA$ be some \fta over $\FF$
recognizing a set of terms $T\subseteq \GTERM$.
One can compute a \fta $\BB$ recognizing $(\to^*_\RR)[T]$ in time polynomial w.r.t.
$\|\RR\|\cdot \|\AA\|^{\Max\{\A(\RR),1\}}$.
\end{theorem}
Our proof consists in computing the ground system
$\SS$ of Section~\ref{s-basic-construction} and to apply Theorem \ref{t-grounddescendants_in_P}.
\begin{proof}
Let us consider the system 
$$\SS:= \{l \tau \rightarrow r\tau \mid  \rrule{l}{r} \in \RR, \tau: \VV \rightarrow Q\}.$$
One can check that
$$\|\SS\| \leq \|\RR\| \cdot |Q|^{\A(\RR)}.$$
By the same arguments as in the case of arities not bigger than $1$, we still get that
\begin{equation}
\label{e-RRvsSSAA-1generalcase}
(\to^*_{\RR})[T]= [Q_f](\to^*_{\CC})\cap \TT(\FF)
\end{equation}
where $\CC:= (\SS \cup \AA)^{-1}$. It is clear that
$$\|\CC\| \leq \|\RR\| \cdot |Q|^{\A(\RR)}+ \|\AA\| \leq \|\RR\|\cdot \|\AA\|^{\A(\RR)}.$$
By  Theorem \ref{t-grounddescendants_in_P}, a \fta $\AA'$ recognizing $[Q_f](\to^*_{\CC})$ 
can be computed in P-time w.r.t. $|Q_f|+\|\CC\|$, thus in P-time w.r.t 
$\|\RR\|\cdot \|\AA\|^{\max\{\A(\RR),1\}}$. Let $\FF'\subseteq \FF $ be the subset of symbols that have at 
least one occurrence either in the transitions of $\AA$ or in the rules of $\RR$. 
By a direct product with the obvious \fta recognizing $\TT(\FF')$
we can compute a \fta $\BB$ recognizing $[Q_f](\to^*_{\CC}) \cap \TT(\FF')$.
The overall computation of $\BB$ takes a P-time w.r.t 
$|\FF'| \cdot \|\RR\|\cdot \|\AA\|^{\max\{\A(\RR),1\}}$. But $|\FF'| \leq \|\RR\|+ \|\AA\|$.
Hence the computation takes a P-time w.r.t 
$$\|\RR\|\cdot \|\AA\|^{\Max\{\A(\RR),1\}}.$$
By (\ref{e-RRvsSSAA-1generalcase}), this automaton $\BB$
recognizes $(\to^*_{\RR})[T]$.
\end{proof}
\subsubsection{Upper-bounds for systems in $\BU^-(k)$}
\label{s-complexity-buk}
The upper-bounds resulting from the use of the precise system $\SS$ from Definition \ref{d-ssystem}
would be unnecessarily high. Therefore we start this subsection by defining a smaller
ground system $\SS_1 \subseteq \SS$. Subsequently we sketch  a proof that the refined system $\SS_1$ 
can also simulate  the original system $\RR$. Finally, we derive from this improved construction
an upper-bound on the complexity of constructing a \fta for the set of ancestors of a 
recognizable set of terms.

Let us define the set of subterms of the lhs of $\RR$ by
$$\SLHS(\RR) := \{ t \in \TT(\FF,\VV) \mid \exists \ell \to r \in \RR,\exists C \in {\cal C}_1(\FF),
\ell \equiv_\alpha C[t]\}.
$$
In words: $\SLHS(\RR)$ consists of all subterms of left-handsides of rules of $\RR$, up to
renaming of the variables.
Let us say that a TRS $(\RR,\FF)$ is {\em variable-free} iff it has no variable left-handside
nor variable right-handside.
From now on, and until the statement of Theorem \ref{t-complexityT}, all definitions, lemmas and 
propositions assume that the TRS $\RR$ under consideration is variable-free.
\begin{definition}
\label{d-S1system}
Let $\RR$ be some TRS over the signature $\FF$.
We consider the ground rewriting system $\SS_1$ 
over $\TT((\FF \cup Q)^{\leq k})$ consisting of all the rules of the form:\\
$$
\barre{l}\barre{\tau} \rightarrow r \dbarre{\tau}\;\;\;\;\;\;\;\;(\ref{e-rule-of-SS})
$$
where $\rrule{l}{r}$ is a rule of $\RR$ 
$$
\ma{\barre{l}}=0\;\;\;\;\;\;\;\;(\ref{e-wbu-rule})
$$
and $\barre{\tau}: \VV \rightarrow \TTqkm$ is a marked substitution such that, 
$\forall x \in \Var(l)$
\begin{equation}
x\dbarre{\tau}= x\barre{\tau}\odot M(\barre{l},x)
\label{e-respects-marking-rule}
\end{equation}
and, there exists substitutions $\tau_i: \VV \rightarrow \SLHS(\RR)$ for $1 \leq i \leq k-1$ and
$\tau_k: \VV \rightarrow Q$ such that
\begin{equation}
\label{e-tau-is-cascade}
\tau= \tau_1\circ \tau_2 \cdots \circ \tau_i \circ \cdots \circ \tau_k.
\end{equation}
\end{definition}
\begin{lemma}
\label{l-SS1-includedin-SS}
$\SS_1 \subseteq \SS$
\end{lemma}
\begin{proof}
Conditions (\ref{e-rule-of-SS},\ref{e-wbu-rule}) are those imposed on the rules of $\SS$ in
Definition \ref{d-ssystem}. For every $x \in \VV$,  $\tau_i(x)$  has a depth smaller or equal to $\dmax$ implying that the additional condition (\ref{e-short-substitution}) of 
Definition \ref{d-ssystem} also holds.
\end{proof}
\begin{lemma}[lifting $\SS_1 \cup \AA$ to ${\cal R}$]$\;\;$\\
\label{l-S1liftedtoR}
Let $\barre{s},\barre{s'}, \barre{t}\in \TTqkm$ such that $\barre{s'}$ is $\m$-increasing.
If $\barre{s}' \to^*_\AA \barre{s}$ and $\barre{s} \to^*_{\SS_1 \cup \AA}\barre{t}$ then, there exists a term $\bar{t'} \in \TTqkm$ 
such that\\
$\barre{s}' \torok^*_\RR \barre{t'}$ and $ \barre{t'}\to^*_\AA \bar{t}$.
\end{lemma}
\begin{proof}
By Lemma \ref{l-SS1-includedin-SS}, $\SS_1 \subseteq \SS$.
We assume that $\RR$ has no variable lhs so that, by remark \ref{r-Anonstandard-allowed-forlifting}, Lemma \ref{l-SliftedtoR} holds.
These two lemmas imply the above lemma.
\end{proof}
\begin{lemma}[projecting $\RR$ on $\SS_1 \cup {\cal A}$]$\;\;$\\
\label{l-RprojectstoS1bis}
Let $\barre{s}, \barre{t}\in \TT(\FF^{\leq k}), q \in Q^{0}$.
If $\barre{s} \torok^*_\RR \barre{t}$ is a $\bu^-(k)$ derivation, $\barre{s}$ is $\tm$-increasing 
and $\barre{t}\to^*_{\AA} q$ 
then, 
there exists a term $\barre{s}'\in \TT((\FF \cup Q)^{\leq k})$ such that
$$\barre{s} \to^*_{\AA}\barre{s}'\to^*_{\SS_1 \cup \AA} q.$$ 
\end{lemma}
\begin{figure}[htb]
\centering
\input projectS1.pstex_t
\caption{Lemma~\ref{l-RprojectstoS1bis}}
\label{project-on-S1}
\end{figure}
The proof of this lemma will be given only ten pages later, at a point where sufficient technical 
preparation will have been achieved.\\
In order to prove this lemma, by induction over the length of derivation $\barre{s} \torok^*_\RR \barre{t}$, we introduce a notion of decomposition of a term relative to a derivation 
(modulo ${\RR \cup \AA}$) that starts from
this term. Each component of the decomposition is called a {\em cascade} and the full 
decomposition is called a {\em bunch of cascades}.\\
We  shall examine, in the sequel, marked derivations of the following form:
\begin{equation} 
\barre{D}: \barre{t}_0 \toro_\RR\to^*_{\AA} \cdots \toro_\RR\to^*_{\AA}\barre{t}_i\toro_\RR\to^*_{\AA}  \cdots \toro_\RR\to^*_{\AA}\barre{t}_\ell
\label{general_RA_derivation}
\end{equation}
where the $i$-th step 
\begin{equation}
\barre{t}_i\toro_\RR\to^*_{\AA} \barre{t}_{i+1}
\label{stepi_RA_derivation0}
\end{equation} starts with an application of a rule 
$l_{i} \to r_{i} \in \RR$:
\begin{equation}
\barre{t}_i=\barre{C}_i[\barre{\ell}_i \barre{\sigma}_i] 
\toro_\RR \barre{C}_i[r_i \dbarre{\sigma}_i]  \to^*_{\AA} \barre{t}_{i+1}
\label{stepi_RA_derivation}
\end{equation}
For every $i \in [0,\ell]$, we denote by $D_i$ the subderivation 
\begin{equation*}
\barre{D}_i\;:\;\;\barre{t}_i\toro_\RR\to^*_{\AA}  \cdots \toro_\RR\to^*_{\AA}\barre{t}_\ell.
\label{taili_RA_derivation}
\end{equation*}
(for $i=\ell$, $\barre{D}_i$ is the derivation of length null starting from $\barre{t}_\ell$).
These marked derivations $\barre{D},\barre{D}_i$ are mapped, by removal of the marks, 
to derivations $D,D_i$ for the system $\RR \cup \AA$. The notion of residual that we use below  
is defined w.r.t. to these derivations. 

\begin{figure}[htb]
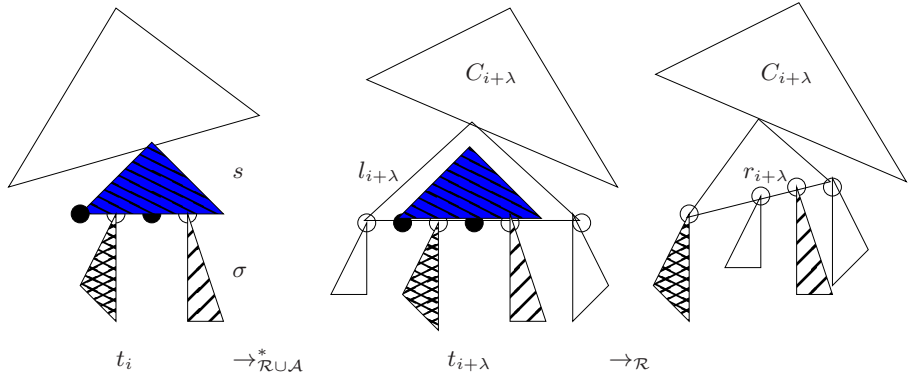

\centering
\input cascadeC2.pstex_t
\caption{A cascade: condition C2}
\label{f-cascade-C2}
\end{figure}

\begin{figure}[htb]
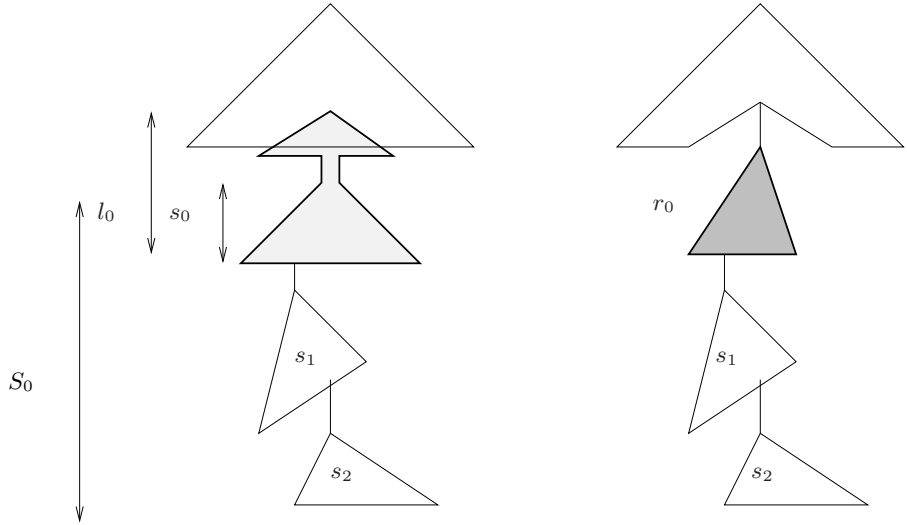

\centering
\input cascade.pstex_t
\caption{A cascade of level $3$: first step}
\label{f-cascade}
\end{figure}

\begin{figure}[htb]
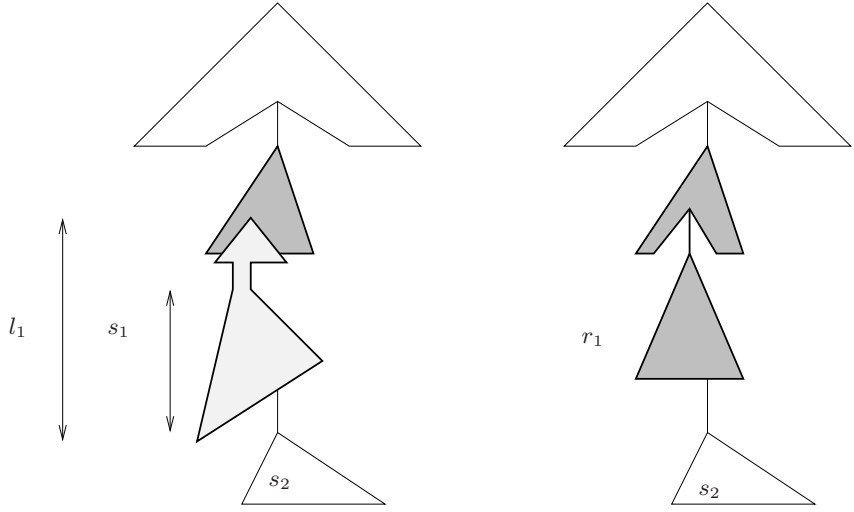

\centering
\input cascade2.pstex_t
\caption{A cascade of level $3$: second step}
\label{f-cascade2}
\end{figure}

\begin{figure}[htb]
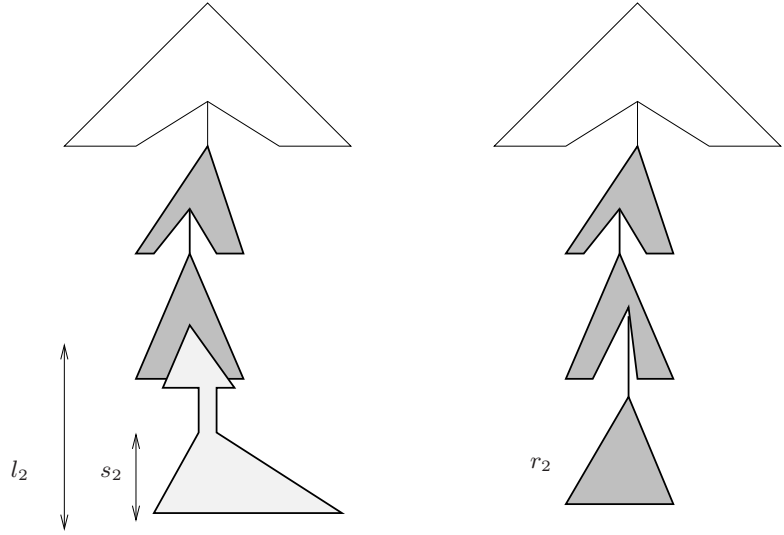

\centering
\input cascade3.pstex_t
\caption{A cascade of level $3$: third step}
\label{f-cascade3}
\end{figure}

\begin{figure}[htb]
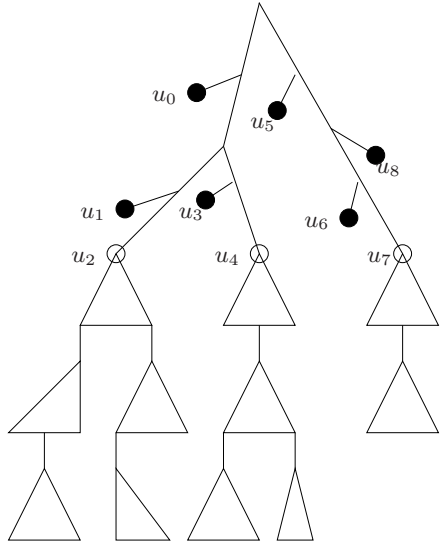

\centering
\input bunchof-cascades.pstex_t
\caption{A bunch of cascades}
\label{f-bunchof-cascade}
\end{figure}

Let us introduce a notion of {\em cascade}: intuitively, a cascade is a subterm such that, all its internal nodes will contribute to a lhs of rule in the future. More precisely, a cascade of 
level $h$ can be layered into at most $h$ comparable occurrences which will be used, successively, from top to bottom, in the rest of the derivation.
\begin{definition}[Cascade]
\label{d-cascade}
Let $\barre{D}$ be some derivation (modulo $\RR \cup \AA$) of the form (\ref{general_RA_derivation}-\ref{stepi_RA_derivation}). 
We define inductively, for pairs $(h,i) \in \N \times [0,\ell]$, 
the notion of {\em cascade of level $h$ w.r.t. $\barre{D}_i$}.
Let $\barre{S} \in \TT((\FF \cup Q)^{\N})$ such 
that $\barre{S}$ is a subterm of $\barre{t}_i$ and let $h \in \N$\\

\noindent C1- If $\barre{S} \in (\FF_0 \cup Q)^\N$, then $\barre{S}$  is a cascade of level 
$h$ w.r.t. $\barre{D}_i$\\

\noindent C2- If $\barre{S}$ has the form $\barre{S}=\barre{s} \barre{\sigma}$ 
(for some term $\barre{s} \notin \VV$ and substitution $\barre{\sigma}$) and $\exists \lambda \geq 0$ fulfilling the conjunction of conditions $C_{2.1},C_{2.2},C_{2.3}$  below, then $\barre{S}$  
is a cascade of level $h+1$ w.r.t. $\barre{D}_i$\\

\noindent C2.1 the occurrence of $s$ in S has a residue in 
${t}_{i +\lambda}$ which is a subterm
of the occurrence of $l_{i+\lambda}$  ( used in the $(i+\lambda)$-th step of $D$ )\\

\noindent C2.2 for every variable $x \in \Var(l_{i+\lambda}) \cap \Var(r_{i+\lambda})$, 
occurring at the frontier of the given occurrence of $s$ in $l_{i+\lambda}$,
$\exists y \in \Var(s)$, $x \barre{\sigma}_{i +\lambda}$ is a residue of $y \sigma$ and
$x \barre{\sigma}_{i +\lambda}$ is a cascade 
of level $h$ w.r.t. $\barre{D}_{i+\lambda}$,\\

\noindent C2.3 for every variable $x \in \Var(l_{i+\lambda}) \setminus  \Var(r_{i+\lambda})$, 
occurring at the frontier of the given occurrence of $s$ in $l_{i+\lambda}$, $\exists y \in \Var(s)$, $x \barre{\sigma}_{i +\lambda}$ is a residue of $y \sigma$ and
$y \barre{\sigma} \in (\FF_0 \cup Q)^\N$.\\

\noindent We call {\em source} of the cascade $\barre{S}$ the given occurrence of the factor 
$\barre{s}$ in $\barre{t}_i$.
\end{definition}
We illustrate on several figures the notion of cascade. A black node on such a figure indicates
a node labelled by an element of $(\FF_0 \cup Q)^\N$.
Figure \ref{f-cascade-C2} illustrates the general features of case C2.\\
The three figures \ref{f-cascade}-\ref{f-cascade3} sketch a cascade of level 3, where 
only rewriting steps in $\RR$ are used; each one of these sketches the positions of redex 
and contractum for one rewriting step.\\
Note that, by Definition \ref{d-cascade}:\\
- every cascade of level $h$ w.r.t. $\barre{D}_i$ is also a cascade of level $h+1,h+2,\ldots$
w.r.t. $\barre{D}_i$;\\
- if $\barre{D}$ is a marked derivation of length one, 
$$\barre{D}: \bar{C}[\bar{l} \bar{\sigma}] \toro_\RR \bar{C}[r \dbarre{\sigma}],$$
\begin{itemize}
\item every ground subterm of the given occurrence of $\bar{l}$ is a cascade of level 
$1$ w.r.t. $\barre{D}$
(because C2.1 holds while the universally quantified  conditions C2.2, C2.3. are 
trivially true);
\item the given subterm $\bar{l}\bar{\sigma}$ is a cascade of level $1$ w.r.t. $\barre{D}$ iff 
$\bar{\sigma}$ maps every variable $x \in \Var(l)$ into $(\FF_0 \cup Q)^\N$.
\end{itemize}
\begin{example}
Let us consider the alphabet $\FF := \{a,b,c,\#_0\}$ where $a,b,c$ have arity $1$ and $\#_0$
has arity $0$. Let $\RR$ consist of the rules
$$ a \to cb\;\;bcb \to ba \;\;ab \to ba $$
Let $\AA$ be a \fta with set of states $Q := \{q,r\}$ and set of rules
$$ \#_0 \to q\;\;\forall s \in Q, as \to r,\;\; bs \to s,\;\;cs \to s$$
Let $\bar{D}$ be the derivation:
\begin{eqnarray*}
\lfloor a \rfloor cbbbq & \toro_\RR &  \lfloor c b \rfloor c^1b^1 b^1b^1q^1 
 = c \lceil b c^1b^1 \rceil b^1b^1q^1 \\
& \toro_\RR & c \lceil b a\rceil b^2b^2q^2 
 = c b \lfloor ab^2\rfloor b^2q^2 \\
& \toro_\RR & cb \lfloor ba\rfloor b^3q^3 
 = cb b \lceil ab^3 \rceil q^3 \\
& \toro_\RR & cbbbaq^4 \to_\AA^* r.
\end{eqnarray*}
$q^3$ is a cascade of level $0$ w.r.t. to the derivation $cb b a b^3q^3 \toro_\RR \to_\AA^* r$\\
$b^3q^3$ is a cascade of level $1$ w.r.t. $cb b a b^3q^3 \toro_\RR \to_\AA^* r$\\
$b^2b^2q^2$ is a cascade of level $2$ w.r.t. $cb ab^2b^2q^2 \toro_\RR^2\to_\AA^* r$\\
$c^1b^1b^1b^1q^1$ is a cascade of level $3$ w.r.t. $cb c^1b^1b^1b^1q^1 \toro_\RR^3 \to_\AA^* r$\\
$cbbbq$ is a cascade of level $3$ w.r.t. $acbbbq \toro_\RR^4 \to_\AA^* r$\\
$acbbbq$ is a cascade of level $4$ w.r.t. $acbbbq \toro_\RR^4 \to_\AA^* r$
\end{example}
\begin{example}
Let us adapt the example above to symbols with larger arity.
Let $\FF := \{A,B,C,\#_0\}$ where $A$ has arity $2$,  $B,C$ have arity $1$ and $\#_0$
has arity $0$. Let $\RR$ consist of the rules
$$ A(x,y) \to CBy\;\;BCBx \to BA(A(\#_0,\#_0),x) \;\;A(x,By) \to BA(BB\#_0,y)$$
Let $\AA$ be a \fta with set of states $Q := \{q,r\}$ and set of rules
$$ \#_0 \to q\;\;\forall s,t \in Q, A(s,t) \to r\;\;Bs \to s\;\;Cs \to s$$
Let $\bar{D}$ be the derivation:
\begin{eqnarray*}
A(q, CBBBq )& \toro_\RR &  CBC^1B^1 B^1B^1q^1 
 \\
& \toro_\RR& CBA(A(\#_0,\#_0),B^2B^2q^2) 
\to_\AA^*   CBA(r,B^2B^2q^2)\\
& \toro_\RR & CBBA(A(\#_0,\#_0),B^3q^3)
\to_\AA^*  CBBA(r,B^3q^3)\\
& \toro_\RR &  CBBBA(BB\#_0,q^4)\to_\AA^* r.
 \end{eqnarray*}
$q^3$ is a cascade of level $0$ w.r.t. to the derivation  
$CBBA(r,B^3q^3) \toro_\RR \to_\AA^* r$\\
$B^3q^3$ is a cascade of level $1$ w.r.t. $ CBBA(r,B^3q^3) \toro_\RR \to_\AA^* r$\\
$B^2B^2q^2$ is a cascade of level $2$ w.r.t. 
$CBA(r,B^2B^2q^2)(\toro_\RR \to_\AA^*)^2 r$\\
$C^1B^1 B^1B^1q^1$ is a cascade of level $3$ w.r.t. 
$CBC^1B^1 B^1B^1q^1 (\toro_\RR \to_\AA^*)^3 r$\\
$CBBBq$ is a cascade of level $3$ w.r.t. 
$A(q, CBBBq ) (\toro_\RR\to_\AA^*)^4 r$\\
$A(q, CBBBq )$ is a cascade of level $4$ w.r.t. 
$A(q, CBBBq ) (\toro_\RR \to_\AA^*)^4 r$.
\end{example}
\begin{lemma}[Subcascade]
If $\barre{S}$ is a cascade of level $h$ w.r.t. a derivation $\barre{D}$ and $\barre{S}'$ 
is a subterm of $\barre{S}$, 
then $\barre{S}'$ is also a cascade of level $h$ w.r.t. $\barre{D}$.
\end{lemma}
\noindent This can be proved by induction on $h$.\\
When such a situation occurs, $\barre{S}'$ is called a {\em subcascade} of $\barre{S}$ 
w.r.t. $\barre{D}$.\\
\begin{definition}[Null transversal]
\label{d-null-transversal}
Let $\barre{t} \in \TT((\FF \cup Q)^{\N})$
and let $U=(u_0,u_1,\ldots,u_m)$ be a transversal of $\barre{t}$.\\
The transversal $U$ is said {\em null} iff 
$$\forall i \in [0,m],  \ma{\barre{t}/u_i}=0.$$
\end{definition}
In words: every node of $U$ has a {\em null} mark.
\begin{definition}[Bunch of cascades]
\label{d-bunch-cascades}
Let $\barre{D}$ be a derivation fo the form (\ref{general_RA_derivation}-\ref{stepi_RA_derivation}).\\
A term $\barre{t} \in \TT((\FF \cup Q)^{\N})$ is called a {\em bunch of cascades} w.r.t. 
derivation $\barre{D}$ iff,\\
$\barre{t} = \barre{t}_0$ and $\Pos(\barre{t})$ has a transversal 
$U=(u_0,u_1,\ldots,u_m)$ such that\\
(BC1) every subterm $\barre{t}/u_i$ is a cascade w.r.t. $\barre{D}$\\
(BC2) $U$ is a null transversal\\
(BC3) either $\barre{D}$ has null length or the (marked) occurrence of 
$l_{0}$ which is used in the first step of 
$\barre{D}$ is the source of one of the cascades $\barre{t}/ u_i$.
\end{definition}
Figure \ref{f-bunchof-cascade} represents a bunch of cascades where $m=8$, the cascades at 
nodes $u_0,u_1,u_3,u_5,u_6,u_8$ have level $0$, at nodes $u_2,u_4$ level $3$, at node $u_7$ 
level $2$.
\begin{definition}
A marked term $\barre{s}$  is said {\em tamely increasing} ($\tm$-increasing in short) iff, 
for every $u,v \in \Pos(\barre{s}), i,j \in \N$, both conditions (TM1),(TM2) below are fulfilled:\\
(TM1) $u \preceq v \Rightarrow \m(\barre{s}/u) \leq \m(\barre{s}/v)$\\
(TM2) $u \cdot i \in  \Pos(\barre{s}) \mbox { and } u \cdot j  \in  \Pos(\barre{s})
\Rightarrow \m(\barre{s}/u \cdot i) =\m(\barre{s}/u \cdot j).$
\label{d-tmincreasing}
\end{definition}
\noindent In words: $\bar{s}$ has marks which increase from root to leaves and which are 
equal on brothers.
\begin{lemma}
\label{l-toAA_toRR_preserve_tm_increasing}
Assume that $\RR$ is a variable-free TRS, $\AA$ is a \fta and $\bar{s}$ is $\tm$-increasing.\\
1- If $\bar{s} \toro_\RR \bar{t}$ is a $\wbu$ marked derivation step, 
then  $\bar{t}$ is $\tm$-increasing.\\
2- If $\bar{s} \to_\AA \bar{t}$ then  $\bar{t}$ is $\tm$-increasing.\\
\end{lemma}
\begin{proof}
Let $\RR$, $\AA$ and $\bar{s}$ fulfill the hypotheses of the lemma.\\
1- Suppose that $\bar{s} \toro_\RR \bar{t}$ by a $\wbu$ derivation step of the form (\ref{marqued_onestep}) where the position of $\Box$ in $C$ is $u_0$.
By Lemma \ref{l-onestep-marks_increase} $\bar{t}$ is $\m$-increasing.
Let $u\cdot i,u\cdot j$ be brother postions of $\bar{t}$.\\
\begin{itemize}
\item If $u\cdot i,u \cdot j$ are non-variable positions of $C$, then, by hypothesis on $\bar{s}$ they have the same mark.
\item If $u \cdot i $ is a non-variable position of $C$ and $u \cdot j$ is the position of $\Box$ 
in $C$, then, since the derivation-step is $\wbu$ the mark of $u \cdot j$ in $\bar{s}$ is null,
hence the mark of $u \cdot i$ in $\bar{s}$ is null (as a brother of $u \cdot j$), 
hence the mark of $u \cdot i$ in $\bar{t}$ is null. The mark of $u \cdot j$ in $\bar{t}$ is the
mark of the root of the rhs, which is null (because $\RR$ has no variable rhs).
\item If $u \cdot i,u \cdot j$ are positions in  $r \dbarre{\sigma}$ i.e. there exists $v \in \N^*$ such that
$u \cdot i= u_0 \cdot v \cdot i,u \cdot j = u_0 \cdot v \cdot j$,
then, either they are brother positions in $r$ and they are both null, or they are brother 
positions in $x \dbarre{\sigma}$ (for some variable $x$); in this last case, 
by hypothesis on $\bar{s}$ the corresponding
positions in $\bar{s}$ have the same mark $m$, hence they are both  marked by 
$\max(m,M(\barre{C}[\barre{l}],x))$ in $\bar{t}$.
\end{itemize}
2- Since the binary relation $\to_\AA$ does not modify the marks of the nodes, the preservation
property is true.
\end{proof}
\begin{lemma}[projecting one step of $\RR$ on a cascade]$\;\;$\\
\label{l-onestepR-projectstoS1}
Let 
$$\barre{D}_1: \barre{t'}_1 \toro_\RR \to^*_{\AA} \barre{t'}_2 \toro_\RR \to^*_{\AA}\cdots \toro_\RR \barre{t}'_i \toro_\RR \to^*_{\AA}\cdots
\toro_\RR \to^*_{\AA}\barre{t}'_\ell,$$  
and let 
$\barre{t}_0, \barre{t}_1 \in \TT(\FF^{\leq k})$  be $\tm$-increasing terms such that\\ 
1- $\barre{t}_0 \torok_\RR \barre{t}_1 \mbox { is } \wbu$\\ 
2- $\barre{t}_1\to^*_{\AA}\barre{t}'_1$ and\\
3- $\barre{t}'_1$ is a bunch of cascades w.r.t. $\barre{D}_1$.\\
Then,  there exists a term $\barre{t}'_0 \in \TT((\FF \cup Q)^{\leq k})$  such that 
$$\barre{t}_0 \to^*_{\AA}\barre{t}'_0 \torok_\RR \to^*_{\AA}\barre{t}'_1,$$
and $\barre{t}'_0$ is a bunch of cascades w.r.t. the derivation $\barre{D}_0$ obtained from $\barre{D}_1$ by 
extension on the left by the step $\barre{t}'_0 \toro_\RR \to^*_{\AA}\barre{t}'_1$.
\end{lemma}
\noindent{Diagram}:\\ 
$$
\begin{array}{cccccccc}
&\barre{t}_0 & \torok_\RR & \barre{t}_1&&&&\\
&\;\;\Big\downarrow \stackrel{*}{\AA} && \;\;\Big\downarrow \stackrel{*}{\AA}&&&&\\
\barre{D}_0:&\barre{t}'_0 & \torok_\RR\to^*_{\AA} & \barre{t'}_1&  \toro_\RR\to^*_{\AA} & \cdots &
\toro_\RR\to^*_{\AA}& \barre{t}'_\ell
\end{array}
$$
\begin{sketch}$\;\;$\\
Let $l_0 \to r_0$ be the rule used in the one-step derivation 
$\barre{t}_0 \toro_\RR \barre{t}_1$ and let $y_0 \in \Pos(t_0)$ be the position 
of the occurrence of $l_0$ used in this rewriting step.
Let $U=(u_0,\ldots,u_i,\ldots,u_m)$ be the null transversal provided by Definition \ref{d-bunch-cascades}
applied on $\barre{t}'_1$.
Let us construct a term $\barre{t}''_1$ and a term $\barre{t}'_0$ such that 
$$\barre{t}'_0 \toro_\RR \barre{t''}_1\to^*_{\AA}\barre{t'}_1$$
and $\barre{t}'_0$ fulfils the properties announced by the lemma.
We define
\begin{equation}
P := \CL(\Pos(\barre{t_1}') \cup y_0 \cdot \Pos (r_0), \Pos(\barre{t_1}))
\label{e-domain-tsecondsub1}
\end{equation}
We then define $\barre{t}''_1: P \rightarrow (\FF \cup Q)^\N$ by,
for every $u \in P$:\\
\begin{equation}
\mbox { if } u \in \Internal(P),\;\;\barre{t}''_1(u) := \barre{t}_1(u);\;\;\;\;
\mbox { if } u \in \Leaves(P),\;\;\barre{t}''_1(u) := \rho_1(u);\;\;\;\; 
\label{e-labels-tsecondsub1}
\end{equation}
where $\rho_1$ is the partial run of the automaton $\AA$ associated with the computation
$\barre{t}_1 \to_\AA^* \barre{t}'_1$.
Let $t'_0$ be the unmarked term obtained from $t''_1$ by applying the rule $\rrule{l_0}{r_0}$
``backwards'' at position $y_0$:
$$t'_0 =C[l_0\tau]_{y_0},\;\;t''_1 =C[r_0\tau]_{y_0}$$
where the substitution $\tau$ is defined by:\\
- if $x \in \Var(l_0) \cap \Var(r_0)$, $\tau(x)= t´_1/(y_0\cdot pos(r_0,x))$.\\
- if $x \in \Var(l_0) \setminus \Var(r_0)$, $\tau(x)= \rho_0(y_0\cdot pos(l_0,x))$.\\
where $\rho_0$ is any run of the automaton $\AA$ over the 
term $\barre{t}_0/y_0\cdot pos(l_0,x)$.
Finally, let $\barre{t}'_0$ be the marked term obtained from the domain and labels of $t'_0$ (on one hand) and the marks of $\barre{t}_0$ (on the other hand).
One can check that:\\
\begin{equation}
\barre{t}_0 \to_\AA^* \barre{t}'_0\torok_\RR\barre{t}''_1 \to^*_{\AA} \barre{t'}_1.
\label{e-t0t1-close-the-diagram}
\end{equation}
We distinguish two cases, according to the relative position of the root $y_0$ of the given occurrence of $r_0$  and of the transversal $U$.\\
{\bf Case 1}: $y_0$ is above at least one $u_i$ (see Figure \ref{f-one-step-left1}).\\
Let us suppose that $y_0 \preceq u_i, y_0 \preceq u_{i+1},\ldots, y_0 \preceq u_{i+p}$
and $\forall j \in [0,i-1]\cup [i+p+1,m], y_0 \not\preceq u_j$ (see Figure \ref{f-one-step-left1}).
\begin{figure}[htb]
\centering
\scalebox{0.80}{
\input one-step-left1.pstex_t
}
\caption{Lemma \ref{l-onestepR-projectstoS1}, Case 1}
\label{f-one-step-left1}
\end{figure}
\begin{figure}[htb]
\centering
\scalebox{0.80}{
\input one-step-left2.pstex_t
}
\caption{Lemma \ref{l-onestepR-projectstoS1}, Case 2}
\label{f-one-step-left2}
\end{figure}
Let
$$U_0 := (u_0,\ldots,u_{i-1},y_0,u_{i+p+1},\ldots,u_m).$$
$U_0$ is a transversal of $\barre{t}'_0$.
Condition (BC1) is clearly fulfilled by the $u_j$, for $j \in [0,i-1] \cup [i+p+1,m]$. 
Since the transversal $U$ of term $\barre{t}'_1$ was fulfilling condition (BC2)
and $\bar{t}'_1$ is $\m$-increasing,
no leaf of the occurrence of $r_0$ can be strictly above $U$.
Hence $\barre{t}'_1/y_0$ consists of an occurrence of $r_0$ followed by some subcascades 
$\barre{S}_1,\barre{S}_2,\ldots, \barre{S}_\mu$ 
of the cascades $\barre{t}'_1/u_{j}$ (for $j \in [i,i+p]$)
and, possibly, some new cascades of level $0$.
Thus, $\barre{t}'_0/y_0$ 
consists of an occurrence of a marked version of $l_0$ followed by some subterms which have
the residuals $\barre{S}_1,\barre{S}_2,\ldots, \barre{S}_\mu$  inside $\bar{t_1}'$
and, possibly, some new cascades of level $0$: 
this is a cascade for $\barre{D}_0$. We have thus checked condition (BC1).\\
Let $v \in U_0$.
\begin{itemize}
\item If $v= u_j$ , for some $j \in [0,i-1] \cup [i+p+1,m]$, since $U$ was null,
$\ma{\barre{t}'_1/u_{j}}=0$. But the mark of this position is the same in $\barre{t}'_1$ and
in $\barre{t}'_0$, hence $\ma{\barre{t}'_0/u_{j}}=0$.
\item If $v= y_0$, since it is the position of the root of $l_0$ in $\barre{t}_0$ and this derivation-step is $\wbu$, $\ma{\barre{t}'_0/v}=0$.
\end{itemize}
We have checked that $U_0$ is null (i.e. (BC2)).\\
The occurrence of $l_{0}$ which is used in the first step of $\barre{D}$ is the source
of the cascade at position $y_0$, hence (BC3) holds.\\
{\bf Case 2}: $y_0$ is strictly below one $u_i$ (see Figure \ref{f-one-step-left2}).\\
Let $(z_0,\ldots,z_p,(u_i)^{-1}y_0, z_{p+1},\ldots, z_{p+p'})$ be the smallest transversal of 
$\barre{t}'_0/u_i$ extending the antichain $(u_i^{-1}y_0)$ (see Lemma \ref{l-smallest-transversal}).
Let
$$U_0:=(u_0,\ldots,u_{i-1},u_iz_0,\ldots,u_iz_p,y_0, u_iz_{p+1},\ldots, u_iz_{p+p'},u_{i+1},\ldots,u_m).$$
$U_0$ is a transversal of $\barre{t}'_0$.
Every $\barre{t}'_0/u_j$ for $j \in [0,i-1] \cup [i+1,m]$, is a cascade (because it was a cascade 
of $\barre{t}'_1$ w.r.t. $\barre{D}_1$). The subterm $\barre{t}'_0/y_0$ is a new cascade, 
consisting of a marked version of $l_0$
followed by some subcascades of $\barre{t}'_1/u_i$ and, possibly, some new cascades of level $0$.
Every subterm $\barre{t}'_0/u_iz_\lambda$ , for $\lambda \in  [0,p+p']$ has, 
as residue in $\barre{t}'_1$, the subterm $\barre{t}'_1/u_iz_\lambda$, which is subcascade of 
the cascade $\barre{t}'_1/u_i$ w.r.t. $\barre{D}_1$. We have thus checked condition (BC1).\\
Let $v \in U_0$.\\
\begin{itemize}
\item If $v =u_j$ , for some $j \in [0,i-1] \cup [i+1,m]$, since $U$ was null and the mark
of $u_j$ is the same in $\barre{t}'_0$ and in $\barre{t}'_1$,
$\ma{\barre{t}'_0/v}=0$.
\item If $v=y_0$, since it is the position of the root of $l_0$ in $\barre{t}_0$ and this derivation-step is $\wbu$, $\ma{\barre{t}'_0/v}=0$.
\item If $v =u_iz_\lambda$, by point 3 of Lemma \ref{l-smallest-transversal}, 
the father $z_\lambda'$ of $z_\lambda$
fulfills $(u_iz_\lambda'\prec u_iz_\lambda \myand  u_iz_\lambda' \prec y_0$;
since $\barre{t}'_0$ is $\m$-increasing, we obtain that $\ma{\barre{t}'_0/u_iz_\lambda' }=0$
and , since $\barre{t}'_0$ is $\tm$-increasing, we obtain that $\ma{\barre{t}'_0/u_iz_\lambda}=0$.
\end{itemize}
We have checked that $U_0$ is null (i.e. (BC2)).\\
The occurrence of $l_{0}$ which is used in the first step of $\barre{D}$ is the source
of the cascade at position $y_0$, hence (BC3) holds.\\
\end{sketch}\\
\begin{definition}Let $\barre{D}$ be a derivation of the form (\ref{general_RA_derivation}).
A term \\ $S \in \TT((\FF \cup Q)^{\N})$ is called a cascade of {\em exact} level $h$ w.r.t. $\barre{D}$ iff 
$S$ is a cascade of level $h$ w.r.t. $\barre{D}$ and $S$ is not a cascade of level $h-1$ w.r.t. $\barre{D}$.
\end{definition}
Note that, by this definition, every cascade of level $0$ is also a cascade of exact level $0$.
\begin{definition} Let $\barre{D}$ be a derivation of the form (\ref{general_RA_derivation}-\ref{stepi_RA_derivation}).
The marked derivation $\barre{D}$ is $\bu^-(k)$ iff, $\forall i \in [0,\ell[$, 
$\mamin{\barre{l_i}}=0$ and $\mamax{\barre{l_i}} < k$.
\end{definition}
In words: $\barre{D}$ is $\bu^-(k)$ iff the steps of rewriting (modulo $\RR$) are meeting the conditions
of the usual $\bu^-(k)$ condition (note that no condition is required on the steps of rewriting (modulo $\AA$)).
\begin{lemma}
Let $\barre{D}$ be a derivation of the form (\ref{general_RA_derivation}-\ref{stepi_RA_derivation}).
If $\barre{S}$ is a cascade of exact level $h$ w.r.t. $\barre{D}$ and $\barre{D}$ is $\bu^-(k)$, then $h \leq k$.
\label{l-cascades_in_buk}
\end{lemma}
\begin{sketch}
Let $\barre{D}$ be a derivation of the form (\ref{general_RA_derivation}-\ref{stepi_RA_derivation}).
One can prove by induction over $h$ the more general statement:\\
if $\barre{S}=\barre{s}\barre{\sigma} $ (where $s \notin \VV$) is a cascade of exact level $h$ 
(with this decomposition) w.r.t.
$\barre{D}_i$ (for some $i$), and every internal node of $\barre{s}$ has a mark $\geq r$, then 
there exists $\lambda \geq 0$, such that a mark $\geq r+h-1$ occurs in the occurrence 
$\barre{l}_{i+\lambda}$ of lefthand-side
of rule used in the $i+\lambda$-th step of $\barre{D}$ (\ref{stepi_RA_derivation}).\\
Since $\barre{D}$ is $\bu^-(k)$, a mark $\geq k$ cannot occur in $\bar{l}_{i + \lambda}$, hence 
no cascade $\barre{S}$ w.r.t. $\barre{D}$ can have an exact level $\geq k+1$.
\end{sketch}\\
{\bf Proof} of {\bf lemma \ref{l-RprojectstoS1bis}}:\\
Let $\barre{s}, \barre{t}\in \TT(\FF^{\leq k}),\; q \in Q^{0},\; n \in \N$, such that
\begin{equation}
\barre{s} \torok^n_\RR \barre{t}
\label{marked_deriv_lemma_projection}
\end{equation} 
is $\bu^-(k)$, $\barre{s}$ is $\tm$-increasing  and 
$ \barre{t}\to^*_{\AA} q$.\\
By Lemma \ref{l-toAA_toRR_preserve_tm_increasing} all the marked terms in derivation 
(\ref{marked_deriv_lemma_projection}) are $\tm$-increasing.
Let us notice that $q$ is a bunch of cascades for the derivation of null length starting on $q$.
Using inductively Lemma \ref{l-onestepR-projectstoS1}, 
we obtain a term $\barre{s}'$ and derivations 
$$ \barre{s} \to^*_{\AA}\barre{s'}$$
$$\barre{D}: \barre{s'}= \barre{t'}_0 \torok_\RR \to^*_{\AA} \barre{t'}_1 \torok_\RR \to^*_{\AA}\cdots 
\torok_\RR \barre{t}'_i \torok_\RR \to^*_{\AA}\cdots
\torok_\RR \to^*_{\AA}\barre{t}'_\ell=q$$ 
such that every $\barre{t}'_i$ (for $0 \leq i \leq \ell-1$)  is a bunch of cascades for $\barre{D}_i$. 
By Lemma \ref{l-cascades_in_buk} every cascade w.r.t. $\barre{D}_i$ has level $\leq k$. 
Hence the $i$-th step of $\torok_\RR$ in derivation $\barre{D}$
is a step for the relation $\to_{\SS_1}$. It follows that $\barre{D}$ is a derivation
modulo $(\SS_1 \cup \to_{\AA})$.$\Box$ \\
Let us express an upper-bound on the complexity for the construction of
the set of ancestors of a recognizable set of terms. (Recall the number $\A(\RR)$ was introduced in \S \ref{ss-trs}).
\begin{theorem}
\label{t-complexityT2}
Let $k \geq 2$,  let $(\RR,\FF)$ be a finite rewriting system in $\BU^-(k)$, with 
$\A(\RR) \geq 1$ and $\AA$ be some 
\nfta over $\FF$ recognizing a language $T$.
One can compute a \fta $\BB$ recognizing $(\to^*_\RR)[T]$ in time polynomial w.r.t.
$$\|\RR\|^{k \cdot (\A(\RR))^{k-1}} \cdot \|\AA\|^{(\A(\RR))^k}.$$
\end{theorem}
\noindent{Note that}:\\
- for systems with $k \leq 1$, the complexity is analyzed in Theorem \ref{t-complexityT}\\
- for systems with $k \geq 2$ and $\A(\RR)=0$, i.e. ground systems, the complexity is 
covered by Theorem \ref{t-grounddescendants_in_P}.

\begin{sketch}$;\;$\\
\noindent{\bf Step 1}: Let us assume $\RR$ is variable-free (we recall it means that
it has no variable left-handside nor right-handside).\\
By Lemma \ref{l-S1liftedtoR} and Lemma \ref{l-RprojectstoS1bis} 
$$ (\tok^*_\RR)[T] = (\to^*_{\SS^0_1 \cup \AA})[Q_f] \cap \TT(\FF)$$
The construction of $\BB$ consists in computing the ground system
$\SS^0_1$ (obtained by erasing the marks in the system $\SS_1$ introduced by Definition \ref{d-S1system}), to apply the construction of Theorem \ref{t-grounddescendants_in_P}
and, finally, to perform a direct product with a \fta recognizing $\TT(\FF)$.\\
The set of possible substitutions $\tau_1,\ldots,\tau_i,\tau_{k-1}$ in
Definition \ref{d-S1system} has cardinality less or equal to
\begin{equation*}
\|\RR\|^{\A(\RR) + \A^2(\RR)+ \ldots + \A^{k-1}(\RR)}
\label{e-maj-for-tau1tok-1}
\end{equation*}
(within this subsubsection, for every integer $m \geq 0$,  we denote by $\A^{m}(\RR)$ the integer $(\A(\RR))^{m}$;
this removal of parenthese should not lead to any ambiguity since the operation 
$\RR \mapsto \A(\RR)$ cannot be iterated).
The set of possible final substitutions $\tau_k$ has cardinality less or equal to
\begin{equation*}
\|\AA\|^{\A^k(\RR)}
\label{e-maj-for-tauk}
\end{equation*}
Since the number of rules of $\RR$ is less or equal than $\|\RR\|$ we get an upper-bound for 
the number of rules of $\SS_1^0$:
\begin{equation}
\Card(\SS_1^0)\leq \|\RR\|^{1+\A(\RR) + \A^2(\RR)+ \ldots + \A^{k-1}(\RR)} \cdot \|\AA\|^{\A^k(\RR)}
\end{equation}
For every rule $\ell_1 \to r_1 \in \SS_1^0$, 
since $\ell_1 = \ell (\tau_1 \cdots \circ\tau_i \cdots \circ\tau_{k-1} \circ \tau_k)$
and $r_1 = r (\tau_1 \cdots \circ\tau_i \cdots \circ\tau_{k-1} \circ \tau_k)$ for some $\ell \to r \in \RR$,
and for every variable $v \in \VV$, $|\tau_i(v)| \leq \|\RR\|$, we have 
\begin{eqnarray}
|\ell_1| + |r_1| & \leq & \|\RR\| \cdot ( 1 + \A(\RR) + \A^2(\RR)+ \cdots + \A^{k-1}(\RR))\nonumber\\
& \leq & \|\RR\| \cdot k \cdot \A^{k-1}(\RR)
\label{e-maj-l1tor1}
\end{eqnarray}
Multiplying the upper-bound for the number of rules by the upper-bound for the size of each rule, we
obtain
\begin{eqnarray}
\|\SS_1^0\|  & \leq & \|\RR\|^{1+\A(\RR) + \A^2(\RR)+ \ldots + \A^{k-1}(\RR)} 
\cdot \|\AA\|^{\A^k(\RR)} \cdot |\RR\| \cdot k \cdot \A^{k-1}(\RR)\nonumber\\
&\leq & \|\RR\|^{4 \cdot k \cdot \A^{k-1}(\RR)} \cdot \|\AA\|^{\A^k(\RR)}.
\label{e-maj-for-cardS1}
\end{eqnarray}
(we assume that $\RR \neq \emptyset$ hence that $2 \leq \|\RR\|$ in the above majorization:
for $\RR = \emptyset$, anyway, the computation of $\BB$ consists of taking $\BB := \AA$,
which takes no time ).
The construction of the system $\SS_1^0$ is straightforward, thus takes a time polynomial 
in $\|\SS_1^0\|$. The computation, from $\AA$ and $\SS_1^0$,
 of a \fta $\AA'$ recognizing $(\to^*_{\SS_1^0\cup \AA})[Q_f]$ takes a time polynomial
in $\|\AA\| + \|\SS_1^0\|$ by Theorem \ref{t-grounddescendants_in_P}.
By inequality (\ref{e-maj-for-cardS1}),
$\|\AA\| + \|\SS_1^0\| \leq \|\RR\|^{5 \cdot k \cdot \A^{k-1}(\RR)} \cdot \|\AA\|^{\A^k(\RR)}$,
which is a polynomial in  $\|\RR\|^{k \cdot \A^{k-1}(\RR)} \cdot \|\AA\|^{\A^k(\RR)}$.
Let $\FF'\subseteq \FF $ be the subset of symbols that have at 
least one occurrence either in the transitions of $\AA$ or in the rules of $\RR$. 
Finally, $\BB$ is obtained from $\AA'$ by performing the direct-product of $\AA'$ with
a \fta recognizing $\TT(\FF')$.
The overall computation of $\BB$ thus takes a time polynomial in 
$|\FF|' \cdot \|\RR\|^{k \cdot \A^{k-1}(\RR)} \cdot \|\AA\|^{\A^k(\RR)}$.
Since we assumed that $k \geq 2,\A(\RR) \geq 1, \| \RR \| \geq 2$, it is also a polynomial in
$$\|\RR\|^{k \cdot \A^{k-1}(\RR)} \cdot \|\AA\|^{\A^k(\RR)}.$$

\noindent{\bf Step 2}: Let $\RR$ be a general TRS (whith, possibly, some variable lhs
or rhs).\\
The transformation $\RR \mapsto \RR_1$ defined 
in \S \ref{s-variable-lhs}, is a polynomial reduction of the general case to 
the subcase treated in step 1 of this proof (see Lemma \ref{l-RR-embedded-in-RR1}) .
\end{sketch}\\
Note that, for every fixed parameters $k \geq 1$ and $\A(\RR)$, the construction of 
the set of ancestors
of a rational set for some TRS $\RR$ in $\BU^-(k)$ can be achieved in polynomial time.
In general, for a fixed $\A(\RR)$ and variable $k$, the dependency in $k$ is double exponential.
In the case of unary terms we get only an exponential complexity.
\begin{corollary}
Let $k \geq 2$,  
let $\FF$ be a signature with symbols of arity $\leq 1$,
let $\AA$ be some \fta recognizing a language $T \subseteq \TT(\FF)$ and let $\RR$ be a finite rewriting system in $\BU^-(k)$.
One can compute a \fta $\BB$ recognizing $(\to^*_\RR)[T]$ in time polynomial w.r.t.
$$\|\RR\|^{k } \cdot \|\AA\|.$$
\end{corollary}
\begin{proof}
If $\A(\RR)= 1$, just replace $\A(\RR)$ by the integer $1$ in the conclusion of Theorem \ref{t-complexityT2}. If $\A(\RR) =0$, the result follows from Theorem \ref{t-grounddescendants_in_P}.
\end{proof}
\subsubsection{Lower-bound}
\label{s-lower-bound}
We show here that, there exists  a fixed signature $\FF$ and 
two fixed recognizable sets $L_1,L_2$ over $\FF$ such that, 
the accessibility from $L_1$ to $L_2$ for a rewriting system in $\BU^-(1)$ 
is NP-hard. This shows that the upper-bound given by Theorem~\ref{t-complexityT2} 
in the case of a fixed parameter $k$, which is exponential w.r.t. $\A(\RR)$, 
cannot presumably be significantly improved.\\
Let us fix the signature $\FF:= \{f,g,\wedge,\vee,\neg,0,1\}$ where the 
arities are $2,2,2,2,1,0,0$ (for the
symbols in the given ordering). We shall also use the subsignature
$\FF':= \FF \setminus \{f,g\}$.
Let us consider the regular term-grammar $\GG$, over the signature $\FF$, with non-terminals $T_0,T_1,T_2, G_0, G_1, G$ and with set of rules:
\begin{eqnarray*}
\label{e-grammarG}
T_1 & \rightarrow & 1 \label{e-ruleT1}\\
T_2 & \rightarrow & f(T_2,G) + 2 \label{e-ruleT2}\\
G_0 & \rightarrow & g(G_0,0) + 0\label{e-ruleG0}\\
G_1 & \rightarrow & g(G_1,1) + 1\label{e-ruleG1}\\
G   & \rightarrow & G_0 + G_1 \label{e-ruleG}
\end{eqnarray*}
For sets of terms $L_1,L_2$ we abbreviate by $L_1 \to^*_\RR L_2$ the sentence
$\exists t_1 \in L_1,\exists t_2 \in L_2, t_1 \to^*_\RR t_2$.
\begin{theorem}
\label{t-accbu1_NPhard} 
The problem to decide, for a given linear term rewriting system $(\RR,\FF)$
in $\BU^-(1)$, whether
$\Language(\GG,T_1) \to^*_\RR \Language(\GG,T_2)$, is NP-hard.
\end{theorem}
We reduce, in P-time, the problem 3-SAT to the above problem.
Let $\varphi$ be some propositional formula in 3-Conjunctive Normal Form:
$\varphi$ is a formula with 
$n_v$ variables $x_1,x_2,\ldots,x_{n_v}$  of the form
$$
\varphi=\bigwedge_{k=1}^{n_c} \bigvee_{\ell=1}^{3} v_{k,\ell}^{\varepsilon_{k,\ell}}
$$
where $n_c \in \N$,$v_{k,\ell} \in \{x_1,x_2,\ldots,x_{n_v}\}$, $\varepsilon_{k,\ell} \in \{-1,+1\}$
with the convention that $v^{+1}$ (resp. $v^{-1}$) denotes $v$ (resp. $\neg v$).
Let us note that, if every variable occurs in at least one clause and if all the clauses of $\varphi$ are distinct, then $n_v \leq 3n_c \leq 24n_v$. 
Hence, after some mild
normalization (either adding variables that do not occur in any clause or adding copies of 
a clause that already occurs, which can be achieved in P-time), $\varphi$ can be put in 
the form of a formula with 
$n$ variables $x_1,x_2,\ldots,x_{n}$  and $n$ clauses:
\begin{equation}
\label{e-varphi}
\varphi=\bigwedge_{k=1}^{n} \bigvee_{\ell=1}^{3} v_{k,\ell}^{\varepsilon_{k,\ell}}
\end{equation}
where $n \in \N$,$v_{k,\ell} \in \{x_1,x_2,\ldots,x_{n}\}$, $\varepsilon_{k,\ell} \in \{-1,+1\}$.
Let us define a kind of {\em linearization} of $\varphi$ over a set of $3n^2$ new variables
$x_{i,j}$, for $1 \leq i \leq n,1 \leq j \leq 3n.$  
\begin{equation}
\label{e-varphihat}
\hat{\varphi}:=\bigwedge_{k=1}^{n} \bigvee_{\ell=1}^{3} \hat{v}_{k,\ell}^{\varepsilon_{k,\ell}}
\end{equation}
where $\hat{v}_{k,\ell}:=x_{i,j}$ iff 
($v_{k,\ell}=x_{i}$ and 
$\Card\{(k',\ell')\in [1,n]^2 \mid  (k',\ell')\leq_{lex} (k,\ell)\; \& \;v_{k',\ell'}=
v_{k,\ell} \}=j$).
In words: $\hat{v}_{k,\ell}:=x_{i,j}$ when the meta-variable $v_{k,\ell}$ denotes the variable
$x_i$ and it is exactly the $j$-th occurrence (from left to right) of $x_i$ in 
formula (\ref{e-varphi}).
Note that $\hat{\varphi}$ is linear and 
\begin{equation}
\label{e-hatvarphi_leq_varphi}
\varphi=\hat{\varphi}\sigma
\end{equation}for the substitution 
\begin{equation}
\label{e-def_sigma}
\sigma: x_{i,j} \mapsto x_i.
\end{equation}
Let us denote by $x_{i,*}$ the sequence of $3n$ variables $x_{i,1},\ldots,x_{i,3n}$
and by $x_{*,*}$ the sequence of $3n^2$ variables 
$x_{1,1},\ldots,x_{i,j},x_{i,j+1}, \ldots,x_{n,3n}$. 
We define three sequences of terms $(f_m)_{m \geq 1},(g_m)_{m \geq 1},(h_m)_{m \geq 1}$ 
by the following recurrence relations:
$$f_1(x_1) := x_1,\;\;
f_{m+1}(x_1,x_2,\ldots,x_{m+1}) :=  f(f_m(x_1,x_2,\ldots,x_{m}),x_{m+1}),$$
$$g_1(x_1) := x_1,\;\;
g_{m+1}(x_1,x_2,\ldots,x_{m+1}) :=  g(g_m(x_1,x_2,\ldots,x_{m}),x_{m+1}),$$
$$
h_{m}(x_{*,*}) :=  f_{m+1}(2,g_{3m}(x_{1,*}),\ldots,g_{3m}(x_{i,*}),\ldots, g_{3m}(x_{m,*})).$$
We define a fixed ground rewriting system $\PL$ consisting of the rules allowing to evaluate a 
Boolean formula, taken in reverse order:
\begin{eqnarray*}
&& 0 \rightarrow 0 \wedge 0 ,\;\;0 \rightarrow 0 \wedge 1 ,\;\;0 \rightarrow 1 \wedge 0,\;\;1 \rightarrow 1 \wedge 1,\;\;\\
&& 0 \rightarrow 0 \vee 0,\;\;1 \rightarrow 0 \vee 1,\;\;1 \rightarrow 1 \vee 0,\;\;1 \rightarrow1 \vee 1,\;\;\\
&& 1 \rightarrow \neg 0,\;\; 0 \rightarrow \neg 1.
\end{eqnarray*}
(The initials $\PL$ intend to make the reader think of ``Propositional Logic'').
We define the {\em special} rule associated with $\varphi$ by:
\begin{equation}
\label{e-special_rule}
\hat{\varphi}(x_{*,*})\rightarrow h_n(x_{*,*})
\end{equation}
where $\hat{\varphi}$ is some term over $\FF$ expressing the Boolean formula $\hat{\varphi}$
(the $n$-ary meta-symbol $\bigwedge$ can be translated as a left-comb with internal nodes labelled by the binary symbol $\wedge$ and similarly for the ternary symbol $\bigvee$).
We finally define the system $(\FF, \RR_\varphi)$, associated with $\varphi$, by:
$$\RR_\varphi := \PL \cup \{ \hat{\varphi}(x_{*,*}) \rightarrow h_n(x_{*,*}) \}.$$
We cut into several lemmas the proof that $\varphi \mapsto \RR_{\varphi}$
is a valid reduction.
\begin{lemma}
\label{f-satisfiability_iff_tnto1}
For every $b_1,b_2,
\ldots b_n,b \in \{0,1\}$,
$b \to^*_\PL \varphi(b_1,b_2,\ldots,b_n) $ iff\\
$b \to^*_{\RR_{\varphi}}  f_{n+1}(2,g_{3n}(b_1,b_1,\ldots,b_1),g_{3n}(b_2,b_2,\ldots,b_2), \ldots, g_{3n}(b_n,b_n,\ldots,b_n))$.
\end{lemma}
\begin{proof}
\noindent Using the special rule (\ref{e-special_rule}) we get:
\begin{equation}
\label{e-firststep_T2toT1}
\hat{\varphi}\tau \to_{\RR_{\varphi}}
f_{n+1}(2,g_{3n}(b_1,b_1,\ldots,b_1),g_{3n}(b_2,b_2,\ldots,b_2), \ldots, g_{3n}(b_n,b_n,\ldots,b_n))
\end{equation}
where the substitution $\tau$ is defined by $\tau(x_{i,j}) := b_i$.
We can factorize $\tau$ as $\tau:= \sigma \circ \theta$ where
$\sigma$ was defined in (\ref{e-def_sigma}) and $\theta(x_i) := b_i$.
The one-step rewriting (\ref{e-firststep_T2toT1}) thus can be seen as
\begin{equation*}
\label{e-firststep_T2toT1-bis}
(\hat{\varphi}\sigma) \theta \to_{\RR_{\varphi}}
f_{n+1}(2,g_{3n}(b_1,b_1,\ldots,b_1),g_{3n}(b_2,b_2,\ldots,b_2), \ldots, g_{3n}(b_n,b_n,\ldots,b_n))
\end{equation*}
which,  by the identity (\ref{e-def_sigma}) shows that
\begin{eqnarray*}
\label{e-firststep_T2toT1-ter}
\varphi(b_1,b_2,\ldots,b_n)  =  \varphi\theta & \to_{\RR_{\varphi}} &
f_{n+1}(2,g_{3n}(b_1,b_1,\ldots,b_1),g_{3n}(b_2,b_2,\ldots,b_2),\\
&& \ldots, g_{3n}(b_n,b_n,\ldots,b_n)).
\end{eqnarray*}
The lemma follows easily from this last relation.
\end{proof} 
\begin{lemma}
\label{f-tnto1_iff_T2to1}
The following two conditions are equivalent:\\
1- $\{1\}\to^*_{\RR_{\varphi}} \Language(\GG,T_2)$ \\
2- There exist $b_1,b_2, \ldots b_n \in \{0,1\}$, such that:\\
$1 \to^*_{\RR_{\varphi}} f_{n+1}(2,g_{3n}(b_1,b_1,\ldots,b_1),g_{3n}(b_2,b_2,\ldots,b_2), \ldots, g_{3n}(b_n,b_n,\ldots,b_n))$.
\end{lemma}
\begin{proof}
\noindent 1- Suppose condition 1 holds: there exists $t_2 \in \Language(\GG,T_2)$ such that
$1 \to^*_{\RR_{\varphi}} t_2$. Since no rule of $\PL^{-1}$ can be applied on $t_2$,
the derivation must decompose as
$$1 \to^*_{\RR_{\varphi}} t'_2 \to_{\RR_{\varphi}} t_2$$
where the last step uses the special rule (\ref{e-special_rule}).
Note that $t_2$ has exactly one occurrence of the constant $2$ while $t'_2$ has no occurrence
of this symbol. Since $1 \to^*_{\RR_{\varphi}} t'_2$, the term $t'_2$  must belong to $\TT(\FF')$.
This implies that the contractum in $t_2$ was $t_2$ itself:
$$t_2:= h_n(x_{*,*})\tau$$
for some substitution $\tau: \{x_{i,j} \mid 1 \leq i \leq n,1 \leq j \leq 3n\} \rightarrow
\TT(\FF')$.
From the fact that $t_2 \in \Language(\GG,T_2)$ we deduce that, $\forall i \in [1,n],\exists b_i \in \{0,1\}$,
$$\tau(x_{i,1}) = \tau(x_{i,2})= \ldots = \tau(x_{i,3n})=b_i.$$
Hence 
$t_2 = f_{n+1}(2,g_{3n}(b_1,b_1,\ldots,b_1),g_{3n}(b_2,b_2,\ldots,b_2), \ldots, g_{3n}(b_n,b_n,\ldots,b_n))$ 
and condition 2 holds.\\ 
2- Suppose condition 2 holds. Since the last term of this derivation belongs to
$\Language(\GG,T_2)$, condition 1 holds.
\end{proof}
For every marked term $\barre{t} \in \TT(\FF^\N)$, we call a path $P' \subseteq \Pos(\barre{t})$ 
a $\FF'$-path
iff all the labels of $P'$ belong to $\FF'$.
Let us consider the following property $\PR(\barre{s})$ of a term $\barre{s} \in \TT(\FF^\N)$:
\begin{eqnarray}
\label{e-def_propertyP}
\forall P' \subseteq \Pos(s), &\mbox{ if }& (P' \mbox{ is a } \FF'-\mbox{path and } 
\Min\{\ma{\barre{s}/u}\mid u \in P'\}=0) \nonumber\\
& \mbox{ then }&
\Max \{\ma{\barre{s}/u}\mid u \in P'\}=0.
\end{eqnarray}
Every $\wbu$ rewriting-step of $\toro_{\RR_\varphi}$ preserves $\PR$ in the following sense
\begin{lemma}
\label{f-Pispreserved}
For every $\barre{s},\barre{t} \in \TT(\FF^\N)$  if
($\PR(\barre{s})$ and 
$\barre{s} \toro_{\RR_\varphi} \barre{t}$ is a $\wbu$-rewriting step), 
then $\PR(\barre{t})$.
\end{lemma}
\begin{proof}
\noindent Let us consider a $\wbu$ rewriting step 
$$\barre{s}=\barre{C}[\barre{l}\barre{\sigma}]\toro_{\RR_\varphi}\barre{C}[r \dbarre{\sigma}] 
=\barre{t}.$$
- If the rule used belongs to $\PL$, $\barre{l}=b$ for some $b \in \{0,1\}$ and $r$ is a 
term  over $\FF'$ with only null marks; thus every $\FF'$-path $Q$ of $\barre{t}$  must
either be included in $\barre{C}$ (case 1) or is obtained from a path $Q'$ of $\barre{s}$
by replacing its maximal element (labelled by $b$) by two elements (labelled by symbols of
$\FF'$)(case 2).
In case 1 $\Max \{\ma{\barre{s}/u}\mid u \in Q\}=0$ because this was true in $\barre{s}$.
In case 2 $\Max \{\ma{\barre{s}/u}\mid u \in Q\}=0$ because the labels of $Q'$ were null and the 
labels of the two new nodes are also null.
Hence $\PR$ is preserved;\\
- If the special rule is used: since $r$ has no label in $\FF'$, every  $\FF'$-path $Q$ of $\barre{t}$ must be either included in $\barre{C}$ (case 3) or included in $x\dbarre{\sigma}$ for some variable
$x$ of $r$ (case 4). In case 3 $\Max \{\ma{\barre{s}/u}\mid u \in Q\}=0$ because this was true in $\barre{s}$
and in case 4, $\Min\{\ma{\barre{s}/u}\mid u \in P'\}\geq 1$.
Hence $\PR$ is preserved.
\end{proof}
\begin{lemma}
\label{f-RRphiisBU}
For every Boolean formula $\varphi$, the system $\RR_{\varphi}$ is linear and $\BU^-(1)$.
\end{lemma}
\begin{proof}
\noindent 
It is clear that $\RR_{\varphi}$ is linear.
Let us consider a $\wbu$ marked derivation (modulo $\RR_{\varphi}$): $s \toro^n \barre{t} \toro \barre{t'}$ with $s \in \TT(\FF)$.\\
Let $\rrule{l}{r}$ be the rule of $\RR_{\varphi}$ which is used in the last step of the associated unmarked derivation.\\ 
- If $\rrule{l}{r} \in \PL$, since the step is $\wbu$, the root of $l$ has a null mark and no other mark appears in $l$ since it has depth 1; \\
- If $\rrule{l}{r}$ is the special rule, since every branch of $l$ is labelled by a word
in $\FF'^*\VV$ and $\barre{t}$ fulfills $\PR$,
and $\barre{l}$ has a root marked $0$, all internal nodes of $\barre{l}$ have the
mark $0$; hence $\mamax{\barre{l}}=0$.\\
By induction over the integer $n$ we can thus prove that, for every $s \in \TT(\FF)$, every $\wbu$ marked derivation $s \toro^n \barre{t}$ is $\bu^{-}(1)$.
\end{proof}
\noindent Let us prove  Theorem \ref{t-accbu1_NPhard}. By Lemmas (\ref{f-satisfiability_iff_tnto1}-\ref{f-tnto1_iff_T2to1})
$\varphi$ is satisfiable iff $\Language(\GG,T_1) \to^*_{\RR_{\varphi}} \Language(\GG,T_2)$. Moreover
$\RR_{\varphi}$ is computable in Polynomial time from $\varphi$ and, by Lemma~\ref{f-RRphiisBU},
it belongs to $\BU^-(1)$. Hence $\varphi \mapsto \RR_{\varphi}$ is a Polynomial-time reduction 
of the satisfiability problem for Boolean formulas in 3-CNF to the problem under consideration.

\section{Testing the Bottom-up property}
\label{s-testing}
We investigate here the question whether the properties $\BU(k)$ (resp. $\BU^-(k)$) are decidable, 
or not.
We concentrate first on the case of semi-Thue systems: 
\begin{itemize}
\item we establish a 
criterium (i.e. a Necessary and Sufficient Condition) for the property $\BU(k)$, in the case of length-increasing semi-Thue systems (Proposition \ref{p-bu-criterium}) 
\item we show that the property $\BU(k)$ is decidable in some non-trivial subclass of length-increasing semi-Thue systems (Proposition~\ref{p-bu1_decidable}) and 
that it is undecidable for general length-increasing semi-Thue systems (Theorem~\ref{t-bu_words_undecidable}).
\item we deduce the undecidability of the property $\BU(k)$ for term rewriting systems
(Theorem~\ref{t-bu_undecidable}).
\end{itemize}
\subsection{A criterium for semi-Thue systems}
\label{s-criterium_for_noteriansemiThuesystems}
\paragraph{More notation for derivations}
The general notion of derivation which was given in \S \ref{sets_binrel} for general binary relations $\to$, turns out not to be precise enough for an analysis in the case where the binary relation is defined through combinatorial means, as is the case for derivations induced by semi-Thue 
systems 
or term rewriting systems (see remark \ref{rem-on-derivations}).
We thus borrow from \cite{Cre-Ott94,Laf95} a more precise notion of derivation, 
some useful notation and a notion of equivalence over derivations.

We assume some semi-Thue system $\RR$ over an alphabet $Y$ is given.
For every rule $R= \rrule{l}{r}$ and words $v,w \in Y^*$, we note $\partial^+((v,R,w)):= vrw,\partial^-((v,R,w)):= vlw$. 
We call derivation any non-empty sequence of triples of the form 
\begin{equation}
D=((v_1,R_1,w_1)\ldots, (v_i, R_i, w_i),\ldots ,(v_n, R_n, w_n))
\label{word_derivation}
\end{equation} 
such that, for every $1\leq i \leq n-1, \partial^+(v_i,R_i,w_i)= \partial^-(v_{i+1},R_{i+1},w_{i+1})$
and also the triples
\begin{equation}
D_v:= (v,\ID,\varepsilon)
\label{trivial_word_derivation}
\end{equation} 
where $\ID$ is a special symbol that we view as the {\em Identity} rule.
We extend the notations $\partial^{\alpha}$ by defining
for $D$ given in (\ref{word_derivation}): 
$$\partial^+(D):= \partial^+(v_n,R_n,w_n),\;\;\partial^-(D):= \partial^-(v_1,R_1,w_1)$$
and for $D_v$ given in (\ref{trivial_word_derivation}): 
$$\partial^+(D_v):= v,\;\;\partial^-(D_v):= v.$$ 
We define an equivalence $\approx$ on derivations by
$$D \approx D' \Leftrightarrow (\partial^+(D) = \partial^+(D')\myand \partial^-(D) = \partial^-(D'))$$
i.e. $D,D'$ are equivalent when they have same starting word and same ending word.
The length $\ell(D)$ is defined as $n$ for the derivation (\ref{word_derivation}) and
$0$ for the derivation (\ref{trivial_word_derivation}).
Given derivations $D,D'$ such that $\partial^+(D)=\partial^-(D')$, their composition
$D \otimes D'$  is just their concatenation (when they both have non-null length),
$D$ when $\ell(D')=0$, and $D'$ when $\ell(D)=0$.\\
The words of $Y^*$ act on the right and on the left over derivations:
for $D$ defined by (\ref{word_derivation}) we set
$$D \cdot v:= ((v_1,R_1,w_1v)\ldots, (v_i,R_i,w_iv),\ldots ,(v_n, R_n, w_nv))$$
and $v \cdot D$ is defined similarly;
$D_u \cdot v:= D_{uv}$ and $v \cdot D_u:= D_{vu}$.\\
One can easily check that, for every derivations $D,D',F,F'$, words $u,v$ and
 signs $\alpha \in \{+1,-1\}$:
$$ u(D \otimes D')v=uDv \otimes uD'v,$$
$$ \partial^\alpha (uDv)=u \partial^\alpha (D)v,$$
$$ \partial^+ (D \otimes D')=\partial^+ (D'),\;\;
\partial^- (D \otimes D')=\partial^- (D).$$
From these formulas it follows easily that,  
$$(D \approx D' \myand F \approx F') \Rightarrow  D \otimes F  \approx D' \otimes F'$$
$$D \approx D' \Rightarrow  uDv \approx uD'v.$$
These two last {\em compatibility} properties will be widely (though implicitly) used in our proofs.
We call a derivation $D$ {\em right-minimal} (r-minimal, for short) iff its only decomposition as $D=D' \cdot v'$ is
the trivial one: $D'=D, v' = \varepsilon$.
\paragraph{Some basic properties}
Let $\RR$ be any semi-Thue system over some alphabet $Y$.
\begin{lemma}For every derivation $D$ and word $u \in Y^*$,
$D$ is $\bu(k)$ iff $D \cdot u$ is $\bu(k)$.
\label{l-right_word_context}
\end{lemma}
\begin{sketch}
$\Leftarrow$ is clear.\\
$\Rightarrow$: In the marked derivation, w.r.t $\hat{\RR}$ (see \S \ref{words_as_terms}),
 associated with
$D$, the letter $\#$ has marks in $[0,k]$. In the marked derivation (w.r.t $\hat{\RR}$)
 associated with
$D\cdot u$, all the positions of the suffix $u\#$ will have the same mark which is the same integer
as before, hence belongs to $[0,k]$. 
\end{sketch}
\begin{lemma}For every derivations $D_1, D_2$, if $D_1 \otimes D_2$ is $\bu(k)$,
then $D_1$ and $D_2$ are $\bu(k)$.
\label{l-factors_of_buk}
\end{lemma}
\begin{sketch}
Suppose that $D_1 \otimes D_2$ is $\bu(k)$. Let us set
$\ell(D_1)=\ell_1,\;\ell(D_2)=\ell_2$.\\
The fact that $D_1$ is $\bu(k)$ too is straightforward.\\
For every $i \in [0,\ell_2]$,
the mark of the $jth$ position of the $i$th word of $D_2$ is smaller than the mark of the $jth$ position of the $(\ell_1+i)$th word of $D_1 \otimes D_2$. Thus, the hypothesis ensures that all the marks of
the marked derivation associated with $D_2$ are in $[0,k]$.
\end{sketch}
\begin{lemma}For every derivations $D_1, D_2$, word $u_1 \in Y^*$ and rule $R_1 \in \RR$,
 if $D_1 \otimes u_1 \cdot R_1$ is $\bu(k)$ and $D_2$ is $\bu(k)$, then
$D_1 \otimes u_1 \cdot R_1\otimes D_2$ is $\bu(k)$.
\label{l-composing_buk}
\end{lemma}
\begin{sketch}
The derivation $D_1 \otimes u_1 \cdot R_1$  is $\bu(k)$, hence weakly bottom-up.
Hence all the positions of $u_1$ in the lhs and rhs of the last step  of this derivation have a null
mark.
All the positions of $\partial^+(R_1)$ in the rhs of the last step also have a null mark, 
by definition of the marking process. Finally, the word $\partial^+(D_1 \otimes u_1 \cdot R_1)$ 
has only 
null marks in the marked derivation (w.r.t. $\hat{\RR}$) associated with $D_1 \otimes u_1 \cdot R_1$.
Let us consider the unique marked derivation $\hat{D}$ (w.r.t. $\hat{\RR}$) associated with the derivation
$D_1 \otimes u_1 \cdot R_1 \otimes D_2$:
$$\hat{D}: \bar{w}_0\#,\bar{w}_1 \#^{m_1}, \ldots, \bar{w}_n \#^{m_n}$$
- its part labelled over $Y$, $(\bar{w}_0,\bar{w}_1, \ldots, \bar{w}_n)$
 is obtained just by concatenating the corresponding marked derivation
associated with $D_1 \otimes u_1 \cdot R_1$ and the marked derivation associated with $D_2$
(where the final letter $\#$ has been erased); in particular this proves that every step
fulfils definition \ref{d-marked-wbu_derivation}, hence that $\hat{D}$ is $\wbu$;
it proves also that all the marks of the words $\bar{w}_0,\bar{w}_1, \ldots, \bar{w}_n$
belong to $[0,k]$;\\
- every mark $m_{p}$, for $0 \leq p \leq \ell_1$, where $\ell_1 := \ell(D_1 \otimes u_1 \cdot R_1)$,
belongs to $[0,k]$, because $D_1 \otimes u_1 \cdot R_1$ is $\bu(k)$;\\
- one can prove by induction over $p$, for $\ell_1 \leq p \leq n$, that every mark $m_{p}$  
belongs to $[0,k]$: $m_{\ell_1} \in [0,k]$ and, if $p > \ell_1$, $m_p$ is the maximum of 
$m_{p-1}$ and of the mark on the $\#$ in the corresponding word of the marked derivation (w.r.t.
$\hat{\RR}$) associated with $D_2$.\\
Hence $\hat{D}$ is $\bu(k)$, which entails that $D_1 \otimes u_1 \cdot R_1\otimes D_2$ is $\bu(k)$.
\end{sketch}
\paragraph{A criterium}
\begin{figure}[htb]
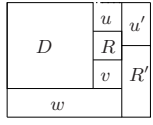

\centering
\scalebox{0.80}{
\input minimal_overlap.pstex_t
}
\caption{A minimal $k$-right-overlap}
\label{minimal_overlap}
\end{figure}
\begin{definition}[Minimal right-overlap]
Let $k \geq 1$ and 
let $\RR$ be a semi-Thue system over some alphabet $Y$.
Let us call {\em minimal $k$-right-overlap} a 7-tuple $(D,R,R',u,v,u',w)$
such that $D$ is  
a r-minimal derivation, $R,R'$ are rules of $\RR$ and $u,v,u',w$ are words in $Y^*$, 
fulfilling:\\
1- $\partial^+(D) = \partial^-(uRv)$,\\
2- $\partial^+(uRvw)= \partial^-(u'R')$,\\
3- $D \otimes uRv$ is $\bu(k)$,\\
4- $0< |w| \leq |vw| < |\partial^-(R')|$.\\ 
The minimal $k$-right-overlap is said {\em resolved} iff
there exists a $\bu(k)$ derivation 
$D'\approx Dw \otimes uRvw \otimes u'R'$.
\label{f-overlapping_laststep}
\end{definition}
(See Figure~\ref{minimal_overlap}).
\begin{lemma}Let $k \geq 1$ and
let $\RR$ be a length-increasing semi-Thue system over some alphabet $Y$. Let us suppose that 
all the minimal $k$-right-overlaps of $\RR$ are resolved.\\
Then, for every $\bu(k)$ derivation $D$, rule $R' \in \RR$ and words $u',v' \in Y^*$, 
if 
\begin{equation}
\partial^+(D) = \partial^-(u'R'v'),
\label{e-hypo-addarule}
\end{equation}
then 
\begin{equation}
\mbox{ there exists a }
\bu(k)-\mbox{derivation }D'\approx D \otimes u'R'v'.
\label{e-conclusion-addarule}
\end{equation}
\label{l-adding_one_rule}
\end{lemma}
\begin{proof}
\begin{figure}[htb]
\centering
\scalebox{0.80}{
\input bu_composition1.pstex_t
}
\caption{Adding one rule}
\label{f-bu_composition1}
\end{figure}
\begin{figure}[htb]
\centering
\scalebox{0.80}{
\input bu_composition23.pstex_t
}
\caption{Adding one rule}
\label{f-bu_composition23}
\end{figure}
We prove that every $(D,R',u',v')$ fulfills the implication 
$((\ref{e-hypo-addarule}) \Rightarrow (\ref{e-conclusion-addarule}))$,
by Noetherian induction over the pair $(|u'|,\ell(D))$, using the lexicographic ordering
on $\N \times \N$.\\

Let us consider some $(D,R',u',v')$ such that $D$ is $\bu(k)$ and the hypothesis (\ref{e-hypo-addarule}) holds. One of cases $0-3$  below must occur
(see Figures~\ref{f-bu_composition1}-\ref{f-bu_composition23}).\\
{\bf case 0}: $\ell(D)=0$\\
In this case we can choose $D'=D \otimes u'R'v'=u'R'v'$ which is $\bu(k)$.\\
{\bf case 1}: $\ell(D)\geq 1$, $D=D_1v_1$ where $D_1$ is a right-minimal $\bu(k)$-derivation and 
$|\partial^+(D_1)| \geq |\partial^-(u'R')|$\\
Since $D_1$ has non-null length, is right-minimal and $\bu(k)$, it decomposes as
\begin{equation}
D_1= D'_2 \otimes u_1R_1 \otimes D''_2
\label{e-decomposition_D1}
\end{equation} 
and the assumed inequality on the boundaries implies that 
$v'=v''v_1$ for some word $v''$.
The tuple $(D''_2,R',u',v'')$ fulfills hypothesis (\ref{e-hypo-addarule})
and $(|u'|,\ell(D''_2))= (|u'|,\ell(D)-1)< (|u'|,\ell(D))$.
By induction hypothesis, there exists some $\bu(k)$-derivation 
$D_3 \approx D''_2 \otimes u'R'v''$.
Let us choose
$$D' := (D'_2 \otimes u_1R_1 \otimes D_3) v_1.$$
Lemma \ref{l-factors_of_buk} applied on decomposition (\ref{e-decomposition_D1}) shows that 
$D'_2 \otimes u_1R_1$ is $\bu(k)$. By Lemma \ref{l-composing_buk} 
$D'_2 \otimes u_1R_1 \otimes D_3$ is $\bu(k)$ and by Lemma \ref{l-right_word_context}
we get that $(D'_2 \otimes u_1R_1 \otimes D_3) v_1$ is $\bu(k)$.
Hence 
$D'$ is a $\bu(k)$ derivation such that $D' \approx D$.\\
{\bf case 2}: $D= (D_1\otimes uRv)wv'$ where $D_1$  is a right-minimal $\bu(k)$-derivation 
and $0< |w| \leq |vw| < |\partial^-(R')|$ .\\
The 7-tuple $(D_1,R,R',u,v,u',w)$ is a minimal $k$-right-overlap,
hence it is resolved: there exists a $\bu(k)$ derivation 
$D_2\approx D_1w \otimes uRvw \otimes u'R'$.
Choosing $D' := D_2v'$ we obtain the conclusion (\ref{e-conclusion-addarule}).\\
{\bf case 3}: $D= (D_1 \otimes uRv)w $  where $D_1$  is a right-minimal $\bu(k)$-derivation 
and $ |u \partial^-(R)| \leq |u'|$ .\\
The inequality on the boundaries implies that ``$R$ and $R'$ can be exchanged'' i.e. that
there exists words $\alpha, \beta \in Y_1^*$ such that
$uRvw \otimes u'R'v' \approx \alpha R'v' \otimes u R \beta.$
Thus
\begin{equation}
(D_1 \otimes uRv)w \otimes u'R'v' \approx D_1w \otimes \alpha R' v' \otimes u R \beta.
\label{e-exchange_last_rules}
\end{equation}
Note that, as every rule of $\RR$ is length-increasing, $|\alpha| \leq |u'|$.
The 4-tuple $(D_1w,R',\alpha,v')$ fulfills hypothesis (\ref{e-hypo-addarule}),
and $(|\alpha|,\ell(D_1w))< (|u'|,\ell(D))$. By induction hypothesis
there exists a 
$\bu(k)$-derivation
\begin{equation} 
D_2 \approx D_1w  \otimes \alpha R'v'.
\label{e-firsthomotopy}
\end{equation}
The 4-tuple $(D_2,R,u,\beta)$ fulfills hypothesis (\ref{e-hypo-addarule}),
and $(|u|,\ell(D_2))< (|u'|,\ell(D))$ (because $|u|< |u'|$).
By induction hypothesis
there exists a 
$\bu(k)$-derivation
\begin{equation} 
D' \approx D_2  \otimes uR\beta.
\label{e-secondhomotopy}
\end{equation}
Combining the hypothesis of case 3 with equivalences (\ref{e-exchange_last_rules})(\ref{e-firsthomotopy})(\ref{e-secondhomotopy}) we obtain that 
$D'\approx D \otimes u'R'v'$, as required.
In all cases we have proved the announced implication: 
$$((\ref{e-hypo-addarule}) \Rightarrow (\ref{e-conclusion-addarule}))$$
\end{proof}

\begin{proposition}
Let $k \geq 1$ and
let $\RR$ be a length-increasing semi-Thue system over some alphabet $Y$. The system $\RR$
is $\BU(k)$ iff all its minimal $k$-right-overlaps are resolved.
\label{p-bu-criterium}
\end{proposition}
\begin{proof}
{\bf($\Rightarrow$)}\\
Suppose that $\RR$ is a semi-Thue system over some alphabet $Y$ and that it is $\BU(k)$.
Since {\em every} derivation must be equivalent to some $\bu(k)$ derivation, it is clear
that every minimal $k$-right-overlap is resolved.\\
{\bf($\Leftarrow$)}\\
Let $\RR$ have all its minimal $k$-right-overlaps resolved.
Let us prove, by induction over $\ell(D)$, that, for every derivation $D$, there exists some $\bu(k)$ derivation $D' \approx D$.\\
{\bf Basis}: $\ell(D)=0$.\\
In this case $D$ is $\bu(k)$, hence we can choose $D'=D$.\\
{\bf Induction step}: $\ell(D)=n+1$ for some $n \geq 0$.\\
$D$ has a decomposition of the form:
$$D = D_1 \otimes u'R'v'$$
for some derivation $D_1$, with length $\ell(D_1)=n$, some rule $R' \in \RR$ and some words 
$u',v' \in Y^*$.
By induction hypothesis, there exists some $\bu(k)$-derivation $D'_1$ such that
$$D_1 \approx D'_1.$$
By Lemma \ref{l-adding_one_rule}, there exists some $\bu(k)$-derivation $D'$
such that
$$D' \approx D'_1 \otimes u'R'v'.$$
It follows that $D \approx D'$, as required.
\end{proof}
\subsection{Decidable/undecidable cases for semi-Thue systems}
\label{s-undecidability_for_semiThuesystems}
\paragraph{A decidable case}
\begin{proposition}
Let $k \geq 1$.
The property $\BU(k)$ (resp. $\BU^-(k)$)
is decidable for length-increasing semi-Thue systems fulfilling the additional condition below:\\
{\bf C}: $\RR$ has no right-linear recursion i.e. there is no finite sequence of rules
$(l_i \rightarrow r_i)_{1 \leq i \leq n}$ such that, for every $1 \leq i \leq n-1$, $l_{i+1}$
is a suffix of $r_{i}$, $n \geq 2$ and $l_1=l_n$.
\label{p-bu1_decidable}
\end{proposition}
\begin{proof}(sketch)$\;\;$\\
Let $\RR$ be a length-increasing semi-Thue system fulfilling condition ${\bf C}$.
By proposition \ref{p-bu-criterium} a necessary and 
sufficient condition
for $\RR$ to be $\BU(k)$ is that all its minimal $k$-right-overlaps  are resolved.
By condition ${\bf C}$ this set of minimal $k$-right-overlaps is finite and constructible.
Hence the above necessary and sufficient condition is testable.
\end{proof}
\paragraph{Undecidable cases}
We treat first the case of the property $\BU(1)$.\\
\begin{proposition}
It is undecidable whether a finite length-increasing semi-Thue system $\RR$ is $\BU(1)$ (resp. $\BU^-(1)$). 
\label{p-bu1_words_undecidable}
\end{proposition}
Our proof will use the following variant of the universality problem for context-free grammars:\\
{\bf Input}: A context-free grammar $G=\langle Z,N,P\rangle$ where $Z=\{z_1,z_2\}$ is the terminal alphabet, $n$ is a strictly positive integer, $N=\{S_1,\ldots,S_n\}$ is the non-terminal alphabet and $P \subseteq N \times {\cal P}((Z \cup N)^+)$ is the finite set of rules.\\
{\bf Question}: $\forall s \in Z^*, S_1 \to^*_P sS_1?$\\
We call this problem the Modified Universality Problem ($\MUP$ in short). It
follows easily from the undecidability of the classical universality problem for 
context-free grammars (\cite[Theorem 8.11 p. 203]{Hop-Ull79}) that the above problem is undecidable.\\

Let us consider an instance $G$ of $\MUP$. We introduce some fresh symbols $S_0,a_1,b$ 
not in $Z \cup N$ and define the alphabet $Y_1 := Z \cup N \cup \{S_0,a_1,b\}$.
Let $\RR$ be the semi-Thue system over $Y_1$ whose set of rules consists of the union of $P$ with
the three new rules:
$$S_0 \to z_1S_0,\;S_0 \to z_2 S_0,\;\ S_0 a_1 \to S_1 a_1.$$
We call $R_{z1},R_{z2},R_{01}$ (in the above enumeration order) these new rules.
We decompose in two lemmas the proof that $G \mapsto \RR$ is a reduction of $\MUP$ to the 
problem whether a semi-Thue system is $\BU(1)$.
\begin{lemma}If $\RR$ is $\BU(1)$, then, for every $s \in Z^*, S_1 \to^*_P sS_1$.
\label{l-bu1-implies-universal}
\end{lemma}
\begin{proof}
Suppose $\RR$ is $\BU(1)$. Let $s \in Z^*$. Consider the following derivation:
\begin{eqnarray*}
S_0 a_1 b & \to_\RR^*  sS_0a_1b & \mbox{ using } R_{z1},R_{z2}\\
          & \to_\RR  sS_1a_1b & \mbox{ using } R_{01}.
\end{eqnarray*}
The associated marked derivation is
\begin{eqnarray*}
S_0 a_1 b & \to_\RR^*  sS_0a_1^1b^1 & \\
          & \to_\RR  sS_1a_1b^2.&
\end{eqnarray*}
The only possible $\bu(1)$ derivation with same boundary in the system $\RR$ would be the composition of the first step
\begin{eqnarray*}
S_0 a_1 b & \to_\RR   S_1a_1b & \mbox{ using } R_{01}
\end{eqnarray*}
with a derivation
\begin{equation}
S_1 \to^*_P  sS_1
\label{e-S1-deriveP-sS1}
\end{equation}
in the right-context $a_1b$. The existence of derivation (\ref{e-S1-deriveP-sS1}) is thus ensured.
\end{proof}
\begin{lemma}If for every $s \in Z^*, S_1 \to^*_P sS_1$, then $\RR$ is $\BU(1)$.
\label{l-universal-implies-bu1}
\end{lemma}
\begin{proof}
Let us suppose that
\begin{equation}
\forall s \in Z^*, S_1 \to^*_P sS_1.
\label{G_isuniversal}
\end{equation}
Let us consider some minimal $1$-right-overlap $(D,R,R',u,v,u',w)$ of the system $\RR$.
The only possible value for $R'$ is $R_{01}$ while $R$ might be either $R_{z1}$ or $R_{z2}$.
It follows that $w=a_1$ and $v=\varepsilon$. Since $D \otimes uRv$ is $\bu(1)$, it has the form
$$ D \otimes uRv:\;\; S_0 \to^*_P sS_0$$
for some $s \in Z^+$.
Hence
$$Dw \otimes uRvw \otimes u'R':\;\; S_0a_1 \to^*_P sS_0a_1 \to_{R_{01}}s S_1a_1.$$
By hypothesis (\ref{G_isuniversal}) there exists also a derivation $D_1$ 
of the form $D_1: S_1 \to^*_P s S_1$. Let us choose
$$D':= R_{01} \otimes D_1.$$ 
Since $D': S_0a_1 \to^*_\RR s S_1a_1$ is $\bu(1)$, the only minimal right-overlap is resolved.
By Proposition \ref{p-bu-criterium}, it follows that $\RR$ is $\BU(1)$.
\end{proof}
Let us prove now Proposition~\ref{p-bu1_words_undecidable}.
\begin{proof}
By Lemma~\ref{l-bu1-implies-universal} and Lemma~\ref{l-universal-implies-bu1}, $G \mapsto \RR$ is a 
many-one reduction of $\MUP$ 
to the problem whether a finite length-increasing semi-Thue system is $\BU(1)$ or not.
This last problem is thus undecidable.
\end{proof}
We treat now the case of $\BU(k)$ for an arbitrary $k \geq 1$.
\begin{theorem}
For every $k \geq 1$ , it is undecidable whether a finite length-increasing semi-Thue system $\RR$ is $\BU(k)$ (resp. $\BU^-(k)$) or not. 
\label{t-bu_words_undecidable}
\end{theorem}
\noindent (Note that for semi-Thue systems, since every variable of a lhs of rule must appear in the
corresponding rhs, the properties $\BU(k)$, $\BU^-(k)$ are equivalent).
\begin{proof}(sketch)
Let $k \geq 1$.
Given an instance $G$ of $\MUP$, we construct an alphabet
 $Y_k:= Z \cup N \cup \{S_{0,1}, \ldots, S_{0,k},a_1,a_2,\ldots,a_k,b\},$ and a semi-Thue system 
$\RR$ over $Y_k$ 
consisting  of the union of $P$ with the two new rules :
$$S_{0,1} \to z_1 S_{0,1},\;\;S_{0,1} \to z_2 S_{0,1},$$
and the $k$ additional rules:
$$S_{0,1}a_1 \to S_{0,2}a_1,\;\; S_{0,2}a_1a_2 \to S_{0,3}a_1a_2,
\ldots,\;\; S_{0,k}a_1a_2\cdots a_k \to S_1a_1a_2 \cdots a_k.$$
One can check, by arguments similar to those used in the proof of 
Proposition~\ref{p-bu1_words_undecidable} that $\RR$ is $\BU(k)$ iff $\forall s \in Z^*, S_1 \to^*_P sS_1$. 
Hence the property $\BU(k)$ is undecidable.
\end{proof}

\subsection{Undecidability for term rewriting systems}

\begin{theorem}
For every $k \geq 1$, the problem to
determine whether a finite linear term rewriting system $(\RR,\FF)$ is $\BU(k)$ (resp. $\BU^-(k)$)
or not, is undecidable.
\label{t-bu_undecidable}
\end{theorem}
\begin{proof}
For every finite semi-Thue system $\RR$ the corresponding term rewriting system 
$\hat{\RR}$ (defined in \S \ref{words_as_terms}) is finite and linear. Moreover, by point 2 of definition \ref{d-property-P},  
$\RR$ is $\BU(k)$ iff $\hat{\RR}$ is $\BU(k)$. 
Hence this theorem is a straightforward corollary 
of Theorem~\ref{t-bu_words_undecidable}. 
\end{proof}

\section{Strongly Bottom-up systems}
\label{s-strongly-bottom-up}

Since the $\BU(k)$ conditions are, as such, undecidable (Theorem~\ref{t-bu_undecidable}),
we are lead to define some stronger but {\em decidable} conditions.
We study in \S \ref{ss-sbusystems} the \emph{strongly bottom-up} ( $\SBU$ for short) restriction.
We introduce in \S \ref{ss-sticking-out-graph} a technical tool that will be used in
\S \ref{ss-sufficient-for-sts} and \S\ref{ss-sufficient-for-trs}  for giving a polynomially decidable condition implying condition $\SBU$. 

\subsection{Strongly bottom-up systems}
\label{ss-sbusystems}
We abbreviate strongly bottom-up to $\sbu$.

\begin{definition}
A system $(\RR,\FF)$ is said \emph{$\SBU(k)$} iff\\
for every derivation $D: s \to^*_\RR t$,
from a term $s\in \TT(\FF)$ to a term $t \in \TT(\FF)$,
$$ D 
\mbox{ is } \wbu \Leftrightarrow D 
\mbox{ is } \bu(k).$$
We denote by $\SBU(k)$ the class of $\SBU(k)$ systems and by
$\SBU = \bigcup_{k \in \N} \SBU(k)$ the class of \emph{strongly bottom-up} systems.
\label{d-sbuk}
\end{definition}
\noindent In other words: instead of requiring that the binary relations $\to^*_\RR$ and 
$\tok^*_\RR$ over $\TT(\FF)$ are equal, 
we require that {\em all} $\wbu$ marked derivations starting on an unmarked term use
only marks smaller or equal to $k$.
The following lemma is obvious.
\begin{lemma}
Every $\SBU(k)$ system is $\BU(k)$.
\end{lemma}
This stronger condition over term rewriting systems is interesting because of the following
property.
\begin{proposition}
For every $k \geq 0$, it is decidable whether a finite term rewriting 
system $(\RR,\FF)$ is $\SBU(k)$.
\label{t-sbu_decidable}
\end{proposition}
\begin{proof}
Note that every marked derivation starting from some $s \in \TT(\FF)$ and 
leading to some
$\barre{t} \in \TT(\FF^\N) \setminus \TT(\FF^{\leq k})$ must decompose as
$$s \toroak_\RR^* \barre{s}' \toro_\RR^*\barre{t}, $$
with $\barre{s}' \in \TT(\FF^{\leq {k+1}})\setminus \TT(\FF^{\leq {k}}).$
A necessary and sufficient condition for $\RR$ to be $\SBU(k)$ is thus that:
\begin{equation}
(\toroak_\RR^*)[\TT(\FF^{\leq {k+1}}) \setminus \TT(\FF^{\leq {k}})]) \cap \TT(\FF) 
= \emptyset.
\label{e-THE_SBUk_criterium}
\end{equation}
By Theorem~\ref{t-kbu-inverse-preserving} the left-handside of equality (\ref{e-THE_SBUk_criterium})
is a recognizable set for which we can construct a \fta; we then just have  to test
whether this \fta recognizes the empty set or not.  
\end{proof}
According to the results of \cite{Kna-Cal99} it seems likely that 
the property [$\exists k \geq 0$ such that $(\RR,\FF)$ is $\SBU(k)$ ]
is undecidable for term rewriting systems.
It is then interesting to look for a decidable sufficient condition.
Our condition is based on a finite graph that we define in next subsection.

\subsection{The sticking-out graph $\SG(\RR)$}
\label{ss-sticking-out-graph}
Let us associate with every Term Rewriting System a {\em graph} whose vertices are the rules
of the system and whose arcs $(R,R')$ express some kind of overlap between 
the right handside of $R$ and the left handside of $R'$.
Every arc has a {\em label} indicating the category of overlap that occurs
and a {\em weight} which is an integer ($0$ or $1$). The intuitive meaning of the weight
is that any derivation step using the corresponding overlap would increase some mark by this weight.
The precise graph is defined below and is directly inspired by the one of \cite{Tak-Kaj-Sek10}, 
though slightly different. 
\begin{definition}
Let $s \in \TERM$, $t \in \TERM \setminus \VV$ and $w \in \Posv(t)$.
We say that $s$ \emph{sticks out of} $t$ \emph{at} $w$ if
\begin{enumerate0}
\item $\forall v \in \Pos(t)$ s.t. $\nopos \preceq v \prec w$, $v \in \Pos(s)$ and 
$s(v)= t(v)$.
\item $w \in \Pos(s)$ and $\sto{s}{w} \not\in \GTERM$.
\end{enumerate0}
If in addition $\sto{s}{w} \not\in \VV$ then
$s$ \emph{strictly} sticks out of $t$ at $w$.
\end{definition}


\begin{definition}
Let $\RR = \{ l_1 \to r_1, \ldots, l_n \to r_n \}$ be a system.
The \emph{sticking-out graph} is the directed graph $\SG(\RR) = (V,E)$ where 
$V = \{ 1, \ldots , n\}$ and $E$ is defined as follows:
\begin{enumerate}[label=\emph{\alph*)}]
\item if $l_j$ strictly sticks out of a subterm of $r_i$ at $w$, $i \stackrel{(a)}{\to} j \in E$;
\item if a strict subterm of $l_j$ strictly sticks out of $r_i$ at $w$, $i \stackrel{(b)}{\to} j \in E$;
\item if a subterm of $r_i$ sticks out of $l_j$ at $w$, $i \stackrel{(c)}{\to} j \in E$;
\item if $r_i$ sticks out of a strict subterm of $l_j$ at $w$, $i \stackrel{(d)}{\to} j \in E$.
\end{enumerate}
\end{definition}

\begin{figure}[htb]
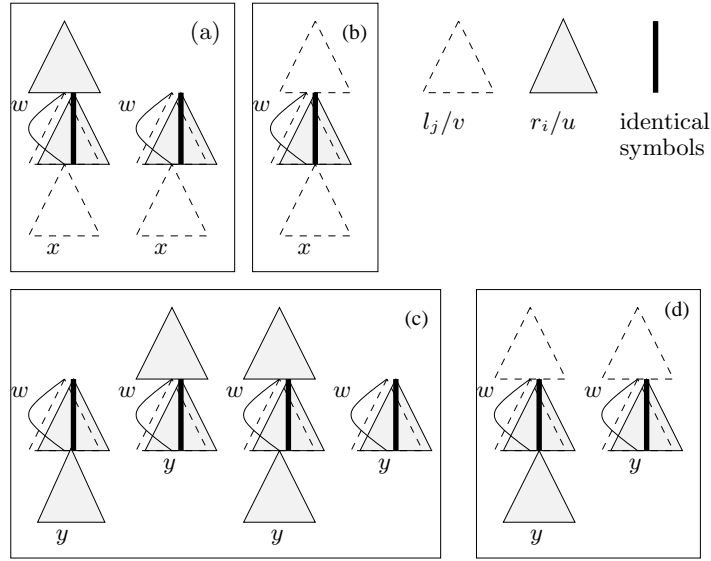

\centering
\input superpositions.pstex_t
\caption{Sticking-out cases}
\label{f-superpositions}
\end{figure}

Figure~\ref{f-superpositions} shows all the possibilities in the four categories
$(a), (b), (c), (d)$.

\begin{example}
The graph of the system $\RR_0 = \{ \mF(\mF(x)) \to \mF(x) \}$ contains one vertex
and two loops labeled $(d)$ and $(a)$.\\
It can be shown with an \emph{ad hoc} proof that $\RR_0 \in \BU$ (actually
in $\BU^-(1)$).
We have already seen in Example~\ref{ex-not-sbu} that $\RR_0 \not\in \SBU$.
\end{example}

\begin{example}
\label{ex-not-epr}
The graph of system $\RR_4 = \{ \mG(\mF(\mG(x))) \to \mF(x) \}$ contains one vertex and
a simple loop labeled $(b)$. $\RR_4$ is not inverse recognizability preserving:
$$(\to^*_{\RR_4})[\{f(a)\}]= \{ g^n(f(g^n(a))) \mid n \geq 0 \},$$
which is not recognizable.
\end{example}
The \emph{weight} of each arc of $\SG(\RR)$ is defined by:
\begin{itemize0}
\item arcs $(a)$ or $(b)$ have weight 1,
\item arcs $(c)$ or $(d)$ have weight 0.
\end{itemize0}
\noindent

\noindent
The \emph{weight of a path} in the graph is the sum of the weights of its arcs.
The \emph{weight of a graph} is the maximal weight of a path in the graph; it is infinite
if the graph contains a cycle with an arc of weight $1$.

\begin{figure}[htb]
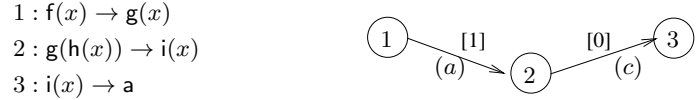

\centering
\input sg-r2.pstex_t
\caption{The sticking-out graph of $\RR_1$}
\label{f-sg-r2}
\end{figure}

The sticking-out of $\RR_1$ of Example~\ref{ex-1} is given in Figure~\ref{f-sg-r2}.

\subsection{A sufficient condition for semi-Thue systems}
\label{ss-sufficient-for-sts}
Let us fix a semi-Thue system $\RR$ over an alphabet $Y$.
The main result of this subsection is that, if every path of $\SG(\RR)$ has a weight
$\leq k$, then $\RR$ has the property $ \SBU(k+1)$.
We prove some lemmas establishing some links between $\wbu$ derivations, on one hand, and
paths of $\SG(\RR)$, on the other hand.
Again, we use the notation defined in \S \ref{s-criterium_for_noteriansemiThuesystems}
for manipulating derivations.
\begin{figure}[htb]
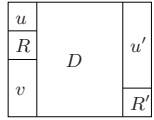

\centering
\scalebox{0.80}{
\input down-derivation.pstex_t
}
\caption{Downwards derivation}
\label{f-down-derivation}
\end{figure}
\begin{lemma}(Downwards derivations)\\
Let $R,R' \in \RR, u,v,u' \in Y^*$ and $D$ a derivation such that
$uRv \otimes D \otimes u'R'$ is a $\wbu$-derivation.
Then, there exists a path from $R$ to $R'$ in $\SG(\RR)$.\\
(See figure \ref{f-down-derivation}).
\label{l-down-derivation-implies-path}
\end{lemma}
\begin{proof}
Let us consider the following property $P(n)$:\\
for every $\wbu$-derivation $(u_iR_iv_i)_{0 \leq i \leq n+1}$ for the system $\RR$,\;
if $|v_{n+1}|=0$ then there exists a path from $R_0$ to $R_{n+1}$ in $\SG(\RR)$.\\
We show by induction over $n$ that, for every $n \in \N$, $P(n)$ holds.\\
{\bf Basis}: $n=0$.\\
We thus have $|v_0|+ |\partial^+(R_0)|+ |u_0|=|\partial^-(R_1)| +|u_1|$.
Since this derivation is $\wbu$ we also have $|v_0|< |\partial^-(R_1)|$ or
$|v_0|=|\partial^-(R_1)|=0$.
From these inequalities it follows that $(R_0,R_1)$ is an edge of
$\SG(\RR)$.\\
{\bf Induction step}: $n \geq 1$\\
We define
$$i:= \min \{j \in [1,n+1]\mid |v_{j}| < |v_0|+|\partial^+(R_0)|\}.$$
Since the given derivation is $\wbu$, $|v_0|<|v_i|+|\partial^-(R_i)|$ or 
($|v_0|=|v_i|$ and $|\partial^-(R_i)|=0$).
Hence  $(R_0,R_i)$ is an edge of $\SG(\RR)$ and $(u_jR_jv_j)_{i \leq j \leq n+1}$ 
is a $\wbu$-derivation fulfilling 
$|v_{n+1}|=0$. By induction hypothesis, there exists a path $p$
from $R_i$ to $R_{n+1}$ in $\SG(\RR)$. The edge $(R_0,R_i)$ 
followed by the path $p$ is a path from $R_0$ to $R_{n+1}$ in $\SG(\RR)$.\\ 
Let $R,R',u,v,u',D$ fulfill the hypothesis of the lemma.
Let us note:
$$u_0 :=u,\; R_0 :=R,\;v_0 := v,\; D:= (u_iR_iv_i)_{1 \leq i \leq n},\;
u_{n+1}:=u',\;R_{n+1}:=R',\;v_{n+1}:=\varepsilon.$$
Applying $P(n)$ to the derivation $(u_iR_iv_i)_{0 \leq i \leq n+1}$, we obtain the conclusion of the lemma.
\end{proof}
\begin{lemma}(Strict Downwards derivations)\\
Let $R,R' \in \RR, u,v,u' \in Y^*$ and $D$ a derivation such that
$uRv \otimes D \otimes u'R'$ is a $\wbu$-derivation and $|v| \geq 1$.
Then, there exists a path with non-null weight from $R$ to $R'$ in $\SG(\RR)$.
\label{l-strict-down-derivation-implies-pospath}
\end{lemma}
\begin{proof}
Let us consider the following property $Q(n)$:\\
for every $\wbu$-derivation $(u_iR_iv_i)_{0 \leq i \leq n+1}$ for the system $\RR$,\;
if $|v_0| \geq 1$ and $|v_{n+1}|=0$, then there exists a path with non-null weight from $R_0$ to $R_{n+1}$ in $\SG(\RR)$.\\
We show by induction over $n$ that, for every $n \in \N$, $Q(n)$ holds.\\
{\bf Basis}: $n=0$.\\
$Q(0)$: we thus have $|v_0|+ |\partial^+(R_0)|+ |u_0|=|\partial^-(R_1)| +|u_1|$.
Since this derivation is $\wbu$ we also have $|v_0|< |\partial^-(R_1)|$.
From these inequalities it follows that $(R_0,R_1)$ is an edge of type (a) or (b) of
$\SG(\RR)$.\\
{\bf Induction step}: $n \geq 1$.\\
We define
$$i:= \min \{j \in [1,n+1]\mid |v_{j}| < |v_0|+|\partial^+(R_0)|\}.$$
{\bf case 1}: $|v_i|\geq |v_0|$.\\
In this case $(R_0,R_i)$ is an edge of $\SG(\RR)$ and
$(u_jR_jv_j)_{i \leq j \leq n+1}$ is a $\wbu$-derivation fulfilling 
$|v_i| \geq 1$ and $|v_{n+1}|=0$. Hence, by induction hypothesis, there exists a path $p$
from $R_i$ to $R_{n+1}$, with non-null weight, in $\SG(\RR)$. The edge $(R_0,R_i)$ 
followed by the path $p$ is a path with non-null weight from $R_0$ to $R_{n+1}$.\\
{\bf case 2}: $|v_i|< |v_0|$.\\
In this case, since the given derivation is $\wbu$, $|v_0|< |v_i|+|\partial^-(R_i)|$.
Hence  $(R_0,R_i)$ is an edge of weight 1 of $\SG(\RR)$. 
By lemma \ref{l-down-derivation-implies-path} there exists a path $p$ from $R_i$ to $R_{n+1}$ in
$\SG(\RR)$. We can conclude as in case 1.\\
From $Q(n)$ we can deduce the lemma.
\end{proof}

\begin{figure}[htb]
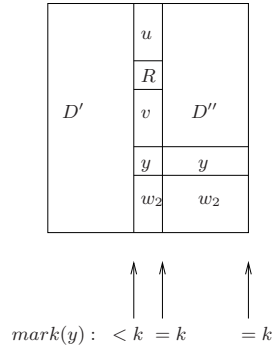

\centering
\scalebox{0.80}{
\input history-mark.pstex_t
}
\caption{History of a mark}
\label{f-history-mark}
\end{figure}

\begin{lemma}{(History of a mark)}\\
Let $D$ be some marked $\wbu$-derivation and let $y \in Y,w_1,w_2 \in Y^*$ 
such that $\partial^-(D)$ is unmarked,
$\partial^+(D)=w_1yw_2$ and the mark of $y$ in the corresponding marked
word is $k>0$.
Then, there exist $u,v \in Y^*, R \in \RR$ and some derivations $D',D''$ such that\\
1- $D = D' \otimes uRvyw_2 \otimes D''yw_2 $\\
2- the mark of $y$ in every step of $D''yw_2$ is $k$ \\
3- the mark of $y$ in $\partial^+(D')$ is $<k$. (See figure \ref{f-history-mark})
\label{l-history-of-mark}
\end{lemma}
\begin{proof}
Let us remark that every derivation $D$ fulfilling the hypothesis of the lemma must have
a length $\ell(D)=n+1$ for some integer $n \geq 0$ (since its result $\partial^+(D)$ has some
non-null mark).
We prove the lemma by induction on this integer $n$.\\
{\bf Basis}: $n=0$.\\
Thus $D= v_1Rv_2 $ for some $v_1,v_2 \in Y^*, R \in \RR$.
Since the given occurrence of $y$ has a non-null mark, it must be a position of $v_2$.
It follows that $D=uRvyw_2$ for some words $u,v \in Y^*$.
Let us define:
$$D':= D_{u\partial^-(R)vyw_2},\; D'':= D_{w_1}.$$
These derivations fulfill conclusions (1-3) of the lemma.\\
{\bf Induction step}: $n \geq 1$.\\
By the same arguments, $D= E \otimes uRvyw_2 $ for some $u,v \in Y^*, R \in \RR$ and some derivation $E$ of length $n$.\\
{\bf Case 1}: The mark of $y$ in $\partial^+(E)= u\partial^-(R)vyw_2$ is $k$.\\
By induction hypothesis $E$ has some decomposition as
$E = E' \otimes u'R'v'yw_2 \otimes E''yw_2 $ such that
the mark of $y$ in every step of $E''yw_2$ is $k$ \\
and the mark of $y$ in $\partial^+(E')$ is $<k$. 
Taking $D':= E'$ and $D'':= E'' \otimes uRv$, the conclusion of the lemma is fulfilled.\\
{\bf Case 2}: The mark of $y$ in $\partial^+(E)= u\partial^-(R)vyw_2$ is $< k$.\\
Taking $D':= E$ and $D'':= D_{u\partial^+(R)v}$, the conclusion of the lemma is fulfilled.\\
\end{proof}
Let $\RR$ be a semi-Thue system and $k \in \N$. We consider the
following property $\Path(k)$: for every $v_1,w_2 \in Y^*,R' \in \RR$ and $\wbu$-derivations $D,E$ such that
\begin{equation}
E=D \otimes v_1R'w_2 \;\myand \;\ma{\last(v_1\partial^-(R'))}=k
\label{e-hypo-pathk}
\end{equation}
there exists a path in $\SG(\RR)$ with weight $\geq k$ and with extremity $R'$.\\
\begin{lemma}
Let $\RR$ be a semi-Thue system. For every $k \in \N$, the property $\Path(k)$ holds.
\label{l-PATH}
\end{lemma}
\begin{figure}[htb]
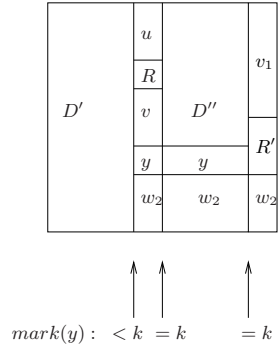

\centering
\scalebox{0.80}{
\input marks-and-weights.pstex_t
}
\caption{From marks (in derivations) to weights (in paths)}
\label{f-marks-and-weights}
\end{figure}
\begin{proof}
We prove by induction over $k \in \N$ the statement\\
$$\forall k \in \N,  \Path(k).$$
{\bf Basis}: $k=0$\\
There exists a path of length $0$, thus of weight $\geq 0$, in $\SG(\RR)$, with extremity $R'$.\\
{\bf Induction step}: $k\geq 1$\\
Let us assume (\ref{e-hypo-pathk}).
Applying  lemma \ref{l-history-of-mark} to the derivation $D$, to the letter
$y= \last (v_1\partial^-(R'))$ and to the words $w_1:= v_1\partial^-(R')y^{-1}$, $w_2$,
we obtain $u,v,R ,D',D''$ such that:\\
$$E=D' \otimes uRvyw_2 \otimes D''yw_2 \otimes v_1R'w_2,$$
the mark of $y$ in every step of $D''yw_2$ is $k$ and the mark of $y$ in $\partial^+(D')$ is $<k$
(see figure \ref{f-marks-and-weights}).
By the definition of a marked rewriting-step, we must have:
$$k= M(\barre{u}\barre{l},x)$$
where $R=\rrule{l}{r}$, $x$ is the variable of $l$ and $\barre{u}$ is the marked word corresponding
to the context where $R$ is applied.
Let us consider $E':=D' \otimes uRvyw_2$. It fulfills 
$$E'=D' \otimes uRvyw_2\;\myand \;\ma{\last(v\partial^-(R))}=k-1$$
By induction hypothesis, there exists a path $p$ in $\SG(\RR)$ with weight $\geq k-1$ and with extremity $R$; by lemma \ref{l-strict-down-derivation-implies-pospath}, there exists a path $q$ with non-null weight from $R$ to $R'$ in $\SG(\RR)$. The concatenation $p \cdot q$ is a path with weight $\geq k$ in $\SG(\RR)$.
\end{proof}
\begin{proposition}
Let $\RR$ be a semi-Thue system and $k \geq 1$.
If $W(\SG(\RR))=k-1$ then $\RR \in \SBU(k)$.
\label{p-weightk_implies_sbuk1}
\end{proposition}
\begin{proof}
Suppose that $\RR \notin \SBU(k)$. This means that some $\wbu$-derivation (w.r.t. $\hat{\RR}$) 
starting from a non-marked (unary) term over $Y \cup \{\#\}$ reaches a marked term with the 
mark $k+1$.
Let $\hat{E}$ be a $\wbu$ derivation (w.r.t. $\hat{\RR}$) with minimal length reaching the mark $k+1$.Let us consider the derivation $E$ (w.r.t. $\RR$) corresponding to $\hat{E}$ (it is obtained from $\hat{E}$ just by erasing all occurrences of the nullary symbol $\#$).
This  derivation $E$ must have a decomposition of the form (\ref{e-hypo-pathk}). 
By lemma \ref{l-PATH} $\Path(k)$ holds, hence
there exists a path in $\SG(\RR)$ with weight $\geq k$. By contraposition,
if $W(\SG(\RR))\leq k-1$ then $\RR \in \SBU(k)$, which proves the proposition.
\end{proof}
\subsection{A sufficient condition for term rewriting systems}
\label{ss-sufficient-for-trs}
\begin{proposition}
Let $\RR$ be a linear system and $k \geq 1$.
If $W(\SG(\RR)) = k-1$ then $\RR \in \SBU(k)$.
\label{p-weightk_implies_sbuk2}
\end{proposition}
\begin{proof}(Sketch)
Let us associate to $\RR$ the semi-Thue system $T$ corresponding to the ``branch-rewriting''
induced by $\RR$: it consists of all the rules 
$$\rrule{u}{v} \in \FF^* \times \FF^* $$
such that there exists a rule $\rrule{l}{r}\in \RR$,
and a variable $x \in \VV$ , such that $ux$  labels a  branch of $l$  and
$vx$  labels a branch of $r$. 
Suppose that the mark $k+1$ appears in a $\RR$-derivation.
Since the marking-mechanism is defined branch by branch, the mark $k+1$ also appears in a $T$-derivation. By Proposition \ref{p-weightk_implies_sbuk1}, there exists
a path in $\SG(T)$  with weight $\geq k$.
Let us fix some total ordering on $\RR$ and define the map $h: T \rightarrow \RR$ by:
$$ h(\rrule{u}{v}) = \rrule{l}{r} $$
iff $\rrule{l}{r}$ is the smallest rule of $\RR$ such that $ux$ (resp. $vx$) labels
a branch of $l$ (resp. $r$) and $x$ is a variable.
This map $h$ is an homomorphism of labelled graphs from $\SG(T)$ to $\SG(\RR)$, i.e. it is compatible with the labels. It follows that it is also compatible with the weights.
Hence there exists a path of weight $\geq k$ in $\SG(\RR)$.
 \end{proof}
\begin{corollary}
\label{c-finite-weight-bu}
Let $\RR$ be a linear system.
If $W(\SG(\RR))$ is finite then $\RR \in \SBU$.
\label{p-weightfinite_implies_sbu}
\end{corollary}

\begin{proposition}
\label{p-lfpo-bu}
$\LFPO^{-1} \subsetneq \SBU$.
\end{proposition}

\begin{proof}
Let $\RR \in \LFPO^{-1}$.
By definition the sticking-out graph of \cite{Tak-Kaj-Sek10} does
not contain a cycle of weight $1$, hence
 from corollary~\ref{c-finite-weight-bu}, $\RR \in \SBU$. So $\LFPO^{-1} \subseteq \SBU$.
$\RR_0 \in \SBU$ but $\RR_0 \not \in \LFPO^{-1}$. We conclude that $\LFPO^{-1} \subsetneq \SBU$.
\end{proof}

\begin{example}
Let $\RR_5 = \{ \mF(\mG(x),\mA) \to \mF(x,\mB) \}$.
$\RR_5 \not\in \LFPO^{-1}$ as $\SG(\RR)$ contains a loop $(a)$ so a loop of weight $1$.
It is easy to show by an ad-hoc proof that $\RR_5 \in \SBU^-(1)$.
However our sufficient condition is not able to capture $\RR_5$.
\end{example}

\begin{corollary}
\label{c-lfpo-bu}
$\LFPO^{-1} \subsetneq \SBU \subsetneq \BU$.
\end{corollary}

\section{Perspectives}
Here are some natural perspectives of development for this work:\\
\begin{enumerate}
\item the method developed here for the sake of showing a property of recognizability
preservation might be used, also, for testing some termination properties; this idea is
implemented in \cite{Dur-Sen-Syl10,Syl10} 
\item it is tempting to extend the notion of \emph{bottom-up} rewriting (resp. system) to left-linear but non right-linear systems. This class would extend the class of  growing systems 
studied in \cite{NT02};this idea is implemented in \cite{Dur-Syl11}.
\item a dual notion of \emph{top-down} rewriting and a corresponding class of
top-down systems should be defined;
this class would presumably extend the class of Layered Transducing 
systems defined in \cite{STFK02}.
\item we know that the condition $\BU(k)$ is undecidable (for every $k \geq 1$)
and that the condition $\SBU(k)$ is decidable (for every $k \geq 1$);
whether the condition $\SBU$ is decidable is thus a natural question;
\item the systems considered in \cite{GHW04} and the systems considered here might be treated 
in a unified manner; such a unified approach should lead to an even larger class of rewriting systems 
with still good algorithmic properties.  
\end{enumerate}
Some work in directions 1,2,3 has been undertaken by the authors.
\paragraph{Acknowledgements}
We thank Marc Sylvestre for his constructive criticism of a previous version of \S \ref{s-basic-construction}, in particular for correcting the construction of subsection \ref{s-variable-lhs}. We also thank the referees of RTA'07 and a referee of JSC for their constructive criticism.

\bibliographystyle{elsarticle-harv}
\bibliography{references}

\end{document}